\documentclass[12pt]{report}
\usepackage{graphicx}
\usepackage{subcaption}
\usepackage{subcaption,graphicx}
\newcommand{\rulesep}{\unskip\ \vrule\ }
\usepackage{amsthm,amssymb,mathrsfs,setspace,booktabs}
\usepackage{mathtools,amsmath,nccmath}
\usepackage{tabularx} 
\usepackage{booktabs} 
\usepackage{caption} 
\usepackage{siunitx} 
\usepackage{adjustbox} 
\usepackage{multirow}
\usepackage{booktabs}
\usepackage{graphicx}
\usepackage{pstricks} 
\usepackage{tikz} 
\usepackage{enumitem}
\usepackage{hyperref} 
\newcommand{\etal}{\textit{et al.}}
\usepackage{appendix}
\usepackage{caption} 
\usepackage[style=numeric,sorting=none]{biblatex}
\usepackage{fancyhdr}
\usepackage{multicol} 
\usepackage{geometry} 
\usepackage{xcolor} 
\hypersetup{
    colorlinks,
    linkcolor={blue!50!black},
    citecolor={blue!50!black},
    urlcolor={blue!80!black}
}
\usepackage{fancyhdr}
\theoremstyle{plain}

\theoremstyle{definition}

\theoremstyle{remark}

\renewcommand{\today}{\ifcase \month \or January\or February\or March\or April\or May%
\or June\or July\or August\or September\or October\or November\or December\fi\:%
\number \year} 

\usepackage{longtable}
\newcommand\listsymbolname{Abbreviations}
\newcommand\listofsymbols[2]{
\btypeout{\listsymbolname}
\addtotoc{\listsymbolname}
    \chapter*{\listsymbolname
      \@mkboth{
          \MakeUppercase\listsymbolname}{\MakeUppercase\listsymbolname}}
\begin{longtable}[c]{#1}#2\end{longtable}\par
    \cleardoublepage
}
\begin{document}
\begin{titlepage}
\enlargethispage{3cm}

\begin{center}
\textbf{\Large Advancing Medical Image Segmentation Through Multi-Task and Multi-Scale Contrastive Knowledge Distillation}\\[10pt]

\vspace*{0.5cm}

{\large \bf Dissertation }

Submitted in partial fulfillment of the requirements of the \\
M.Tech Data Science and Engineering Degree Program \\
\vspace{5mm}
By \\
{ \bf Risab Biswas \\
(2021SC04063) 
} \\

                      \vspace{10mm}
                   {\em  Under the Supervision of} \\ \vspace{3mm}
             {\textbf{Dr. Chaitanya Kaul, Research Associate,}\\ Inference, Dynamics, and Interaction Group, \\ School of Computing, University of Glasgow, \\ Scotland, United Kingdom, G12 8RZ} \\

\vfill
\vfill

\begin{figure}[h]
  \begin{center}
  \includegraphics[height=36mm]{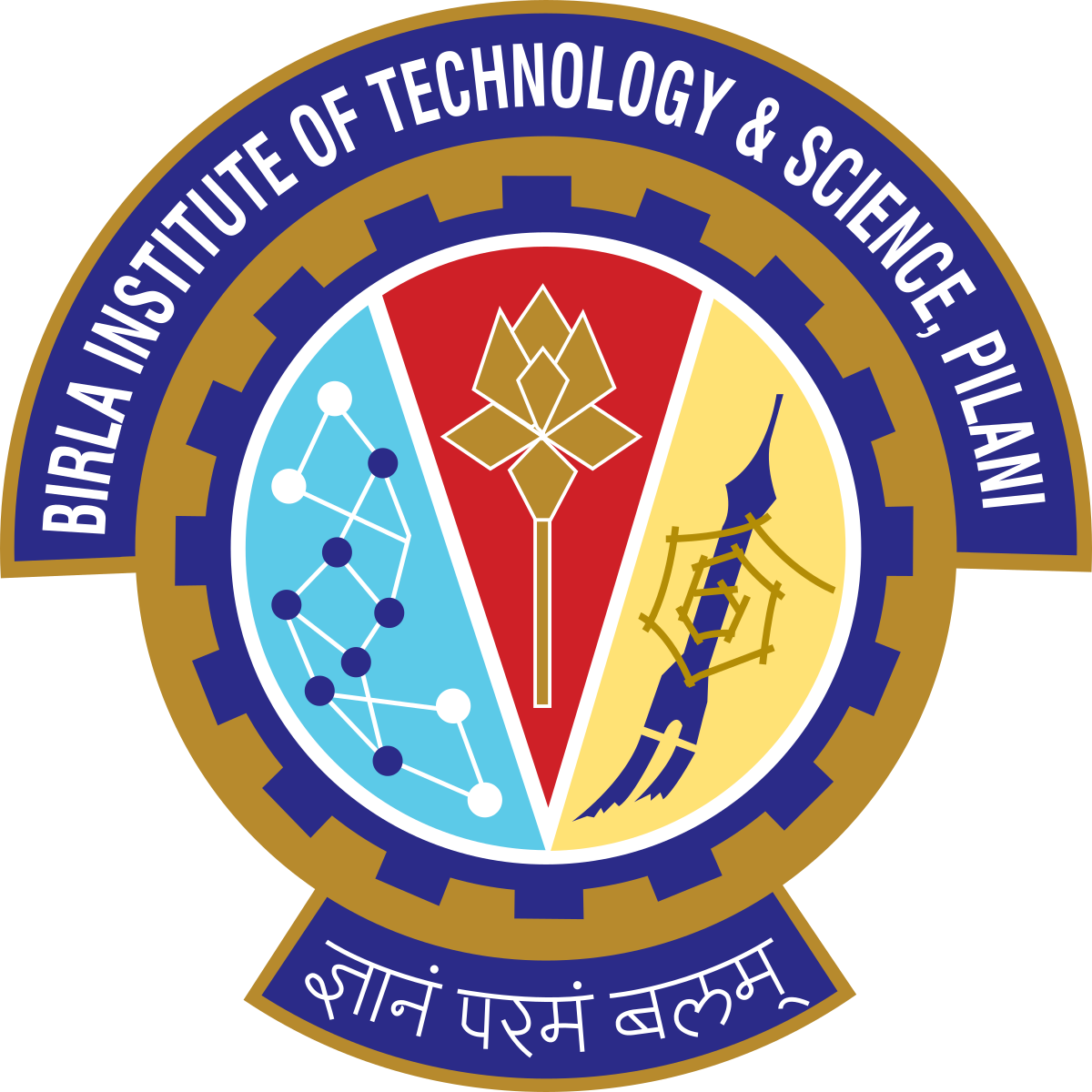}
  \end{center}
\end{figure}
\vspace*{0.2cm}

{\bf BIRLA INSTITUTE OF TECHNOLOGY AND SCIENCE \\Pilani (Rajasthan)\\
INDIA \\
}
{\it \today}
\end{center}
\end{titlepage}

\clearpage

\begin{titlepage}
    \centering
    \vspace*{\fill}
    \Huge
    \textit{Dedication}\\
    \large
    \emph{To my parents, my sister and the love of my life...}\\[2cm]
    \vspace*{\fill}
\end{titlepage}

\begin{center}
{\Large{\bf{DECLARATION}}}
\end{center}

\noindent

I, \textbf{Risab Biswas (Roll No: 2021SC04063)}, hereby declare that, this report entitled \textbf{``Advancing Medical Image Segmentation Through Multi-Task and Multi-Scale Contrastive Knowledge Distillation”} submitted to Birla Institute of Technology and Science, Pilani, India towards the partial requirement of \textbf{Master of Technology} in \textbf{Data Science and Engineering}, is an original work carried out by me under the supervision of \textbf{Dr. Chaitanya Kaul} and has not formed the basis for the award of any degree or diploma, in this or any other institution or university. I have sincerely tried to uphold academic ethics and honesty. Whenever a piece of external information, statement or result is used then, that has been duly acknowledged and cited.

\vspace{4cm} 

\noindent Pilani (Rajasthan) \hfill \textbf{Risab Biswas}

\noindent \today \hfill

\clearpage

\begin{center}
{\large{\bf{\underline{ACKNOWLEDGEMENT}}}}
\end{center}

\noindent
I would like to express my heartfelt appreciation to all those who contributed to the successful completion of my project. First and foremost, I would like to extend my profound gratitude to Dr. Chaitanya Kaul from the School of Computing at the University of Glasgow. His visionary guidance, invaluable feedback, and unwavering support have been instrumental throughout this endeavour. I am incredibly fortunate to have had such a remarkable mentor for my master's thesis. Chaitanya ensured that I thoroughly understood the project, patiently explaining concepts and generously dedicating time to address my inquiries every week. Our discussions have not only impacted my professional growth but also enriched my personal life. His mentorship has significantly shaped me as a researcher, laying a solid foundation for my future endeavours as a prospective PhD student. I must add that he has also shaped me as a better individual and has had a great positive impact on me.

I also extend my gratitude to my examiner, Dr. Sankara Nayaki K, whose insightful suggestions contributed to the refinement of the thesis.  Her experience in the medical imaging domain has helped me bring clarity to this thesis. Her meticulous attention to detail and constructive feedback have not only enhanced the quality of my thesis but have also honed my analytical skills.

I am greatly thankful to Mr. Anupam Nandwana, CEO of the organisation where I work, for helping me manage my workload during my master's degree. Anupam has been a constant source of motivation for me ever since I completed my undergraduate studies. I wish to tell everybody that a great boss like him can transform your professional journey into an inspiring adventure. 

I want to express my gratitude to my parents for their endless support throughout my academic journey. Their encouragement and unwavering faith in my abilities have always motivated me. Moreover, I am deeply thankful to my beloved partner, Koninika, for being my strongest support system during my master's studies.

\vspace{1cm} 

\noindent Pilani (Rajasthan) \hfill \textbf{Risab Biswas}

\noindent \today \hfill

\clearpage
\begin{flushleft}
    \setlength{\parskip}{0pt}
    {\centering{{\large{\bf{\underline{ABSTRACT}}}}} \par}
\end{flushleft}
This thesis aims to investigate the feasibility of knowledge transfer between neural networks for medical image segmentation tasks, specifically focusing on the transfer from a larger multi-task ``Teacher" network to a smaller ``Student" network. In the context of medical imaging, where the data volumes are often limited, leveraging knowledge from a larger pre-trained network could be useful. The primary objective is to enhance the performance of a smaller student model by incorporating knowledge representations acquired by a teacher model that adopts a multi-task pre-trained architecture trained on CT images, to a more resource-efficient student network, which can essentially be a smaller version of the same, trained on a mere 50\% of the data than that of the teacher model. 

To facilitate knowledge transfer between the two models, we devised an architecture incorporating multi-scale feature distillation and supervised contrastive learning. Our study aims to improve the student model's performance by integrating knowledge representations from the teacher model. We investigate whether this approach is particularly effective in scenarios with limited computational resources and limited training data availability. To assess the impact of multi-scale feature distillation, we conducted extensive experiments. We also conducted a detailed ablation study to determine whether it is essential to distil knowledge at various scales, including low-level features from encoder layers, for effective knowledge transfer. In addition, we examine different losses in the knowledge distillation process to gain insights into their effects on overall performance.

\vfill
\clearpage

\tableofcontents
\clearpage
\listoffigures
\listoftables
\newpage

\pagenumbering{arabic}
\setcounter{page}{1}

\chapter{Introduction} 
\label{chapter:introduction}
Medical image segmentation (MIS) is an important part of medical image processing because it provides detailed insights by automatically delineating organs and tumours at the pixel level. MIS presents unique challenges due to the diverse appearance of organic structures, limited data availability, and varying responses to contrast agents. Medical image segmentation is a challenging task, especially when compared to segmenting natural images, due to the presence of tiny lesions that are crucial for diagnosis. Conventional architectures like Fully Convolutional Networks (FCN)~\cite{long2015fully} face difficulties in this domain as they rely on pretraining on ImageNet, which doesn't effectively capture medical nuances. Although fine-tuning FCN for medical segmentation has been attempted, the lack of intermediate feature utilization hampers the segmentation performance. The main reason behind this could be the use of symmetric encoder-decoder architecture by FCN, which may not efficiently capture and fuse multi-scale information. The information flow might not be optimal for precise segmentation. The U-Net~\cite{ronneberger2015u} architecture, transformed MIS by allowing for the extraction of meaningful features critical to accurate diagnosis. Despite their effectiveness, U-Net and its variants have limitations, necessitating further research into enhancements. To improve segmentation accuracy, various approaches have been explored, including attention mechanisms and multi-scale features. In addition, lightweight networks have been developed to balance computational efficiency and accuracy, which is critical for real-time applications. However, challenges such as computational demands and data scarcity continue to exist.

\section{Problem Statement}
Medical image segmentation tasks often face challenges due to limited data volumes and computational resources. This thesis aims to address these issues by exploring the feasibility of knowledge transfer between deep neural networks. The focus is on transferring knowledge from a larger multi-task ``Teacher" network to a smaller ``Student" network to improve the performance of the latter. To facilitate this, a new architecture based on multi-scale feature distillation and contrastive learning is proposed. The goal is to incorporate knowledge representations learned from the teacher model to enhance the performance of the student model.

Let $T$ represent the parameters of the larger multi-task ``Teacher" network, and let $S$ denote the parameters of the smaller ``Student" network. The goal is to transfer knowledge from $T$ to $S$ to improve the performance of the student model in medical image segmentation tasks.

We aim to minimize the following objective function:
\begin{equation}
\text{minimize} \quad [\mathcal{L}_{seg}(S(x), \mathbf{x}, \mathbf{y}_{true}) + \lambda \cdot \mathcal{R}(S))]
\label{equation:segmentation_loss}
\end{equation}
where:
\begin{itemize}
  \item $\mathcal{L}_{seg}$ is the segmentation loss function measuring the loss between the predicted segmentation mask of the student model $S$ and the ground truth mask $\mathcal{y}_{true}$, given the input 2D CT images $\mathbf{x}$.
  \item $\mathcal{R}$ represents a regularization term to prevent overfitting in the student model, and $\lambda$ is a hyperparameter controlling the trade-off between fitting the training data and minimizing model complexity.
\end{itemize}
To improve a student model, we can use multi-scale feature distillation and contrastive learning to transfer knowledge from a teacher model. The student model's parameters should be optimized to minimize segmentation loss and maximize the similarity between its learned representations and the teacher model's representations at multiple scales and layers. While optimizing, we balance minimizing segmentation loss for accurate segmentation results with maximizing similarity between representations, which helps in effective knowledge transfer. This approach allows the student model to learn holistically from the teacher model by leveraging the teacher model's rich representations to enhance the student model's segmentation performance.

Formally, the knowledge transfer process can be expressed as:
\begin{equation}
\text{minimize} \quad \sum_{i=1}^{N} \mathcal{L}_{\text{seg}}(S(\mathbf{x}_i; S), \mathbf{y}_i) + \alpha \cdot R(S) + \sum_{j} \beta_j \cdot \mathcal{L}_{\text{con}}(S, T, j) + \gamma \cdot \mathcal{L}_{\text{PMD}}(S, T)
\end{equation}
where:
\begin{itemize}
  \item $\mathcal{L}_{\text{seg}}$ is the segmentation loss and $\mathcal{L}_{\text{con}}$ is the contrastive loss measuring the discrepancy between the representations learned by the student model $S$ and the teacher model $T$ at scale or layer $j$.
  \item $R(S)$ represents a regularization term to prevent overfitting in the student model $S$.
  \item $\alpha$ and $\beta_j$ are hyperparameters controlling the trade-off between segmentation loss, regularization, and knowledge distillation objectives.
  \item $\gamma$ is the weighting factor for the PMD loss $\mathcal{L}_{\text{PMD}}$, measuring the discrepancy between the prediction maps of the student and teacher networks.
\end{itemize}
Segmentation loss, $\mathcal{L}_{\text{seg}}$ can be either cross-entropy loss or dice loss. Contrastive loss ($\mathcal{L}_{\text{con}}$) and Prediction maps distillation loss ($\mathcal{L}_{\text{PMD}}$) is discussed in details in Chapter~\ref{chapter:Methodology}. 

\section{Motivation}
Medical image segmentation plays a pivotal role in modern healthcare, facilitating accurate diagnosis, treatment planning, and disease monitoring. The precise delineation of anatomical structures from medical images is particularly crucial for organs like the spleen, as it aids clinicians in identifying and analyzing pathological conditions. However, the task of automatic spleen segmentation from CT images is confronted with various challenges that necessitate innovative solutions to enhance accuracy and efficiency. Image variability and anatomical variations pose significant challenges in spleen segmentation from CT images. CT images exhibit inherent variability in contrast, resolution, and acquisition parameters, necessitating segmentation algorithms to be robust across diverse imaging conditions. Moreover, individual anatomical variations in spleen size, shape, and position further complicate the segmentation task. A one-size-fits-all approach is inadequate, demanding methods that can adapt to the unique anatomies of different patients. Pathological conditions affecting the spleen, such as tumours, cysts, and infections, introduce irregularities in organ structure. These abnormalities disrupt the normal appearance of the spleen and hinder the accurate segmentation of diseased organs. Consequently, the ability to accurately segment spleens under pathological conditions is critical for the early detection and monitoring of diseases, thereby influencing patient outcomes. CT images are susceptible to various sources of noise, artefacts, and imperfections arising from the imaging process. These artefacts introduce uncertainties and distortions, affecting the reliability of segmentation algorithms. Addressing noise and artefacts is pivotal for ensuring the robustness of segmentation algorithms, especially in clinical settings where image quality may vary.

The computational demands of processing large volumes of CT data for accurate spleen segmentation using deep learning models present a significant challenge. Developing a lightweight neural architecture capable of handling this computational complexity while providing real-time or near-real-time results is crucial for clinical workflow integration. Thus, the architecture provided in the thesis can be greatly beneficial to overcome the above challenges. By addressing the inherent challenges in automatic spleen segmentation from CT images, this research aims to offer tangible benefits that directly impact clinical workflows and patient care. Automatic spleen abnormality detection offers clinicians multiple advantages. Firstly, it enhances diagnostic accuracy by employing a multi-task learning approach, capturing not only spleen structural details but also its relationship with neighbouring anatomical structures. This comprehensive understanding leads to more accurate identification of anomalies, facilitating nuanced diagnostic assessments. Secondly, the computational efficiency embedded in the methodology enables real-time or near-real-time segmentation results, valuable in time-sensitive clinical scenarios. Thirdly, for chronic diseases or cases requiring continuous monitoring, the approach allows reliable tracking of spleen changes over time, aiding in disease progression assessment and treatment evaluation. Furthermore, automation reduces the workload on radiologists and healthcare professionals, enhancing overall workflow efficiency. Lastly, the consideration of anatomical variations ensures adaptability to diverse patient populations, widening the segmentation tool's applicability in clinical settings.

The proposed architecture aims to offer improvements in efficiency, and adaptability, which contributes to providing clinicians with a powerful tool for better-informed decision-making. This ultimately enhances patient outcomes and the overall quality of healthcare delivery, especially in scenarios with limited computing resources.

\section{Research Aim}
This thesis aims to investigate and enhance the performance of medical image segmentation models, particularly in scenarios with limited data and compute availability, through the utilization of knowledge distillation techniques. The objectives are:
\begin{itemize}
    \item Develop a robust multi-task teacher network capable of simultaneously addressing both segmentation and reconstruction tasks, where the reconstruction task complements the segmentation task.
    \item Implement supervised contrastive learning techniques at multiple scales to facilitate knowledge transfer between the multi-task teacher and student networks, aiming to improve the segmentation accuracy of the lighter student network.
    \item Conduct a comprehensive comparative study between mean squared error (MSE) loss and contrastive learning while also exploring the impact of different sizes and parameters of the student network on segmentation performance.
    \item Investigate various distillation techniques, including predictive map distillation (PMD), and compare their effectiveness with alternative approaches, providing insights into their suitability for knowledge transfer.
    \item Analyze the role of multi-scale knowledge distillation by evaluating distillation at different network layers, elucidating the importance of capturing diverse features for enhanced segmentation performance in medical imaging tasks.
\end{itemize}

\section{Research Hypothesis}
In this thesis, we hypothesize that transferring knowledge from a larger multi-task pre-trained U-Net architecture (referred to as the ``Teacher" network) to a smaller U-Net model (the ``Student" network), trained on half of the data used for the teacher model, will significantly improve the student model's performance in medical image segmentation tasks. Using multi-scale feature distillation and supervised contrastive learning techniques, we expect the integration of knowledge representations acquired from the teacher model to significantly improve the segmentation accuracy of the student model, especially in scenarios with limited training data availability. Despite the student model's smaller size and dataset volume compared to the teacher network, we believe that knowledge distillation will enable it to perform competitively. Specifically, we hypothesize that the architecture proposed in this work, which facilitates knowledge transfer between the teacher and student models, combined with the use of multi-scale supervised contrastive knowledge distillation, will contribute to a significant performance boost in the student model. Furthermore, we hypothesize that contrastive learning mechanisms will play an important role in facilitating effective knowledge transfer during the student model training process. We hope to validate these hypotheses and gain valuable insights into the efficacy of multi-task, multi-scale supervised contrastive knowledge distillation for medical image segmentation tasks through a series of extensive experiments, including detailed ablation studies and analyses. 

\section{Thesis Outline}
In this chapter, we have explained to the reader why the problem is important, and what
it can be used for, and have given the reader an introduction and formulation of the
problem at hand. The outline of the rest of the thesis is:
\begin{itemize}
    \item In \textbf{Chapter~\ref{chapter:lit_rev}}, we review the existing literature on medical image segmentation techniques, knowledge distillation, multi-task learning, and contrastive learning.
    \item In \textbf{Chapter~\ref{chapter:Methodology}}, we present the methodology and architecture. We describe the dataset used for the research and the pre-processing techniques applied. We explain the architecture of the teacher and student models, detailing the multi-task pre-trained UNet architecture and the proposed novel architecture for knowledge transfer. 
    \item  In \textbf{Chapter~\ref{chapter:results}}, the results are presented of the experiments conducted, including confusion matrix, training curves, and statistical testing comparing the student model with and without knowledge distillation.
    \item In \textbf{Chapter~\ref{chapter:discussion}} we summarize and discuss what we set out to do, what has been
    achieved, what problems arose, and propose possible routes for future work. We also identified the limitations of the research methodology and experimental setup and we discussed the future directions and suggested potential areas for future research and improvement. 
    \item Lastly in \textbf{Chapter~\ref{chapter:conclusion}}, we summarize the key findings of the research and their implications for advancing medical image segmentation.
\end{itemize}
\chapter{Literature Review}
\label{chapter:lit_rev}
In this section, we present a comprehensive overview of the existing literature on medical image segmentation. We begin by discussing foundational works in deep learning-based segmentation. These works have laid the groundwork for subsequent developments. Later, we delve into recent trends and innovations, such as attention mechanisms, transformer architectures, and multi-task learning techniques. We examine the relationship between multi-task learning and segmentation, and how learning multiple related tasks simultaneously can enhance model generalization and efficiency. We also discuss the use of contrastive learning techniques for knowledge distillation, which enables the transfer of knowledge from large, complex teacher networks to smaller, more lightweight student networks. Our review emphasizes the importance of developing efficient and lightweight models and effective knowledge transfer techniques, with a focus on contrastive representation learning. Through a thorough analysis of existing literature, we aim to identify key challenges and opportunities in the field, laying the groundwork for our thesis.

\section{Medical Image Segmentation}
\begin{figure*}[ht!] 
    \centering
    \includegraphics[clip=true, trim = 00 00 00 00, width=1.08\linewidth]{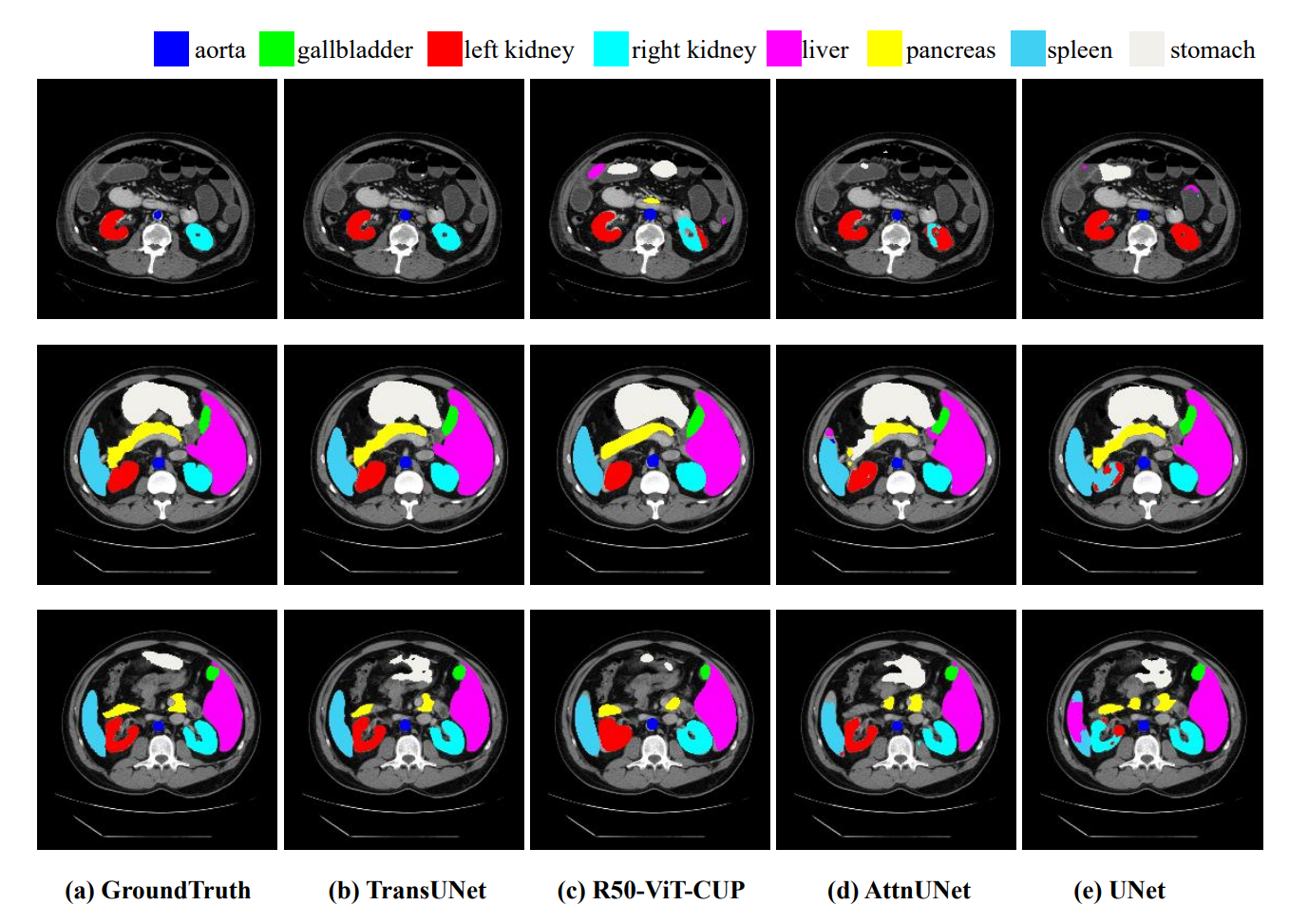} 
    \captionsetup{justification=centering}
    \caption{\centering An example of multi-class medical image segmentation~\cite{chen2021transunet}}
    \label{fig:Example_MSI}
\end{figure*}

\begin{figure*}[ht!] 
    \centering
    \includegraphics[clip=true, trim = 00 00 00 00, width=1.08\linewidth]{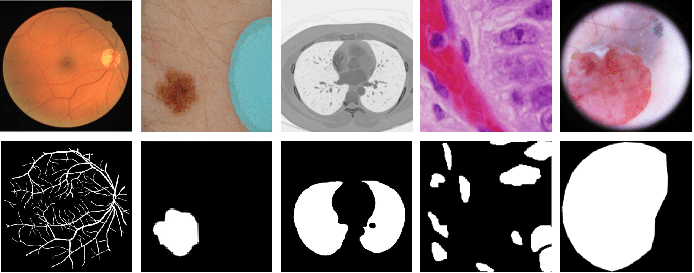} 
    \captionsetup{justification=centering}
    \caption{\centering An example of binary medical image segmentation}
    \label{fig:Example_MSI_binary}
\end{figure*}

Medical image segmentation plays a crucial role in various clinical applications, aiding in diagnosis, treatment planning, and disease monitoring. Typically there exist two kinds of segmentation problems - multi-class and binary. Figure~\ref{fig:Example_MSI} and~\ref{fig:Example_MSI_binary} illustrates the same. Over the years, numerous segmentation techniques have been developed to delineate anatomical structures and pathological regions from medical images accurately. Traditional segmentation methods often relied on handcrafted features and classical machine learning algorithms. However, with the advent of deep learning~\cite{lecun2015deep}, convolutional neural networks (CNNs) have emerged as powerful tools for medical image segmentation due to their ability to automatically learn hierarchical representations from raw data. One of the pioneering works in deep learning-based medical image segmentation is the U-Net~\cite{ronneberger2015u} architecture. 
The U-Net introduced an encoder-decoder structure with skip connections, enabling precise segmentation even with limited annotated data. Since then, various extensions and modifications of the U-Net architecture have been proposed to address specific challenges in medical image segmentation tasks. In the past few years, there has been significant progress in the field of medical image segmentation, with a notable focus on the development of effective architectures. The U-Net family, including variants~\cite{cciccek20163d},~\cite{jin2020ra},~\cite{zhou2018unet++},~\cite{isensee2018nnu} has gained recognition for its ability to tackle various medical imaging problems. This sustained development underscores the importance and effectiveness of U-Net-based approaches in addressing challenges in medical image analysis. The Attention U-Net~\cite{oktay2018attention}, integrates attention mechanisms into the U-Net architecture for medical image segmentation. It employs attention gates to selectively enhance informative features while suppressing irrelevant ones during feature propagation from the encoder to the decoder. This allows the model to focus on relevant regions, improving segmentation accuracy. The Attention U-Net has since become a foundational model in medical image segmentation, inspiring further research into attention-based architectures. FocusNet~\cite{kaul2019focusnet} employs a dual encoder-decoder architecture, where attention gating plays a pivotal role by facilitating the propagation of relevant features. This mechanism enables the transfer of features from the decoder of one U-Net to the encoder of the next U-Net in the sequence. Building upon this foundation, FocusNet++~\cite{kaul2021focusnet++} introduces an attention mechanism within paired convolutions. By directly incorporating attention processes into different filter groups, the latter achieves a more precise and selective feature representation than its predecessor. This innovation allows the model to focus on specific regions of interest within an image, enhancing its capacity to identify intricate information and its overall performance in tasks like image segmentation.
In medical image segmentation, researchers are actively exploring the integration of transformer-based models with convolutional neural networks (CNNs) to improve feature processing capabilities. Notably, techniques like U-Net Transformer~\cite{petit2021u} have evolved, which include multi-head attention mechanisms into skip connections to increase feature representation. TransUNet~\cite{chen2021transunet} is a pioneering transformer-CNN hybrid model developed exclusively for medical image segmentation. This model includes a transformer encoder followed by a cascaded convolutional decoder, which allows for effective segmentation map construction. Similarly, UNETR~\cite{hatamizadeh2022unetr} and Swin UNETR~\cite{hatamizadeh2021swin} use transformers on the encoder side and a convolutional decoder to generate segmentation maps. Swin U-Net~\cite{cao2022swin}, a pure Transformer-based segmentation model, has emerged as a promising approach in medical imaging. Unlike traditional methods relying on pre-trained backbones, Swin U-Net processes image features directly using transformer layers. FCT~\cite{tragakis2023fully} is the first fully convolutional transformer network, which processes the input in two stages: first, it learns to extract long-range semantic dependencies from the input image, and then it learns to capture hierarchical global attributes from the features. FCT is compact, accurate, and robust.

\section{Multi-Task Learning}
Multi-task learning (MTL) is a technique where a model is trained to perform multiple tasks simultaneously. Instead of training each model separately for each task, a single model is trained jointly on all tasks, sharing information among them. The main idea behind multi-task learning is that the knowledge learned from one task can be beneficial for learning other related tasks, leading to improved performance on all tasks involved. By sharing representations across tasks, the model can generalize better and learn more robust features. However, this kind of learning is useful when tasks share some underlying structure or have related objectives. For example, in natural language processing, tasks such as part-of-speech tagging, named entity recognition, and sentiment analysis could benefit from shared knowledge about language syntax and semantics. Similarly, in vision jointly performing image reconstruction and semantic segmentation can mutually benefit from the shared representation learned. Advantages of multi-task learning include - 
\begin{itemize}
    \item \textbf{Improved Generalization:} Learning multiple tasks jointly can help the model generalize better, as it learns to capture more generalized representations that are useful across multiple tasks.
    \item \textbf{Data Efficiency:} MTL can help improve learning efficiency, especially when data for individual tasks is limited. By leveraging information from multiple tasks, the model can effectively utilize the available data more efficiently.
    \item \textbf{Regularization:} Jointly learning multiple tasks can act as a form of regularization, preventing overfitting and improving the model's ability to generalize to new tasks.
\end{itemize}
There are two types of MTL - hard parameter sharing and soft parameter sharing. In our thesis for the teacher network, we employed a multi-task U-Net using hard parameter sharing, where we share a common encoder for both the underlying tasks (reconstruction and segmentation). This common encoder layer is very useful for leveraging shared features between tasks, leading to more efficient learning and better generalization capabilities across tasks. 

\section{Contrastive Learning}
\begin{figure*}[ht!] 
    \centering
    \includegraphics[clip=true, trim = 00 00 00 00, width=1.0\linewidth]{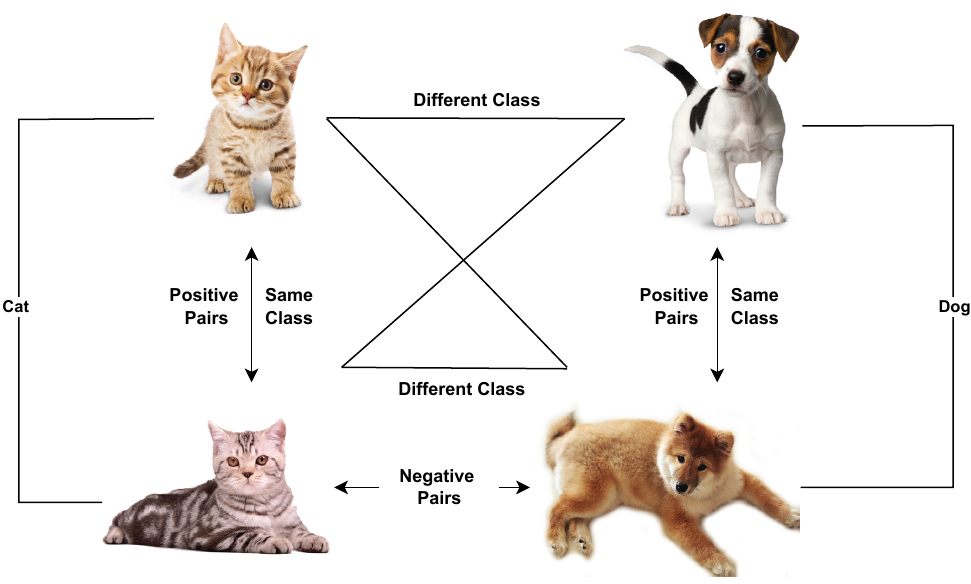} 
    \captionsetup{justification=centering}
    \caption{\centering Representation of Contrastive Pairs}
    \label{fig:Contrastive_Pairs}
\end{figure*}

Contrastive learning is an approach that focuses on extracting meaningful representations by contrasting positive and negative pairs of instances. The idea behind this approach is that when we learn an embedding space, similar instances should be closer together, and dissimilar instances should be farther apart. The contrastive learning process involves selecting an ``anchor" feature from a particular scale of the network, which serves as a reference point. The learning algorithm then distinguishes between instances: those that belong to the same distribution as the anchor, called ``positive" samples, and those belonging to a different distribution, called ``negative" samples, illustrated in Figure~\ref{fig:Contrastive_Pairs}. By framing the learning process as a discrimination task, contrastive learning enables models to capture essential attributes and similarities in the data. This method is particularly useful in image and natural language processing tasks, where it can help improve accuracy and reduce the amount of labelled data required for training.

Positive pairs typically comprise samples with similar features or from the same class. For instance, in image classification, positive pairs could consist of images with similar feature representations or belonging to the same class. Contrastive learning encourages the student model to map these positive pairs closer together in the feature space, capturing their underlying semantic commonalities. On the other hand, negative pairs include samples that are contrary or belong to distinct groups. These pairs serve as contrastive instances, allowing the student model to distinguish between classes and acquire discriminative features. By moving negative pairs farther apart in the feature space, the student model learns to effectively distinguish between different classes. Contrastive learning thus facilitates the extraction of informative features by carefully manipulating positive and negative pairs, improving the student model's understanding of the data and encouraging generalization and robustness to variations in the input domain.

Furthermore, to enhance contrastive learning's efficacy in knowledge distillation, it can be augmented with various strategies such as regularization, data augmentation, and architectural changes. Even with limited computational resources, the student model can achieve state-of-the-art performance by integrating these complementary techniques and improving its representations.

\subsection{Contrastive Loss}
In the learned embedding space, contrastive loss tries to reduce the agreement between negative pairings (instances from separate samples) and increase the confidence between positive pairs (instances from the same sample). The intention is to drive distinct instances apart and bring similar instances closer together. Typically, contrastive loss can be defined as a margin-based loss in which a distance metric, like the cosine similarity or Euclidean distance, is used to assess the similarity of instances. By penalizing positive samples for being too far apart and negative samples for being too close together in the embedding space, the contrastive loss is calculated.

The Information Noise Contrastive Estimation (InfoNCE) loss~\cite{oord2019representation} is a widely used contrastive loss that originates from the noise contrastive estimation framework. It calculates how similar positive and negative pairings are to one another in the learned embedding space. The agreement between positive pairs is maximized and the agreement between negative pairs is minimized when using InfoNCE loss. Considering the contrastive learning problem as a binary classifier is the fundamental principle of InfoNCE loss. The model is trained to distinguish between positive and negative examples given a positive pair and a collection of negative pairings. A probabilistic method, such as the softmax function, is used to quantify how similar two cases are to one another.

\subsection{Contrastive Loss for Knowledge Distillation}
By promoting feature similarity between positive pairs of images and dissimilarity between negative pairs, contrastive loss provides a logical method for knowledge distillation. Negative pairs are made up of image patches with distinct properties, whereas positive pairs can be image patches with identical representations. Even in the absence of explicit supervision, the student model learns to imitate the feature representations learned by the teacher model by employing contrastive loss as a distillation loss. This improves segmentation performance by allowing the student model to capture the semantics and underlying structure of the medical images. To improve the knowledge transfer process even further, the contrastive loss can be used in conjunction with other distillation methods like knowledge distillation from logits or intermediate feature representations. Contrastive loss as a distillation loss provides a versatile and effective framework for training reliable and accurate segmentation models, even with a smaller network. In this thesis,  we used the InfoNCE loss to perform the knowledge distillation between the teacher and student network.

\subsection{Previous Works on Contrastive Learning}
Contrastive learning~\cite{chen2020simple} has emerged as a prominent paradigm in the field of machine learning, offering a powerful framework for representation learning. Khosla~\etal~\cite{khosla2020supervised} introduced ``Supervised Contrastive Learning" as a novel framework for supervised learning tasks, aiming to enhance the discriminative power of learned representations by leveraging the principles of contrastive learning. They proposed a contrastive loss function tailored for supervised settings, where both labelled and unlabeled data are available. This approach extends the benefits of contrastive learning to supervised tasks. For natural images, contrastive learning has also contributed to self-supervised visual representation techniques, including~\cite{grill2020bootstrap},~\cite{he2020momentum}. In medical imaging, most recently, contrastive learning has been used in a semi-supervised framework by Qianying~\etal~\cite{liu2023multi}. They proposed a local contrastive framework defined over multi-scale feature maps that offers a more robust approach for enhancing visual representation. The primary motivation of this thesis is to explore the possibility of multi-scale knowledge distillation using contrastive learning. To the best of my knowledge, this is the first investigation on the impact of multi-scale contrastive feature distillation from a larger network to a smaller network for the medical image segmentation task. 

\section{Knowledge Distillation}
Deep neural networks have grown popular in various applications, including object detection in images and text generation utilizing GPT models. However, these models frequently have significant computing requirements and are difficult to deploy on devices with limited resources, resulting in increased inference times. In response to these challenges, lightweight models have evolved as a solution, which is especially useful for applications like medical imaging in resource-limited settings. Deploying such models on low-power computing systems, such as mobile or edge devices, is critical for real-time processing, particularly in remote areas with limited access to advanced infrastructure. These lightweight models maximize resource utilization, resulting in lower latency and memory needs.
\begin{figure*}[t!] 
    \centering
    \includegraphics[clip=true, trim = 00 00 00 00, width=1.02\linewidth]{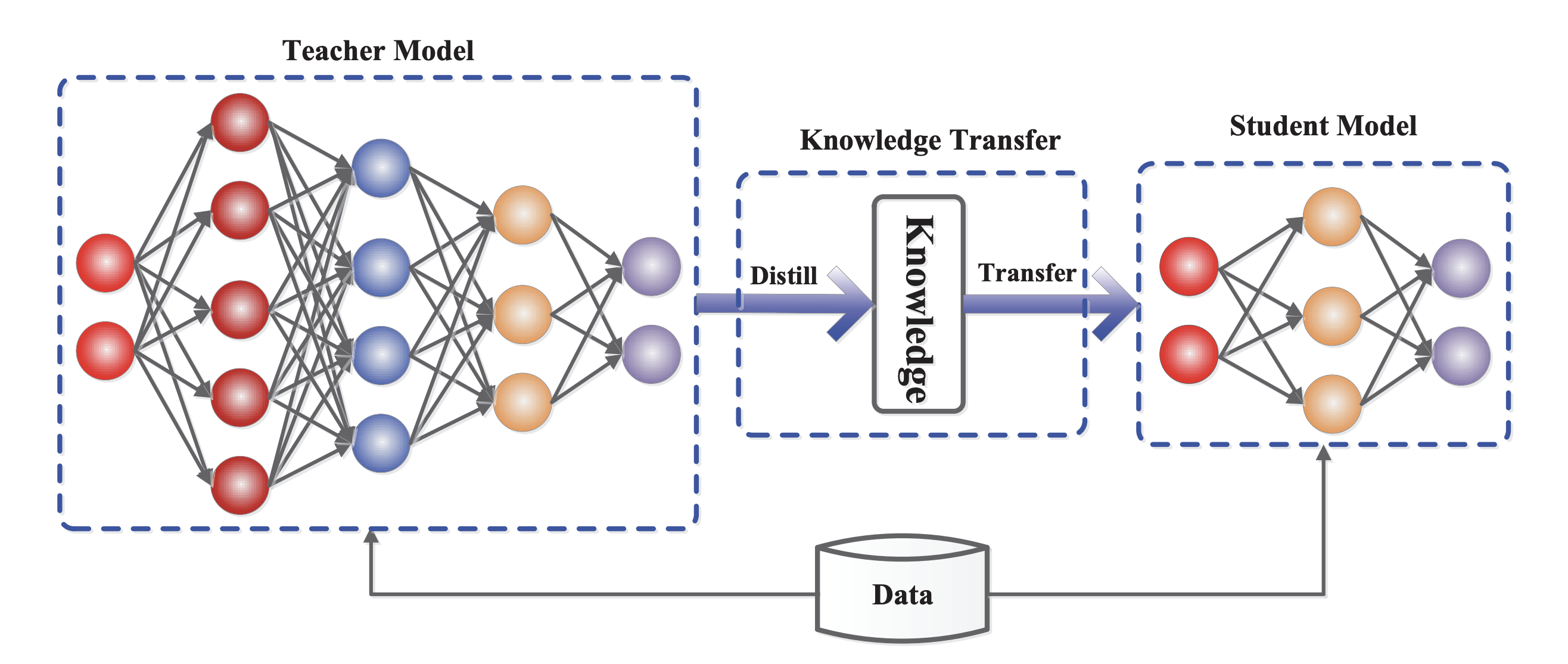} 
    \captionsetup{justification=centering}
    \caption{\centering Teacher-Student Framework for Knowledge Distillation~\cite{Gou_2021}}
    \label{fig:Genric_KD}
\end{figure*}
Despite their advantages, lightweight models may have lower accuracy, potentially affecting essential applications such as clinical diagnosis. Achieving high precision with ultra-lightweight models, which usually have only a few thousand parameters, is a significant challenge. This is when knowledge distillation proves helpful. Knowledge distillation aims to transfer knowledge from a larger model to a smaller model. The size refers to the number of parameters in the model, which is directly proportional to the model's latency. The larger network that transfers knowledge is known as the ``Teacher Network", while the smaller network that gathers the knowledge is known as the ``Student Network". Knowledge distillation allows us to compress the model while maintaining accuracy by using a larger, more complicated teacher network. In essence, it allows us to achieve a compromise between model size and performance, ensuring that even lightweight models produce accurate results.

\subsection{Steps Involved in Knowledge Distillation}
A knowledge distillation mechanism comprises three key components: Knowledge, The distillation algorithm, and the teacher-student architecture. The design of the teacher-student architecture plays a critical role in facilitating an effective transfer of knowledge between the teacher and student models. This architecture often reveals a capacity gap between the large, complex teacher neural network and the smaller, less complex student neural network. The relationship between the teacher and student models is defined by their network structures, enabling effective knowledge sharing. The student network can take various forms, including a simplified version of the teacher network, a quantized version preserving the teacher's structure, a network resembling the teacher model, a network with basic operations, or an optimized and condensed structure. To bridge the model capacity gap, efforts focus on minimizing the disparities between the teacher and student models through various distillation algorithms and optimization techniques. 

\begin{itemize}
\item \textbf{Train the Teacher Network:} During the first step, the teacher network is trained on a dataset using a standard training procedure. Once the teacher network is trained, it is frozen. The teacher network is first trained separately until full convergence. Here, the loss function can be any loss function based on the problem statement. For example, for the medical image segmentation task, it could be the dice loss. 
\item {\textbf{Train the Student Network using Knowledge Distillation:}} The student network could be either a scaled-down version of the teacher model or some other lightweight model. Here, lightweight means fewer parameters than those of the teacher network. During the training of the student network, knowledge is distilled from the frozen teacher model. This knowledge transfer can happen at multiple scales, using a variety of learning techniques. During this process, backpropagation happens only on the student network; we do not train the teacher network again. There are two loss functions defined. One is student loss or task loss, such as segmentation loss, which is specific to the task at hand, and distillation loss.
\end{itemize}

Typically, in medical image segmentation with knowledge distillation, the process involves utilizing either the segmentation output or the features extracted by the teacher network to guide the student network's learning.
\begin{itemize}
\item \textbf{Segmentation Output-based Distillation:}
In this approach, the distillation loss, such as PMD is computed based on the soft segmentation outputs generated by the teacher network. These soft segmentation outputs convey valuable information about the desired segmentation masks, allowing the student network to learn from the teacher's expertise in delineating structures or abnormalities within medical images.
\item \textbf{Feature-based Distillation:}
Alternatively, the teacher network's intermediate features can be utilized to guide the student network's learning process. Instead of directly comparing segmentation outputs, the distillation loss may be computed based on the feature representations extracted by the teacher network. These features capture rich information about image characteristics and structures, enabling the student network to learn informative representations that facilitate accurate segmentation.
\end{itemize}
In both of the above cases, the distillation process aims to transfer knowledge effectively from the teacher network to the student network, enhancing the segmentation performance of the latter.

\subsection{Multi-Scale Knowledge Distillation}
In a multi-scale setting, capturing information at various levels of detail or scale, knowledge distillation can enhance the effectiveness of the knowledge transfer process. When conducting knowledge distillation at multiple scales, the distillation process incorporates information from the teacher network's segmentation outputs along with the features at different resolutions or levels of granularity. This enables the student network to learn from the teacher's learned representation across various scales, thereby improving its ability to accurately segment medical images with diverse structures and complexities. 
In this scenario, a weighted loss function is computed between the distillation losses at various scales and the segmentation loss. The weights assigned to each component of the loss function are determined based on the importance of capturing knowledge at different scales and the desired segmentation performance. By integrating multi-scale knowledge distillation into the segmentation framework, the student network can effectively leverage information from the teacher network across different levels of detail, leading to enhanced segmentation accuracy and robustness across a range of medical imaging scenarios. 

\subsection{Previous Works on Knowledge Distillation}
To improve the deployment of deep learning models on mobile and embedded devices, researchers and engineers are exploring techniques like model compression, quantization, and efficient architecture design. One of the first works on model compression was proposed in~\cite{buciluǎ2006model}. Model compression aims to reduce computational and memory requirements, making the model suitable for deployment on low-compute systems while preserving performance. Since their inception, model compression techniques have been extensively investigated and developed by researchers to address the challenges of deploying deep learning models in real-world applications with limited resources. The process of training a small model using knowledge transferred from a larger one was later formalized and popularized as ``Knowledge Distillation" by Hinton~\etal~(2015)~\cite{hinton2015distilling}. Since its introduction, many researchers have utilized knowledge distillation for a variety of tasks across different domains~\cite{zhang2019your, tung2019similarity, tian2019contrastive}. Recent advancements such as~\cite{he2019knowledge, liu2019structured, xie2018improving, wang2020intra, 9115859} have witnessed a surge in knowledge distillation methods tailored for addressing semantic segmentation challenges. These methods explore diverse approaches to distil structural information that can theoretically enhance segmentation tasks. Knowledge distillation has demonstrated successful performance across various domains, prompting its adoption in medical imaging. In recent years, researchers have explored the application of knowledge distillation techniques to enhance the efficiency and accuracy of medical imaging tasks such as classification~\cite{ho2020utilizing}, segmentation tasks~\cite{wang2019segmenting, li2020towards, }. Recent works such as~\cite{qin2021efficient}~employ a distillation architecture capable of extracting information from existing medical image segmentation networks and transferring it to a lightweight network known as the student network. To enhance the efficiency and effectiveness of the distillation process, they designed a distillation technique that focuses on encoding and distilling the importance of semantic region information. Specifically, it calculates a collection of inter-class contrasts between different tissue regions, termed the region contrast map, from intermediate feature maps using ground truth segmentation masks as guidance. By leveraging the region contrast map, the network effectively guides the distillation process, ensuring that the student network captures essential semantic information crucial for accurate segmentation. Inspired by this, in this thesis, we created a teacher-student knowledge distillation framework using contrastive learning where a larger teacher network transfers knowledge at a multi-scale to a much smaller student network trained on a lesser number of samples. 

\section{Summary}
In this chapter, we reviewed the current literature in the field of medical image segmentation, knowledge distillation and contrastive learning. The literature review offers valuable insights into the advancement of medical image segmentation techniques. It highlights the shift from traditional methods to deep learning-based approaches, especially Convolutional Neural Networks (CNNs). U-Net is a pioneering architecture that has established the groundwork for accurate segmentation. Later developments, such as Attention U-Net and transformer-CNN hybrids, have further improved segmentation precision. Multi-task learning (MTL) and contrastive learning are two powerful techniques that have been used to enhance segmentation performance and knowledge transfer. MTL, especially with hard parameter sharing, can help models handle multiple tasks at the same time by using shared information for better generalization. On the other hand, contrastive learning provides a structured framework for knowledge distillation, which allows the transfer of knowledge from larger teacher networks to smaller student networks. Previous techniques have tried to create an efficient knowledge distillation architecture. 

However, limitations persist with existing techniques, particularly concerning limited data and computational resources. While current approaches strive to develop lightweight models, they often rely on heavy teacher networks. In response to this challenge, our thesis proposes an optimal approach. We introduce lightweight teacher and student models and prioritize the development of a more efficient knowledge transfer technique. Rather than relying solely on large models or extensive datasets, we emphasize the utilization of contrastive representation learning for optimal knowledge transfer. Our approach involves developing a robust multi-task teacher network, implementing contrastive learning for effective knowledge transfer, and conducting a comparative study between different loss functions. We also explore various distillation techniques and analyze the role of multi-scale knowledge distillation in enhancing segmentation accuracy. We aim to improve medical image segmentation and enable accurate segmentation models for clinical applications.

\chapter{Methodology}
\label{chapter:Methodology}
This chapter provides a detailed description of the methodology used to address the research problem. It begins with a detailed discussion of the dataset used for the study, as well as the preprocessing techniques involved in processing a CT image in section~\ref{section:Dataset and Preprocessing} to ensure its suitability for the task at hand.  Following the discussion of data preparation, attention is focused on the Multi-Task Teacher Network's architecture and operational mechanisms in section~\ref{section:Multi-Task Teacher Network}. The chapter then digs into the design and details of the student networks. This component is responsible for distilling knowledge from the teacher network, with an emphasis on achieving compact representation without sacrificing performance. The architectural complexities and training strategies used for the student Network are discussed in detail in section~\ref{section:Student Network}.

Multi-Scale Knowledge Distillation, which is a critical component of our approach is discussed in detail in section~\ref{section:Multi-Scale Knowledge Distillation}. The section also provides both a graphical and mathematical framework to understand the architecture intuitively. This framework defines two core components: Prediction Maps Distillation Loss and Multi-Scale Contrastive Loss in sub-section~\ref{subsection:Prediction Maps Distillation Loss} and \ref{subsection:Multi-Scale Contrastive Loss}, respectively. These subsections describe the mechanisms by which knowledge is transferred between the teacher and student networks, across multiple levels of information abstraction.
\section{Dataset and Preprocessing}
\label{section:Dataset and Preprocessing}
This section will discuss the data set used and how the dataset is preprocessed before feeding it to the neural network. In our project, we focus on the preprocessing of the spleen dataset from the medical image decathlon~\cite{simpson2019large}, consisting of 61 3D CT volumes of the spleen. The type of segmentation that we will be focusing on in this thesis is called as ``Binary Segmentation", wherein we segment the area of interest from the background. This is similar to segmenting tumours, lesions or in our case - organs of interest(Spleen). The dataset is divided into a training set with 41 CT volumes and a testing set with 20 CT volumes, each accompanied by a corresponding segmentation mask representing the ground truth for the Spleen. The dataset is publicly accessible through the link provided in the previous section. The primary challenge posed by this dataset is the considerable variation in foreground size. Given that the dataset comprises 3D CT volumes and not directly usable 2D images, the crucial initial step involves transforming these volumes into 2D images, or NumPy arrays. This conversion enables the subsequent feeding of data into the deep-learning model.

\begin{figure}[ht]
  \subcaptionbox*{Spleen Images}[.45\linewidth]{%
    \includegraphics[width=\linewidth]{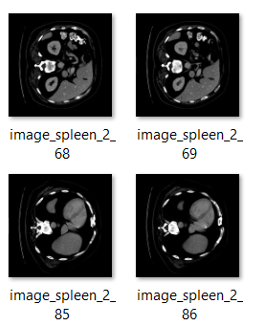}%
  }%
  \hfill
  \hfill
  \rulesep
  \hfill
  \subcaptionbox*{Spleen Masks}[.45\linewidth]{%
    \includegraphics[width=\linewidth]{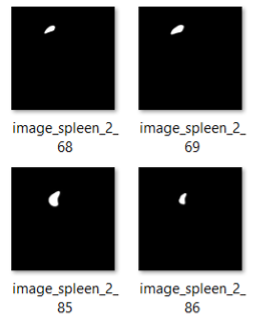}%
  }
  \caption{Processed 2D slices and their corresponding ground truth mask.}
\end{figure}
To implement this conversion, the provided Python code utilizes the NiBabel library to handle Neuroimaging file formats, specifically the Neuroimaging Informatics Technology Initiative (NIfTI) format. Each 3D CT volume is processed by slicing it into 2D images, which are then saved as PNG files. Now, let us discuss the conversion process of the 3D CT volumes to 2D slices that can be fed to the model. For each subject, the NIfTI file is loaded using the NiBabel library in Python, and its pixel data is extracted as a NumPy array. We can use this Numpy array directly as well instead of saving the 2D images. However, we chose to save them to get more insights into the data and as a quality check step to ensure the data is not corrupted during the conversion from nifty to 2D image. The maximum pixel value in the 3D CT volume is determined, and a multiplier is calculated to scale the pixel values to the range [0, 255]. The 3D volume is then iterated through its slices, and for each slice, a corresponding 2D image is generated and saved in the output directory. The image filenames follow the pattern -image\_subject\_slicenumber.png. If a subject's NIfTI file is not found in the specified directory, the processing for that subject is skipped. By executing the preprocessing code, the data preprocessing step ensures that the original 3D CT volumes are appropriately transformed into a format suitable for input into the deep learning model, facilitating subsequent training and evaluation. 

\section{Multi-Task Teacher Network}
\label{section:Multi-Task Teacher Network}
\begin{figure*}[ht!] 
    \centering
    \includegraphics[clip=true, trim = 00 00 00 00, width=1.08\linewidth]{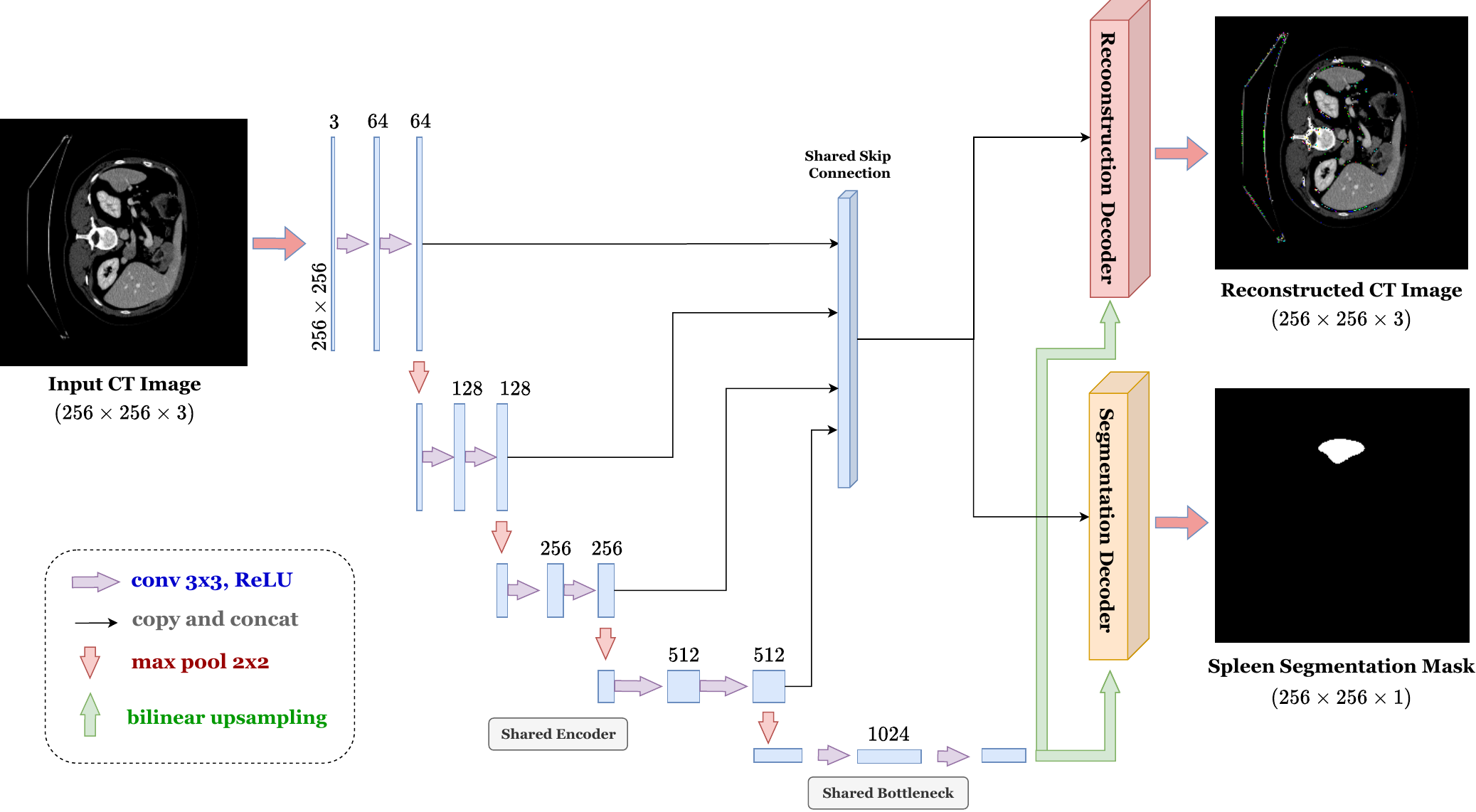} 
    \captionsetup{justification=centering}
    \caption{\centering Illustration of a Multi-Task Teacher Model}
    \label{fig:Teacher_Arch}
\end{figure*}
We trained two teacher models T1 and T2 which are one a multi-task pre-trained U-Net and a multi-task TransUNet respectively. The architecture of T1 is illustrated in Figure~\ref{fig:Teacher_Arch}, and T2 possesses a similar architecture with one shared encoder and two decoders for two tasks. The U-Net has an encoder-decoder structure with skip connections. These skip connections make it easier to pass detailed spatial information from the encoder to the decoder, resulting in more precise segmentation. The multi-task U-Net is made up of a shared encoder and two decoder heads: one for reconstruction and one for segmentation. The teacher model is trained on related tasks so that it can capture a wide range of features and representations useful for segmentation.

Given an input image $\mathbf{x} \in \mathbf{X}$, with dimensions $C \times H \times W$, where $C$ is the number of channels and $H$ and $W$ are the height and width, respectively. The encoder ($\text{Enc}$) maps it to a latent feature space $\mathbf{z} = \text{Enc}(\mathbf{x})$, where $\mathbf{z} \in \mathcal{Z}$. The decoder consists of two branches: one for segmentation and one for reconstruction. For segmentation, the decoder ($\text{Dec}_{\text{seg}}$) maps the latent features $\mathbf{z}$ to the segmentation label space $\mathbf{y}_{\text{seg}} = \text{Dec}_{\text{seg}}(\mathbf{z})$, where $\mathbf{y}_{\text{seg}} \in \mathcal{Y}_{\text{seg}}$.
The segmentation dice loss is defined as:
\begin{equation}
\mathcal{L}_{\text{seg}}(\mathbf{y}_{\text{seg}}, \mathbf{y}_{\text{true}}) = 1 - \frac{2 \sum_{i} y_{\text{seg}}^{(i)} y_{\text{true}}^{(i)}}{\sum_{i} y_{\text{seg}}^{(i)} + \sum_{i} y_{\text{true}}^{(i)}}
\end{equation}
where, $\mathbf{y}_{\text{true}}$ represents the ground truth segmentation mask.

For reconstruction, the decoder ($\text{Dec}_{\text{rec}}$) maps the latent features $\mathbf{z}$ to the reconstruction label space $\mathbf{y}_{\text{rec}} = \text{Dec}_{\text{rec}}(\mathbf{z})$, where $\mathbf{y}_{\text{rec}} \in \mathcal{Y}_{\text{rec}}$. The mean squared error (MSE) loss is used for reconstruction, defined as:
\begin{equation}
\mathcal{L}_{\text{rec}}(\mathbf{y}_{\text{rec}}, \mathbf{x}) = \frac{1}{n} \sum_{i=1}^{n} (y_{\text{rec}}^{(i)} - x^{(i)})^2
\end{equation}
where, $n$ is the number of pixels and $x^{(i)}$ represents the $i$th pixel intensity of the input image.

The total loss for the teacher model is defined as the segmentation loss plus the reconstruction weight multiplied by the reconstruction loss:
\begin{equation}
\mathcal{L}_{\text{Teacher}} = \mathcal{L}_{\text{seg}}(\mathbf{y}_{\text{seg}}, \mathbf{y}_{\text{true}}) + \lambda_{\text{rec}} \cdot \mathcal{L}_{\text{rec}}(\mathbf{y}_{\text{rec}}, \mathbf{x})
\end{equation}
Where $\lambda_{\text{rec}}$ is the weight assigned to the reconstruction loss, adjusting the trade-off between segmentation and reconstruction during training. 
Next, we will look into the details of our student network, $S$.

\section{Student Network}
\label{section:Student Network}

\begin{figure*}[ht!] 
    \centering
    \includegraphics[clip=true, width=1.05\linewidth]{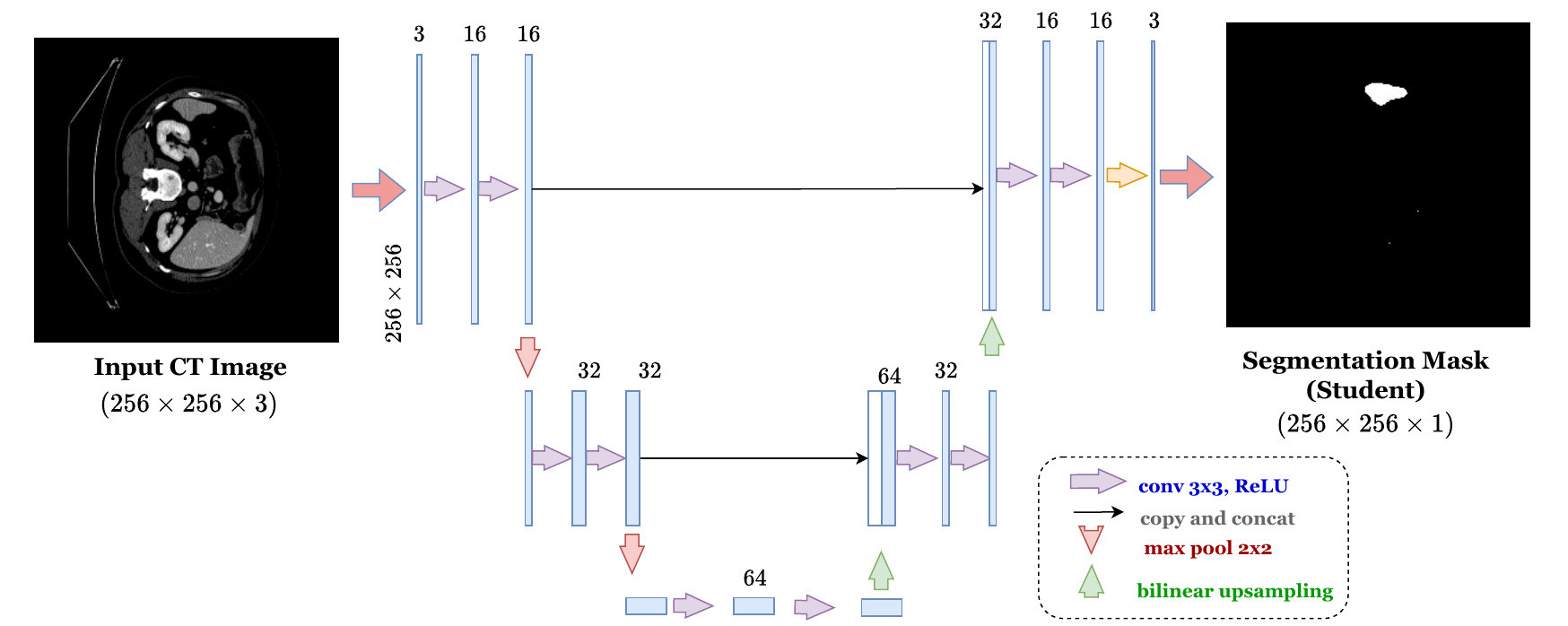} 
    \captionsetup{justification=centering}
    \caption{\centering Illustration of the Student Model}
    \label{fig:Student_Arch}
\end{figure*}
Similar to the teacher model, in this thesis we trained two student models S1 and S2. The student model shown in Fig~\ref{fig:Student_Arch}, a simplified version of the teacher model, is significantly smaller in scale and is trained on only 50\% of the data compared to the teacher model. This version of the student is termed as S1 in the thesis. Whereas a slightly bigger model than S1 but a smaller one than T1 is termed as S2. While the teacher model, comprises four encoder blocks with channels ranging from $64$ to $512$, a bottleneck convolutional block with $1024$ channels, and extensive decoder blocks for both segmentation and reconstruction tasks, the student model S1 is a compact implementation that trims down the architecture substantially. It consists of only two encoder blocks with reduced channels, a bottleneck block with $64$ channels, and two decoder blocks with further decreased channels. Consequently, the student model bears a much smaller footprint compared to its teacher counterpart.

The architecture of the student model can be mathematically defined as follows:
\begin{align*}
&\text{Encoder Block}_1: \quad E_1(\mathbf{x}) = \text{ReLU}(\text{Conv}_1(\mathbf{x})) \quad \text{where,} \quad \mathbf{x} \in \mathbb{R}^{3 \times H \times W} \\
&\text{Encoder Block}_2: \quad E_2(\mathbf{x}) = \text{ReLU}(\text{Conv}_2(\mathbf{x})) \quad \text{where,} \quad \mathbf{x} \in \mathbb{R}^{16 \times \frac{H}{2} \times \frac{W}{2}}
\end{align*}
\[
\text{Bottleneck Block}: \quad B(\mathbf{x}) = \text{ReLU}(\text{Conv}_3(\mathbf{x})) \quad \text{where,} \quad \mathbf{x} \in \mathbb{R}^{32 \times \frac{H}{4} \times \frac{W}{4}}
\]
\begin{align*}
&\text{Decoder Block}_1: \quad D_1(\mathbf{x}) = \text{ReLU}(\text{Conv}_4(\mathbf{x})) \quad \text{where,} \quad \mathbf{x} \in \mathbb{R}^{64 \times \frac{H}{2} \times \frac{W}{2}} \\
&\text{Decoder Block}_2: \quad D_2(\mathbf{x}) = \text{ReLU}(\text{Conv}_5(\mathbf{x})) \quad \text{where,} \quad \mathbf{x} \in \mathbb{R}^{16 \times H \times W}
\end{align*}
\[
\text{Classifier Block}: \quad \text{Output}(\mathbf{x}) = \text{Conv}_6(\mathbf{x}) \quad \text{where,} \quad \mathbf{x} \in \mathbb{R}^{16 \times H \times W}
\]
Here, $\text{Conv}_i$ represents a convolutional operation with appropriate parameters.

However, these constraints are deliberate and intended to explore the trade-off between model simplicity and segmentation performance. Though the student model may not achieve the same results as the teacher model due to its limited ability to capture intricate features and spatial details required for accurate segmentation, the hypothesis is that with the help of knowledge distillation, the student model will perform better than it would without it. Despite its simplicity, the student model is expected to effectively use knowledge distilled from the teacher model, compensating for its limitations and achieving competitive segmentation results in real-world applications. This we will discuss in the next section. 

\section{Multi-Scale Knowledge Distillation} 
\label{section:Multi-Scale Knowledge Distillation} 
The input image $x \in \mathbb{R}^{C \times H \times W}$ is simultaneously fed to both the pre-trained multi-task teacher model and the student model. The pre-trained teacher model captures the feature representation of the input image, which is then distilled to the student model using contrastive learning. By aligning the feature representations between the teacher and student models through contrastive learning, the student model effectively inherits the knowledge encoded in the teacher's representations. This process facilitates knowledge transfer from the teacher to the student, enhancing the student model's ability to perform segmentation tasks. During this, only the student network undergoes training. The pre-trained teacher model guides the student model through knowledge distillation, but the teacher model itself remains static and does not undergo further training. By focusing solely on training the student network, computational resources are efficiently utilized, and the knowledge distilled from the teacher model is effectively transferred to the student network, enhancing its performance in segmentation tasks.
\begin{figure*}[ht!] 
    \centering
    \includegraphics[clip=true, width=1.1\linewidth]{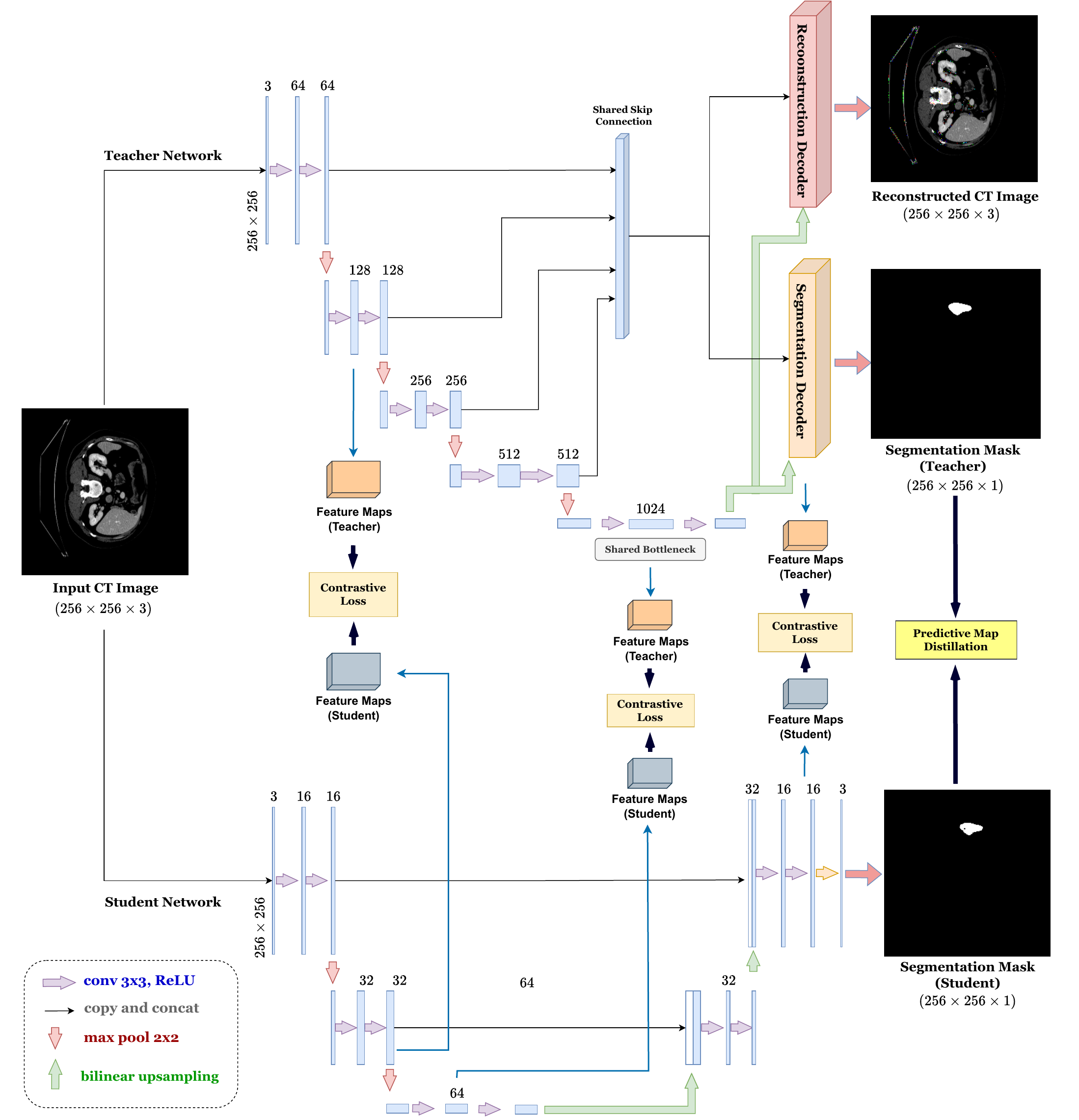} 
    \captionsetup{justification=centering}
    \caption{\centering The overall architecture of our multi-scale contrastive knowledge distillation framework for segmentation.}
    \label{fig:Arch}
\end{figure*}

Local contrastive learning methods typically involve projecting the features of the final layer before the classifier. However, feature maps from earlier layers contain more coarse geometric knowledge, such as organ shapes, whereas later feature maps focus on finer details. Both aspects are critical for segmentation tasks, which require understanding the relationships between the global and the local features. As a result, we focus on a multi-scale approach that uses features from various layers of the feature extractor. Specifically, we extract features of $k$ different scales from $k$ layers of extractors and then use separate projectors for each scale. This emphasizes the importance of combining features from different layers to capture the global and local information required for effective segmentation. Let's discuss each of these scales in our architecture in detail. 

\textbf{Encoder Scale:} The encoder scale extracts low-level features and abstract representations from the input image. These features store knowledge concerning edges, textures, and other basic visual elements found in the medical image. By aligning encoder features in the teacher and student networks, the student can learn to capture similar low-level features, which are critical for understanding the structural characteristics of the input image.

\textbf{Bottleneck Scale:} The bottleneck scale is a higher-level abstraction. It captures intricate patterns and semantic information, which are critical for understanding the image's overall context. By aligning bottleneck features, the student can learn to capture high-level semantic information, allowing them to make better segmentation decisions.

\textbf{Decoder Scale:} The decoder scale reconstructs the segmented output from the encoded features. The student can refine the segmentation results by incorporating context information and spatial relationships between different image regions. By aligning decoder features, the student can learn to reconstruct segmentation results consistent with those produced by the teacher network, leading to improved segmentation accuracy and spatial coherence.

Each scale (encoder, bottleneck, decoder) uniquely captures different levels of abstraction and semantic information from the input image. Aligning features at each scale facilitates effective knowledge transfer from the teacher to the student, enhancing the student's ability to perform segmentation tasks across multiple levels of abstraction. In the upcoming subsections, we will discuss in depth about how this knowledge distillation is happening. 

\subsection{Prediction Maps Distillation Loss}
\label{subsection:Prediction Maps Distillation Loss}
The Prediction Map Distillation (PMD) loss function, based on the principles outlined in~\cite{hinton2015distilling} transfers knowledge from a teacher network to a student network. It works by comparing the softmax outputs of the student and teacher networks and adjusting for a temperature parameter, $\mathcal{T}$. The PMD loss quantifies the difference in probability distributions by calculating the Kullback-Leibler (KL) divergence between the logarithm of the student's softmax output and the teacher's softmax output. The PMD loss quantifies the discrepancy between their knowledge representations. This discrepancy serves as a measure of the difference in the knowledge encoded by the teacher and learned by the student. By optimizing the student network to minimize this PMD loss, the student gradually acquires knowledge that is consistent with that of the teacher. In essence, the PMD loss acts as a guiding signal during the training process, enabling the student network to distil and absorb the rich information encoded in the teacher's outputs, ultimately enhancing its performance and generalization capabilities. The below equation mathematically explains the PMD loss: 
\begin{equation}
 \mathcal{L}_{\text{PMD}} = \text{KL}\left(\text{log\_softmax}\left(\frac{y}{\mathcal{T}}\right), \text{softmax}\left(\frac{\text{teacher\_scores}}{\mathcal{T}}\right)\right) \times \mathcal{T}^2   
\end{equation}
\subsection{Multi-Scale Contrastive Loss}
\label{subsection:Multi-Scale Contrastive Loss}
Here, we will discuss how we are utilizing contrastive learning to compute the loss between the features from the teacher network and the student network. We compute the contrastive loss between the teacher and student model at various scales such as encoder-to-encoder, bottleneck-to-bottleneck and decoder-to-decoder. We use InfoNCE~\cite{oord2019representation} as the contrastive loss. Mathematically we can define it as -
\begin{equation}
\begin{aligned}
\mathcal{L}_{con} &= -E_X \left[ \log \left( \frac{f_k(x_{t+k}, c_t)}{ \sum_{x_j \in X} f_k(x_j, c_t)} \right) \right]
\end{aligned}    
\end{equation}
\text{where:}
\begin{itemize}
    \item $E_X$ denotes the expectation taken over samples $X$, typically drawn from the dataset.
    \item $x_{t+k}$ is a positive sample representing the data at a time step $t + k$.
    \item $c_t$ represents the context embedding or representation at time step $t$.
    \item $f_k(x_{t+k}, c_t)$ is the similarity score between the positive sample $x_{t+k}$ and the context representation $c_t$.
    \item $\sum_{x_j \in X} f_k(x_j, c_t)$ represents the sum of similarity scores between all samples in the dataset and the context representation $c_t$.
\end{itemize}
InfoNCE encourages the student network to learn representations that are not only semantically meaningful but also consistent with those learned by the teacher network. By enforcing consistency in the feature space, the contrastive loss ensures that the student network captures relevant information from the input images, which is crucial for accurate segmentation. Moreover, capturing subtle details and understanding the semantic content of the CT images are critical for accurate segmentation, the contrastive loss helps align the semantic representations learned by the teacher and the student network. This ensures that both networks focus on capturing similar semantic features, which leads to better generalization and segmentation performance. During the knowledge transfer, the contrastive loss encourages the student network to learn discriminative features. By maximizing the agreement between the features of the teacher and student networks, contrastive learning guides the student to focus on learning features that are relevant for distinguishing different classes or structures in medical images. The contrastive loss used in our experiment also acts as a regularization term in the KD process.
The regularization helps prevent overfitting by penalizing complex models that may memorize the training data. By encouraging the student network to produce features that are similar to those of the teacher network, the loss function imposes a regularization effect that encourages the student to learn more robust and generalizable representations. During the knowledge distillation, efficient interaction of features between the teacher and the student network is also critical. Instead of directly mimicking the teacher's output probabilities, the contrastive loss chosen for our investigation focuses on aligning the internal representations learned by the two networks. Considering, that we are performing bottleneck-to-bottleneck knowledge distillation, our distillation loss can be written as follows - 
\begin{equation}
\mathcal{L}_{con}(B_{T}, B_{S}) = - \mathbb{E}_X \left[ \log \left( \frac{f_k(B_{t}, B_{s})}{\sum_{B_j \in X} f_k(B_j, B_{s})} \right) \right]   
\label{eq:Bottleneck_ConLoss}
\end{equation}
\text{where:}
\begin{itemize}
    \item \( \mathcal{L}_{con}(B_{T}, B_{S}) \) represents the contrastive loss between the bottleneck features of the teacher and student networks.
    \item \( \mathbb{E}_X \) denotes the expectation taken over samples \( X \), typically drawn from the dataset.
    \item \( B_{T} \) represents the bottleneck features of the teacher network.
    \item \( B_{S} \) represents the bottleneck features of the student network.
    \item \( f_k(B_{T}, B_{S}) \) is the similarity score between the bottleneck features of the teacher and student networks.
    \item \( \sum_{B_j \in X} f_k(B_j, B_{S}) \) represents the sum of similarity scores between all bottleneck features in the dataset and the bottleneck features of the student network.
\end{itemize}
Now let's write the overall loss that needs to be minimized during the training of the student model. Let's denote the encoder features of the teacher network as $E_T$, the encoder features of the student network as $E_S$, the bottleneck features of the teacher network as $B_T$, the bottleneck features of the student network as $B_S$, the decoder features of the teacher network as $D_T$, and the decoder features of the student network as $D_S$. The overall objective during knowledge distillation can be represented mathematically as follows: 
\begin{equation}
\begin{aligned}
\mathcal{L}_{\text{Total}} ={} &\mathcal{L}_{\text{seg}} \cdot w_{\text{seg}} + \mathcal{L}_{con}(E_{T}, E_{S}) \cdot w_{\text{enc}} + \mathcal{L}_{con}(B_{T}, B_{S}) \cdot w_{\text{bn}} + \\ 
& \mathcal{L}_{con}(D_{T}, D_{S}) \cdot w_{\text{dec}} + \mathcal{L}_{\text{PMD}}
\end{aligned}
\label{eq:total_loss}
\end{equation}
\text{where:}
\begin{itemize}
    \item \( \mathcal{L}_{\text{Total}} \) represents the total objective during knowledge distillation for the student network.
    \item \( \mathcal{L}_{\text{seg}} \) is the segmentation loss and the corresponding weight, \( w_{\text{seg}} \).
    \item \( \mathcal{L}_{\text{con}}(E_{T}, E_{S}) \) represents the contrastive loss between the encoder features of the teacher and student networks. \( w_{\text{enc}} \) is the weight assigned to the encoder contrastive loss.
    \item \( \mathcal{L}_{\text{con}}(B_{T}, B_{S}) \) represents the contrastive loss between the bottleneck features of the teacher and student networks, \( w_{\text{bn}} \) is the weight assigned to the bottleneck contrastive loss.
    \item \( \mathcal{L}_{\text{con}}(D_{T}, D_{S}) \) represents the contrastive loss between the decoder features of the teacher and student networks, \( w_{\text{dec}} \) is the weight assigned to the decoder contrastive loss.
    \item \( \mathcal{L}_{\text{PMD}} \) represents any additional loss specific to your knowledge distillation process, such as pixel-wise mean squared difference (PMD) loss.
\end{itemize}
The student network is trained to minimize a weighted sum of the losses described above, where $w_*$ are weighting factors used to balance the impact of individual loss terms. It's important to note that the teacher network is solely utilized for knowledge transfer and is maintained with pre-trained weights. Note - Once the student network is trained, the teacher network is no longer involved in the final inference process. The student network, being less computationally expensive, distils the knowledge acquired from the teacher network. In the next chapter, we will discuss our experimental results both qualitative and quantitative.
\section{Summary}
This chapter explained how the research problem is addressed by detailing the processes and techniques used including data preprocessing and creating the architecture for our investigation.
We describe the design intricacies of both the teacher and the student model, highlighting their ability to capture a diverse range of features and representations essential for segmentation. A detailed discussion on PMD and Multi-Scale Contrastive Loss is also provided. 
In this chapter, we discussed the objective function to be optimized and contrastive learning, providing a mathematical intuition towards the objective. We explained the mathematical formulations and operational principles underlying these loss functions, clarifying their role in guiding the training of student networks to effectively distil knowledge from the teacher models.

This chapter presents a detailed explanation of the methodology, which lays the groundwork for subsequent experimental analyses and validation in the context of medical image segmentation. 

\chapter{Experimental Results}
\label{chapter:results}
In this chapter, we present the results of our experiments and conduct a comprehensive analysis of our findings across multiple dimensions. We begin by describing the implementation details of our experimental setup in section~\ref{setion:Implementation Details}, where we also discuss the software and hardware configurations used as well as the model hyperparameters and the choice of optimizer. Next, we describe the evaluation metrics used to assess the performance of our models in section~\ref{section:Evaluation Metrics}. The training curves are provided in section~\ref{section:Visualizing Training Process}. Following this, section~\ref{section:Quantitative Results} and \ref{section:Qualitative Results} present our quantitative and qualitative results from the experiments we performed, respectively. Ablation study to investigate the effect of knowledge distillation at multi-scale is discussed in section~\ref{section:Ablation Studies}. At the end of this chapter in section~\ref{section:Statistical Significance of the Model}, we discussed about the statistical significance of our model as well. 

\section{Implementation Details}
\label{setion:Implementation Details}
The training set to train the teacher network has 2920 images with corresponding ground truth masks and 730 images in the testing set. We chose not to perform any data augmentation, as the goal of this experiment is to understand the implications of knowledge distillation rather than creating a state-of-the-art model. The training environment was set up to ensure reproducibility and consistency during experimentation. All of our experiments are carried out using PyTorch. For training and inference, we employ a single NVIDIA RTX 4070Ti GPU with 12 GB of VRAM. To train the teacher network, we utilized AdamW~\cite{loshchilov2017decoupled} with a learning rate of 1e-4 and a batch size of 8. We trained the teacher network for 200 epochs. The student network is trained over 120 epochs using the RMSProp optimizer during the knowledge distillation process. We initiate the training process for all our models using randomly initialized weights. This approach helps prevent any pre-existing biases and allows the models to learn meaningful representations from the data without prior assumptions or constraints. We have kept the image size at (256$\times$256$\times$3). The teacher model is trained using dice BCE (Binary Cross-Entropy) loss and we used dice loss for training the student model. Request access to code via this \href{https://github.com/RisabBiswas/Knowledge-Distillation-in-Medical-Image-Segmentation}{github}.

\section{Evaluation Metrics}
\label{section:Evaluation Metrics}
We have chosen to use standard evaluation metrics such as Intersection over Union (IoU), precision, recall, and Dice coefficient to measure the quality of our segmentation results. These metrics are widely accepted, unbiased, and easy to understand for quick conclusions.

\textbf{Intersection over Union (IoU):}
IoU measures the overlap between the predicted segmentation and the ground truth, providing a measure of segmentation accuracy. The IoU measures how well the predicted segmentation matches the ground truth, thereby assessing the segmentation task's accuracy. A higher IoU indicates better alignment between the predicted and the ground truth segmentations, demonstrating the model's ability to accurately identify relevant structures or abnormalities in medical images.
\begin{equation}
\text{IoU} = \frac{|\mathbf{y}_{\text{seg}} \cap \mathbf{y}_{\text{true}}|}{|\mathbf{y}_{\text{seg}} \cup \mathbf{y}_{\text{true}}|} \\  
\end{equation}
\textbf{Dice Coefficient:}
The Dice coefficient is also, widely recognized as an important evaluation metric in medical imaging segmentation problems due to its applicability and practicality in assessing volume segmentation task accuracy. The Dice coefficient, denoted as DICE($\mathbf{y}_{\text{seg}}$, $\mathbf{y}_{\text{true}}$), quantifies the similarity between the predicted segmentation and the ground truth segmentation of volumetric tumour masks. The metric function is defined as follows:
\begin{equation}
\text{DICE}(\mathbf{y}_{\text{seg}},\mathbf{y}_{\text{true}}) = \frac{2 \times |\mathbf{y}_{\text{seg}} \cap \mathbf{y}_{\text{true}}|}{|\mathbf{y}_{\text{seg}}| + |\mathbf{y}_{\text{true}}|}
\end{equation}
\textbf{Recall:}
Recall, also known as sensitivity, refers to the proportion of true positive cases correctly identified by the model. In medical image segmentation, recall is significant for detecting all relevant structures or abnormalities, even if it means generating more false positives. A high recall value indicates that the model can accurately capture all instances of the target class, reducing the risk of missing clinically significant findings in medical images.
\begin{equation}
\text{Recall} = \frac{TP}{TP + FN} \\
\end{equation}
\textbf{Precision:}
Precision refers to the percentage of positive cases predicted by the model that were positive. Precision in medical image segmentation is critical for reducing false positives and ensuring the accuracy of detected structures or abnormalities. A high precision value indicates that the model generates fewer false positives, lowering the risk of unnecessary interventions or misinterpretations in clinical practice. Precision enhances recall by emphasizing the model's ability to identify relevant structures while minimizing false positives.
\begin{equation}
\text{Precision} = \frac{TP}{TP + FP}
\end{equation}
here, TP, FP and FN represent true positives, false positives, and false negatives.
$\mathbf{y}_{\text{seg}}$ and $\mathbf{y}_{\text{true}}$ represent the predicted and the ground truth segmentation, respectively. $|\mathbf{y}_{\text{seg}}|$ \text{ represents the areas of the predicted regions}. $|\mathbf{y}_{\text{true}}|$ represents the areas of the ground truth regions, $|$$\mathbf{y}_{\text{seg}}$ $\cap$ $\mathbf{y}_{\text{true}}$$|$ represents the intersection of the predicted and ground truth regions, $|$$\mathbf{y}_{\text{seg}}$ $\cup$ $\mathbf{y}_{\text{true}}$$|$ represents the union of the predicted and ground truth regions.

To evaluate the reconstruction results of the multi-task teacher model, we are using PSNR. PSNR between the reconstructed images ($\mathbf{y}_{rec}$) and the original images ($\mathbf{x}$), you can use the following formula:
\begin{equation}
\text{PSNR}(\mathbf{y}_{rec}, x) = 10 \cdot \log_{10} \left( \frac{{\text{MAX}^2}}{{\text{MSE}(y_{\text{rec}}, x)}} \right)   
\end{equation}
MAX is the maximum possible pixel value of the images (e.g., 255 for 8-bit images). MSE ($\mathbf{y_{rec}}$, $\mathbf{x}$) is the mean squared error between the reconstructed images $\mathbf{y_{rec}}$ and the original images $\mathbf{x}$, calculated as the average of the squared differences between corresponding pixels.

\section{Training Progress Overview}
\label{section:Visualizing Training Process}
We also provide graphs showing the convergence of the networks we created. This is critical for understanding our model's performance and generalization capabilities. The training curve shows how the model's performance improves with each iteration or epoch as it learns from the training data. In contrast, the validation curve shows how well the model generalizes to previously unseen data by evaluating its performance on a separate validation set. Examining these curves gives insight into several aspects of our model's training, such as overfitting and underfitting. If the training curve continues to improve while the validation curve stagnates or worsens, this indicates overfitting, which occurs when the model learns to memorize the training data rather than generalizing to new instances. If both curves plateau at a suboptimal performance level, this could indicate underfitting, which occurs when the model is too simple to capture the underlying patterns in the data. The shape of the curves provides information about the training progress. A decline in the training curve followed by gradual convergence indicates effective learning, whereas a non-declining curve may indicate instability in the training process.
\begin{figure}[htbp]
  \centering
  \begin{subfigure}[b]{0.49\textwidth}
    \centering
    \includegraphics[width=\textwidth]{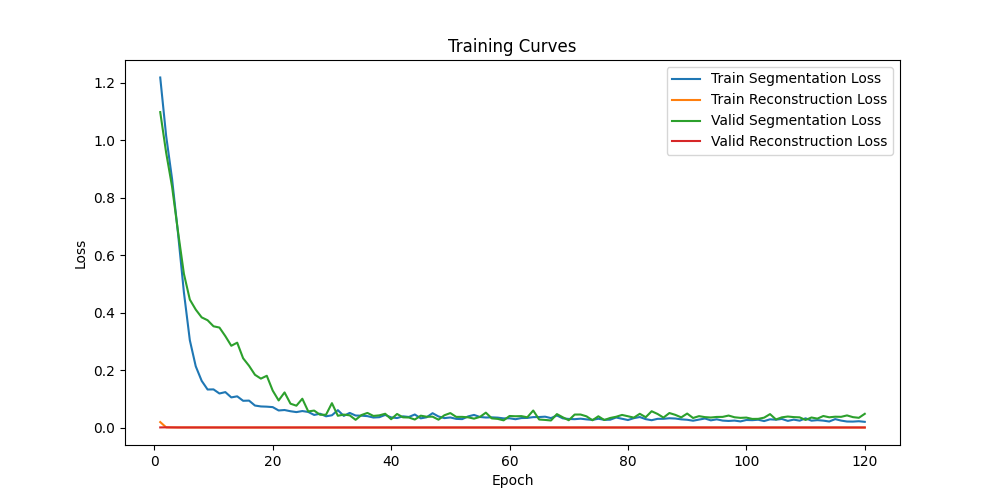}
    \caption{MultiTask UNet (T1)}
  \end{subfigure}
  \hfill
  \begin{subfigure}[b]{0.49\textwidth}
    \centering
    \includegraphics[width=\textwidth]{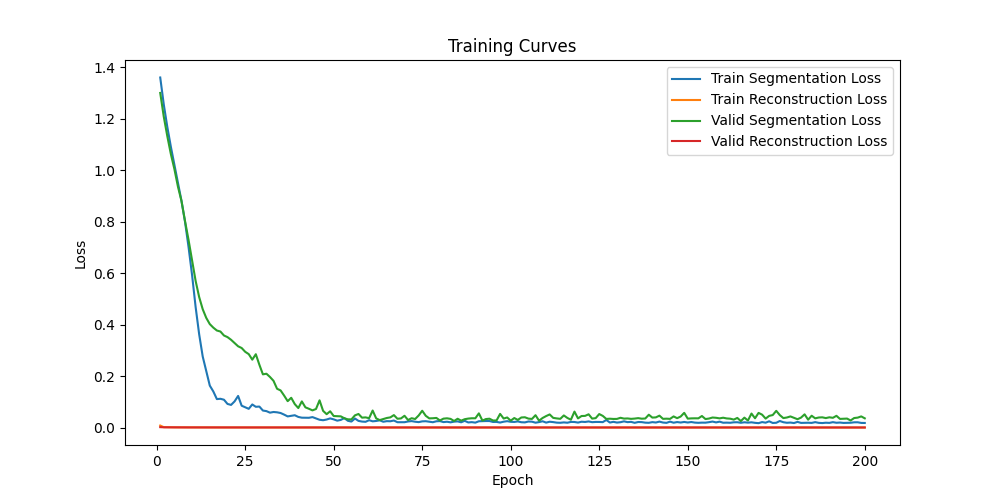}
    \caption{MultiTask TransUNet (T2)}
  \end{subfigure}
  \caption{Training Curves for Teacher Model}
  \label{fig:training_curves_T1&T2}
\end{figure}
As shown in Figure~\ref{fig:training_curves_T1&T2}, the validation loss for the segmentation task remained relatively stable after \~40 epochs for both teacher networks T1 and T2. This stabilization implies that the models' performance did not improve significantly beyond this point, as evidenced by the lack of a significant decrease in validation loss. We deliberately chose not to implement any early stopping mechanisms during the training process. Early stopping typically entails monitoring the validation loss and terminating training when it stops improving or begins to deteriorate, preventing overfitting and conserving computational resources. However, in this case, we decided to let the training continue uninterrupted, despite the plateau in validation loss. This decision was made after considering several factors. First, we wanted to see if the models could improve further with more training epochs, although validation loss had stabilized. Second, early stopping introduces subjectivity in determining the optimal stopping point, which can vary depending on the dataset, model architecture, and training objectives.

For knowledge distillation, where the teacher model serves as a reference for training the student model, we decided against using early stopping for a variety of logical and technical reasons. Knowledge distillation typically involves training the teacher model only once because it serves as a fixed reference during the training of the student model. Unlike traditional training scenarios, where early stopping may be used to avoid overfitting or to save computational resources, knowledge distillation focuses on extracting the most informative knowledge from the teacher model rather than fine-tuning its performance. As a result, early stopping may not be necessary or beneficial in this situation.

\begin{figure}[htbp]
  \centering
  \begin{subfigure}[b]{0.49\textwidth}
    \centering
    \includegraphics[width=\textwidth]{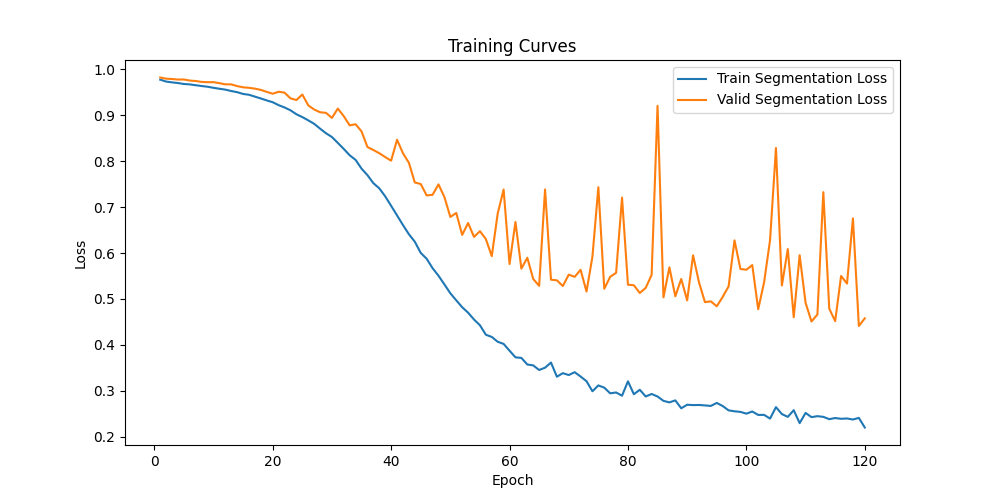}
    \caption{Student Model (S1)}
  \end{subfigure}
  \hfill
  \begin{subfigure}[b]{0.49\textwidth}
    \centering
    \includegraphics[width=\textwidth]{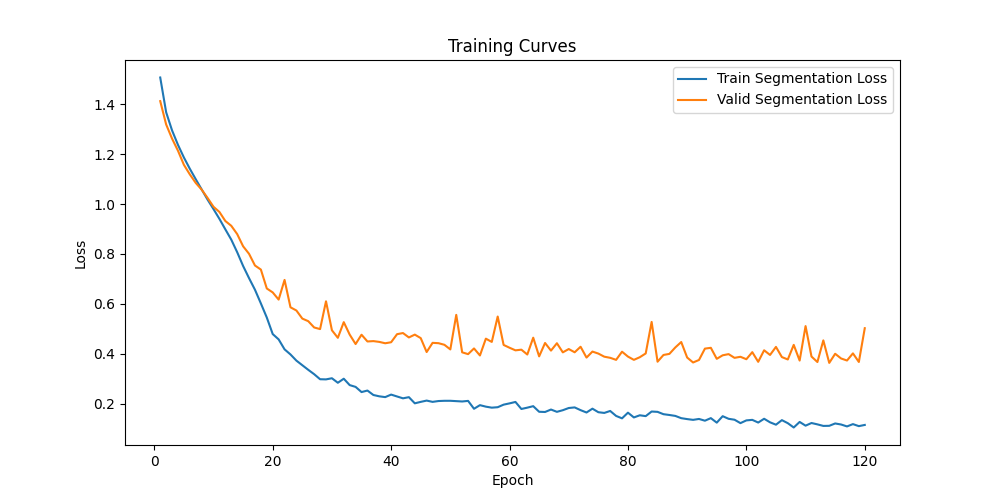}
    \caption{Student Model (S2)}
  \end{subfigure}
  \caption{Training Curves for Student Model}
  \label{fig:training_curves_S1&S2}
\end{figure}

Figure~\ref{fig:training_curves_S1&S2} provides an overview of the training and validation curves for our student models, designated as S1 and S2. As depicted in the training plot, S1, being a smaller model trained on a limited dataset, encountered challenges in achieving smooth convergence. Specifically, the validation loss of S1, illustrated in orange in Figure~\ref{fig:training_curves_S1&S2} (a), exhibits discernible spikes, indicating fluctuations in the optimization process compared to the training loss curve.

Conversely, S2 (Figure~\ref{fig:training_curves_S1&S2} (b)), despite being slightly larger, demonstrates a smoother convergence pattern when trained on the same dataset as S1. This is a good observation to show how the model size matters to the overall performance of the task i.e. segmentation here. As discussed previously, this has been our primary objective throughout this thesis, which is to investigate how we can mitigate this performance shortcoming using techniques such as knowledge distillation and contrastive learning, to make S1 perform closer to S2 without explicitly increasing its size. 
\begin{figure}[htbp]
 \centering
 \begin{minipage}[b]{0.48\textwidth}
  \centering
  \includegraphics[width=\textwidth]{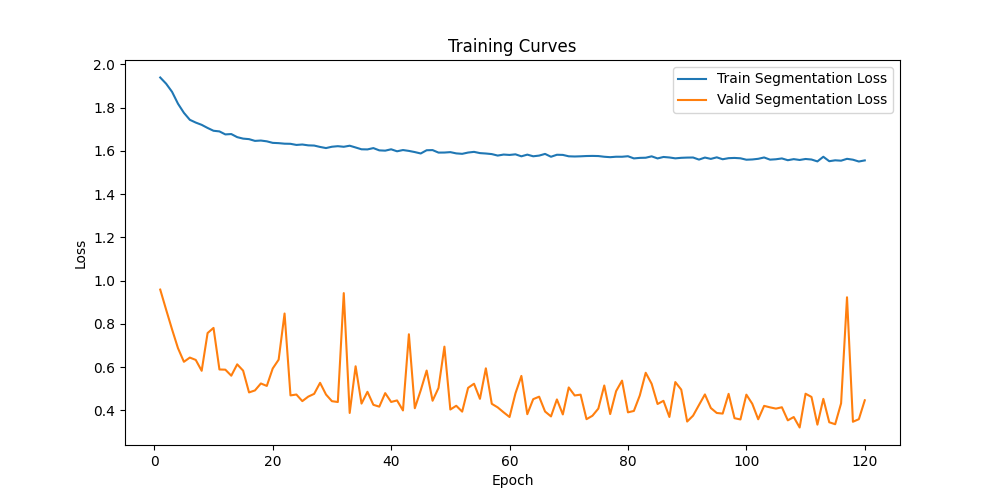}
  \caption*{B$\rightarrow$B}
 \end{minipage}
 \hfill
 \begin{minipage}[b]{0.48\textwidth}
  \centering
  \includegraphics[width=\textwidth]{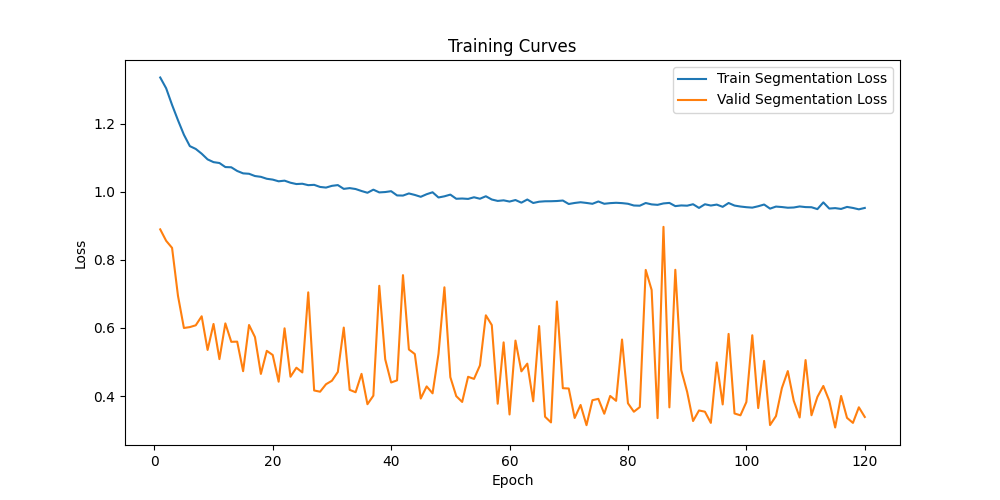}
  \caption*{B$\rightarrow$B + PMD}
 \end{minipage}
 
 \begin{minipage}[b]{0.48\textwidth}
  \centering
  \includegraphics[width=\textwidth]{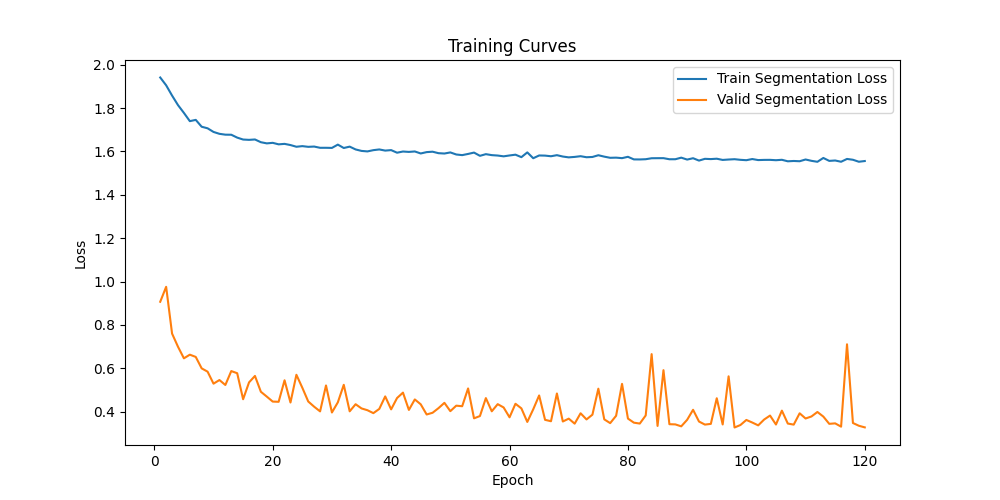}
  \caption*{B$\rightarrow$B + E$\rightarrow$E}
 \end{minipage}
 \hfill
 \begin{minipage}[b]{0.48\textwidth}
  \centering
  \includegraphics[width=\textwidth]{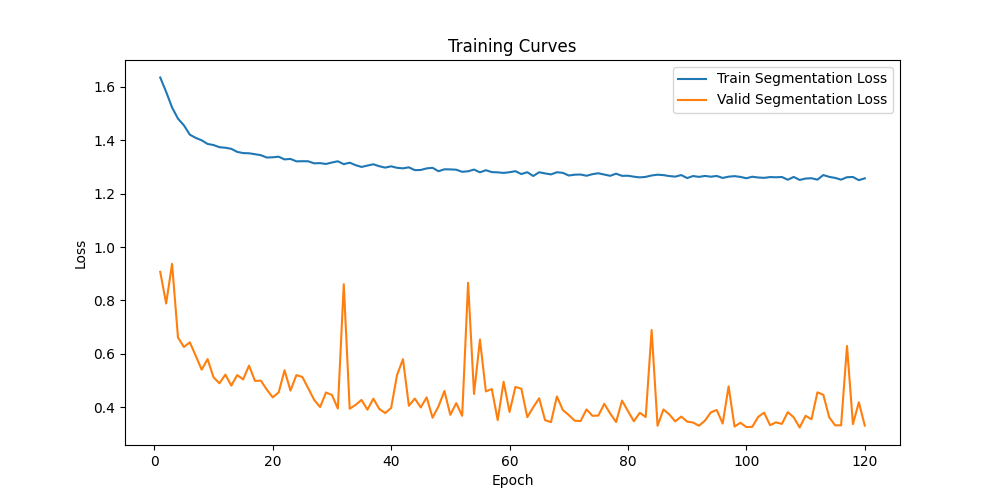}
  \caption*{B$\rightarrow$B + E$\rightarrow$E + PMD}
 \end{minipage}
 
 \begin{minipage}[b]{0.48\textwidth}
  \centering
  \includegraphics[width=\textwidth]{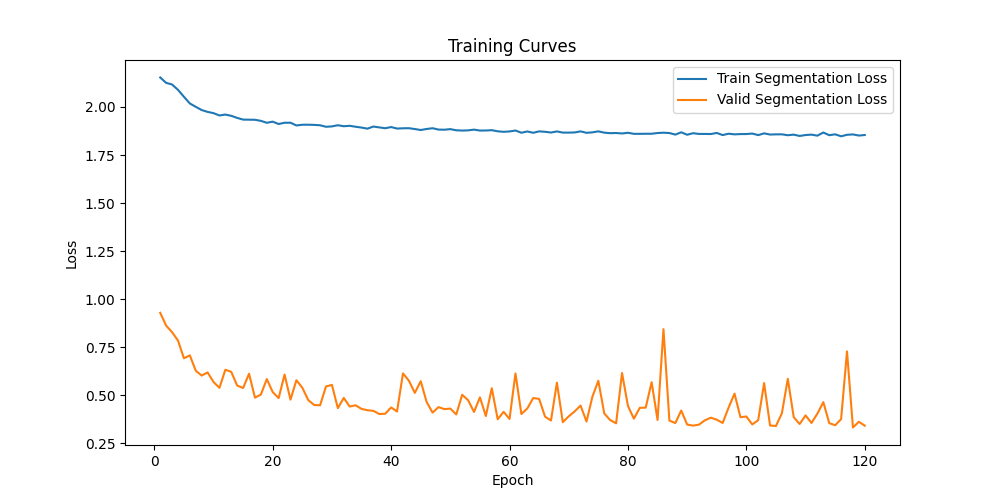}
  \caption*{B$\rightarrow$B + E$\rightarrow$E + D$\rightarrow$D}
 \end{minipage}
 \hfill
 \begin{minipage}[b]{0.48\textwidth}
  \centering
  \includegraphics[width=\textwidth]{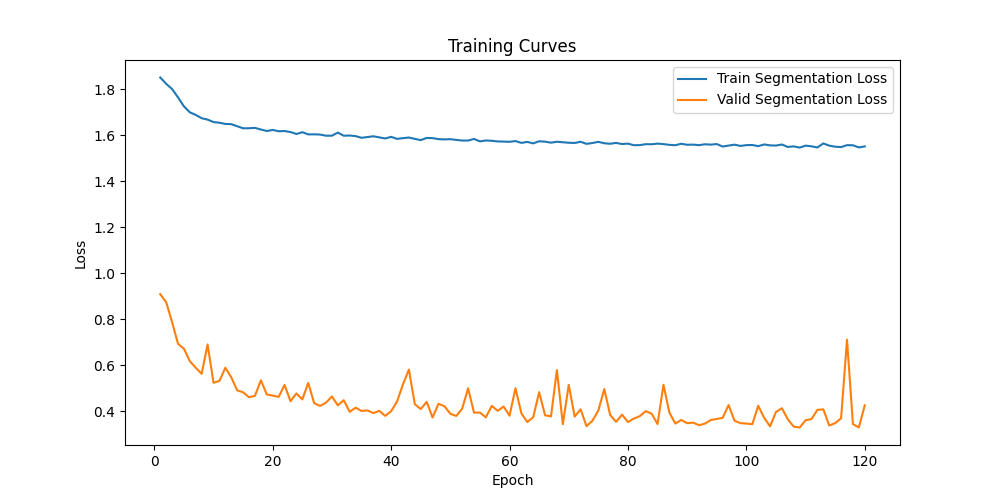}
  \caption*{B$\rightarrow$B + E$\rightarrow$E + D$\rightarrow$D + PMD}
 \end{minipage}
 
 \begin{minipage}[b]{0.48\textwidth}
  \centering
  \includegraphics[width=\textwidth]{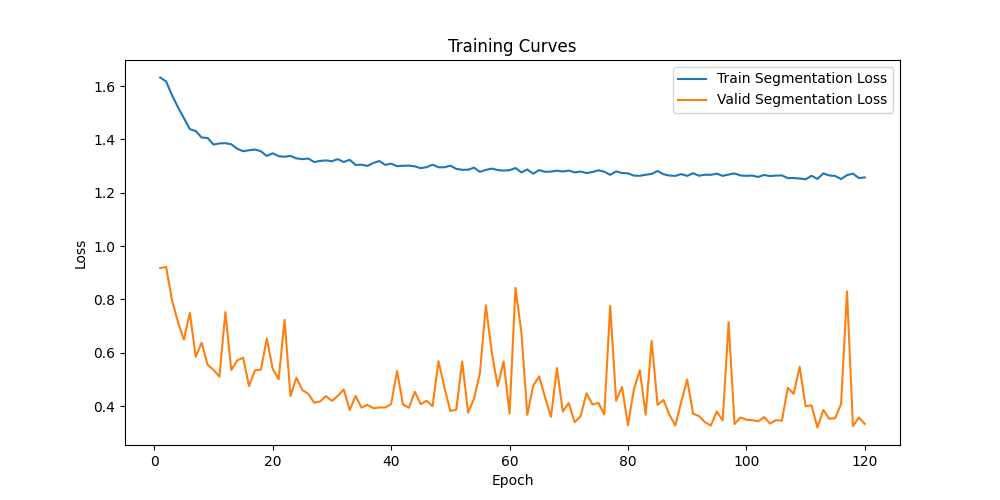}
  \caption*{E$\rightarrow$E}
 \end{minipage}
 \hfill
 \begin{minipage}[b]{0.48\textwidth}
  \centering
  \includegraphics[width=\textwidth]{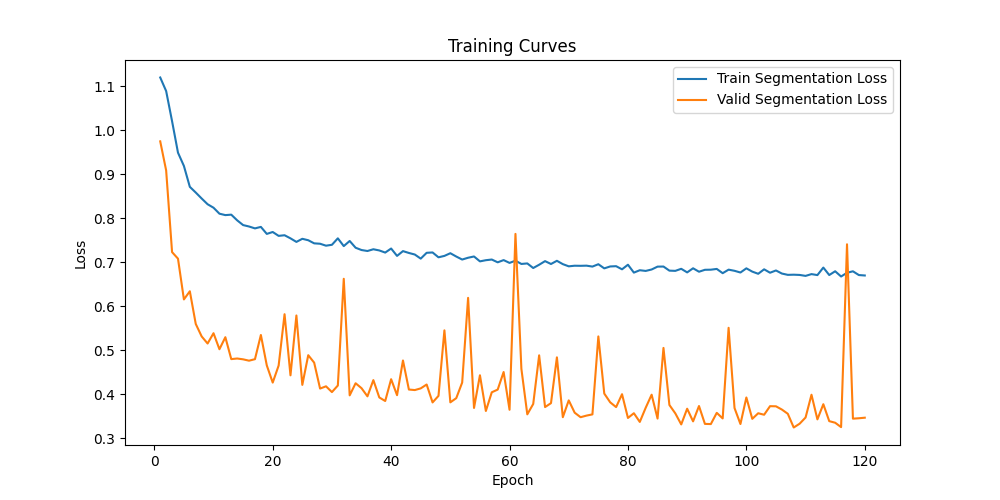}
  \caption*{E$\rightarrow$E + PMD}
 \end{minipage}
 \captionsetup{justification=centering}
 \caption{Training Curves at Various Scales of Knowledge Distillation Between T1 and S1 Using Contrastive Learning.}
 \label{fig:T1-S1_KD_ConLoss_training_curves}
\end{figure}

\begin{figure}[htbp]
 \begin{minipage}[b]{0.48\textwidth}
  \centering
  \includegraphics[width=\textwidth]{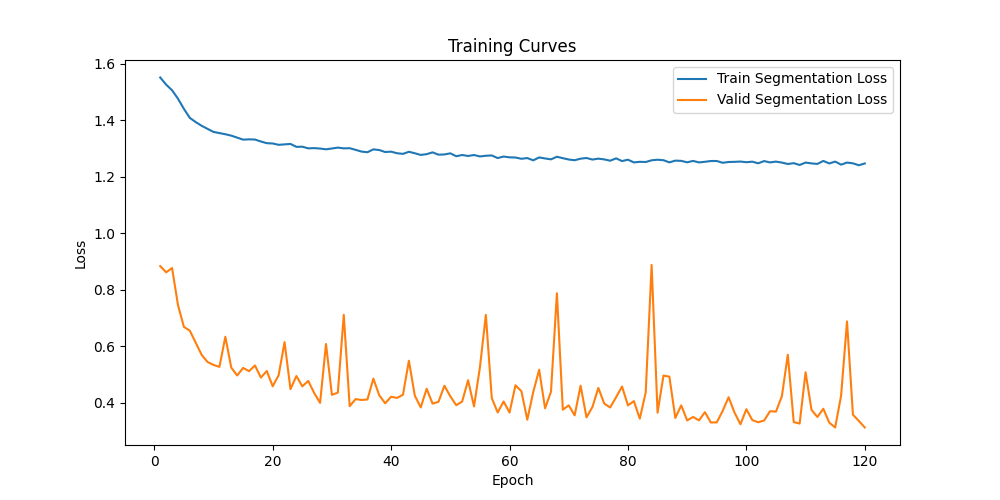}
  \caption*{E$\rightarrow$E + D$\rightarrow$D}
 \end{minipage}
 \hfill
 \begin{minipage}[b]{0.48\textwidth}
  \centering
  \includegraphics[width=\textwidth]{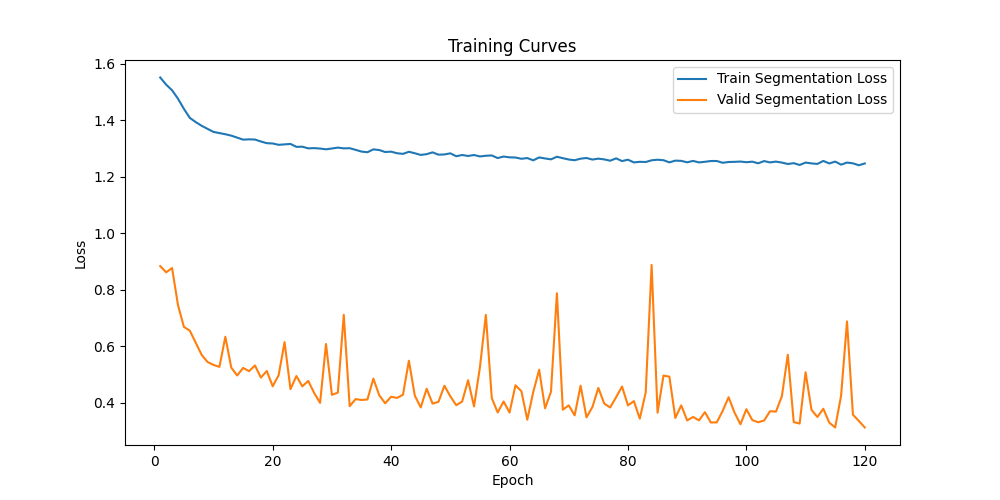}
  \caption*{E$\rightarrow$E + D$\rightarrow$D + PMD}
 \end{minipage}
 
 \begin{minipage}[b]{0.48\textwidth}
  \centering
  \includegraphics[width=\textwidth]{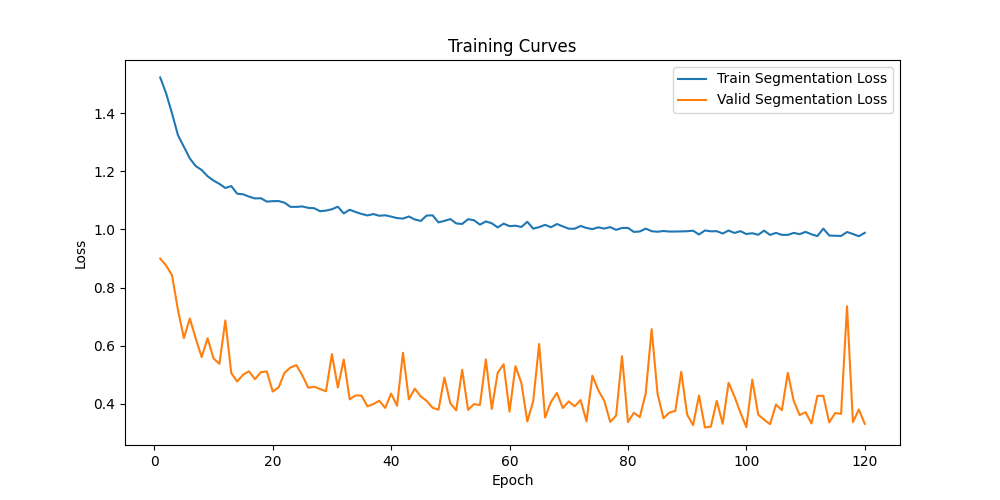}
  \caption*{D$\rightarrow$D}
 \end{minipage}
 \hfill
 \begin{minipage}[b]{0.48\textwidth}
  \centering
  \includegraphics[width=\textwidth]{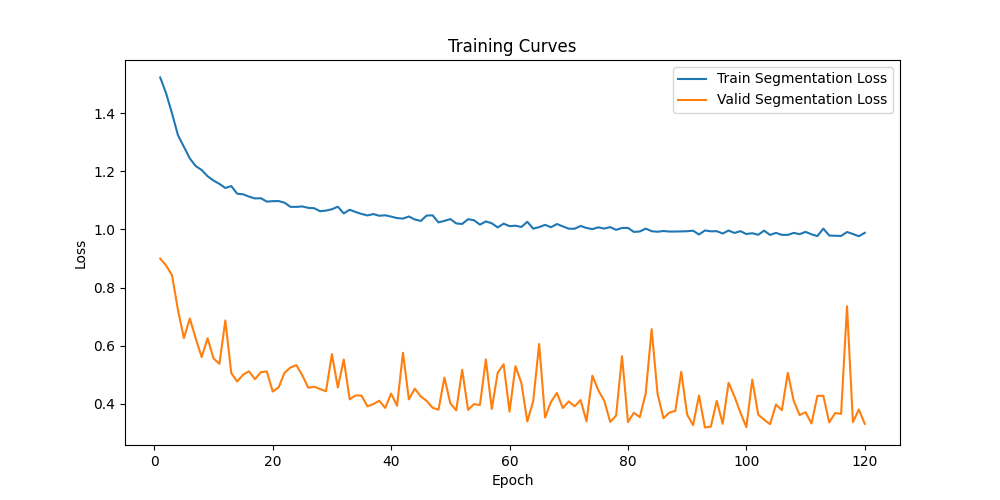}
  \caption*{D$\rightarrow$D + PMD}
 \end{minipage}
 \captionsetup{justification=centering}
 \caption{Training Curves at Various Scales of Knowledge Distillation Between T1 and S1 Using Contrastive Learning.}
 \label{fig:T1-S1_KD_ConLoss_training_curves_2}
\end{figure}
In Figures \ref{fig:T1-S1_KD_ConLoss_training_curves} and \ref{fig:T1-S1_KD_ConLoss_training_curves_2}, we take a look at the dynamics of the training process of S1, when distilling knowledge from T1. These figures provide an in-depth insight into the changes in learning curves across different levels of knowledge distillation, achieved through contrastive learning.

\begin{figure}[htbp]
 \begin{minipage}[b]{0.48\textwidth}
  \centering
  \includegraphics[width=\textwidth]{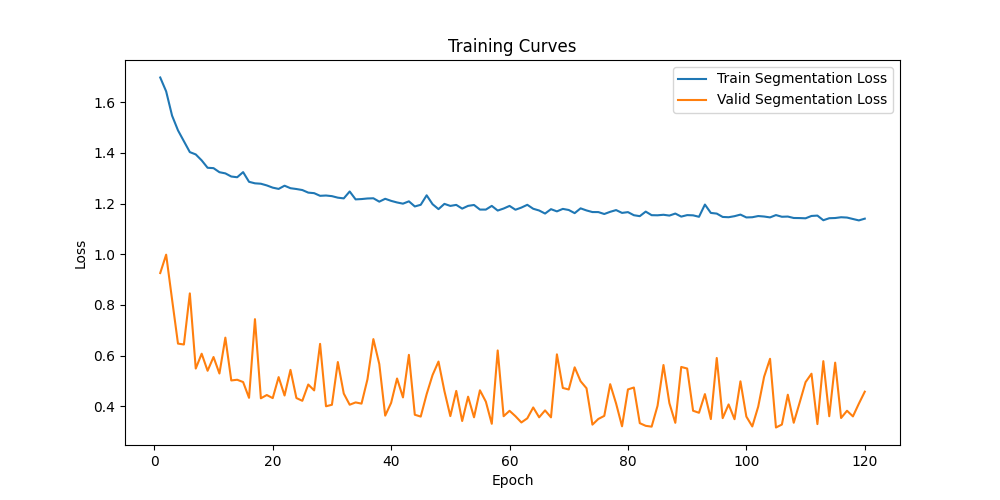}
  \caption*{E$\rightarrow$E + D$\rightarrow$D}
 \end{minipage}
 \hfill
 \begin{minipage}[b]{0.48\textwidth}
  \centering
  \includegraphics[width=\textwidth]{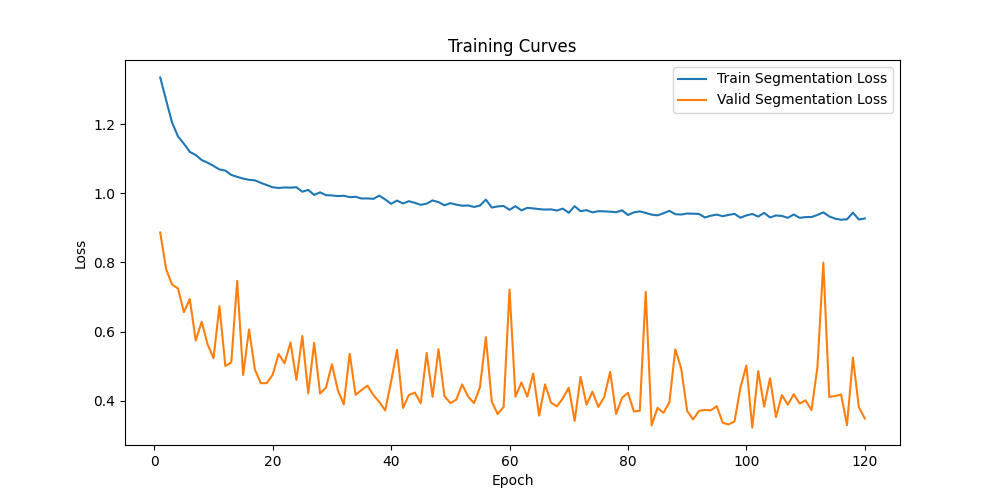}
  \caption*{B$\rightarrow$B + D$\rightarrow$D + PMD}
 \end{minipage}
 \captionsetup{justification=centering}
 \caption{Training Curves at Various Scales of Knowledge Distillation Between T2 and S1 Using Contrastive Learning.}
 \label{fig:T2-S1_KD_ConLoss_training_curves_1}
\end{figure}

\begin{figure}[htbp]
 \centering
 \begin{minipage}[b]{0.48\textwidth}
  \centering
  \includegraphics[width=\textwidth]{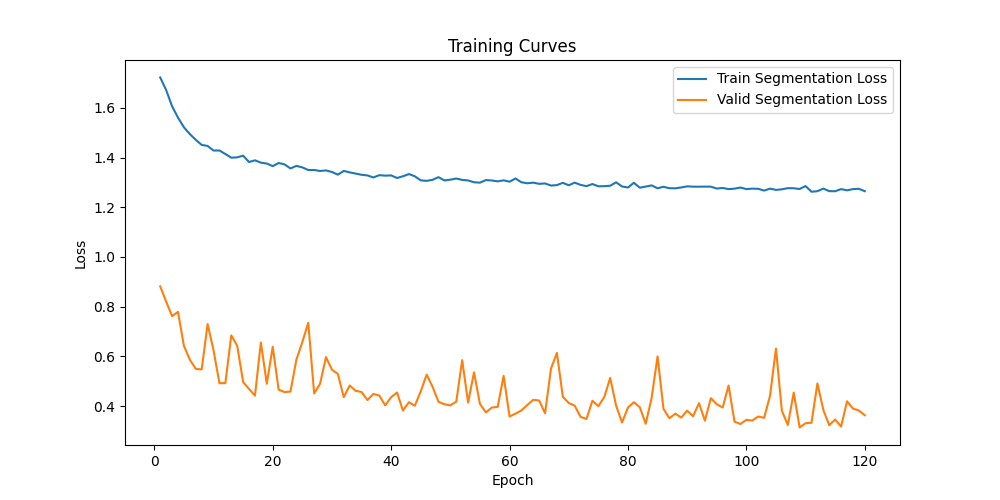}
  \caption*{B$\rightarrow$B}
 \end{minipage}
 \hfill
 \begin{minipage}[b]{0.48\textwidth}
  \centering
  \includegraphics[width=\textwidth]{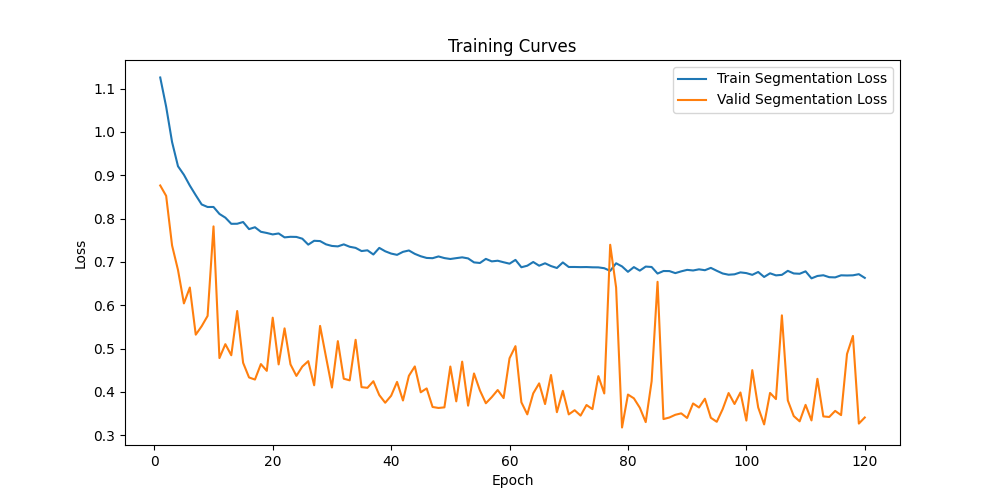}
  \caption*{B$\rightarrow$B + PMD}
 \end{minipage}
 
 \begin{minipage}[b]{0.48\textwidth}
  \centering
  \includegraphics[width=\textwidth]{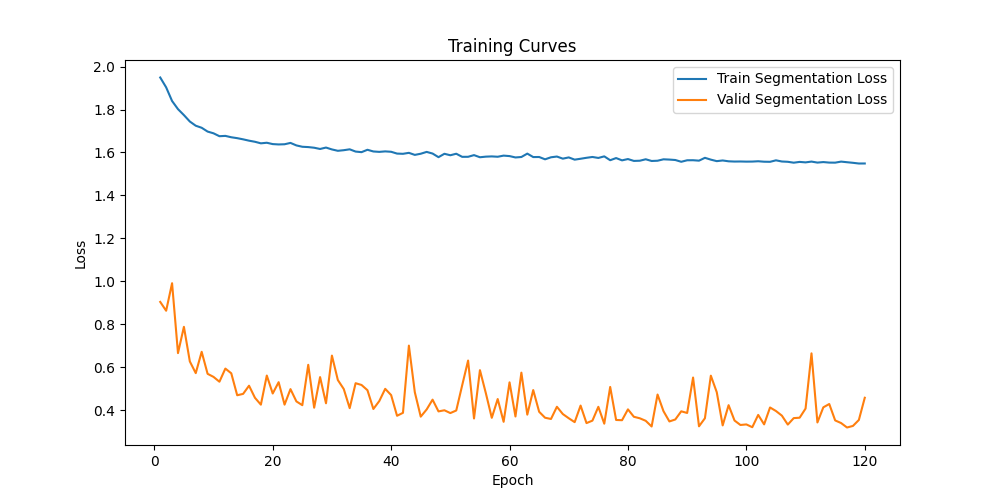}
  \caption*{B$\rightarrow$B + E$\rightarrow$E}
 \end{minipage}
 \hfill
 \begin{minipage}[b]{0.48\textwidth}
  \centering
  \includegraphics[width=\textwidth]{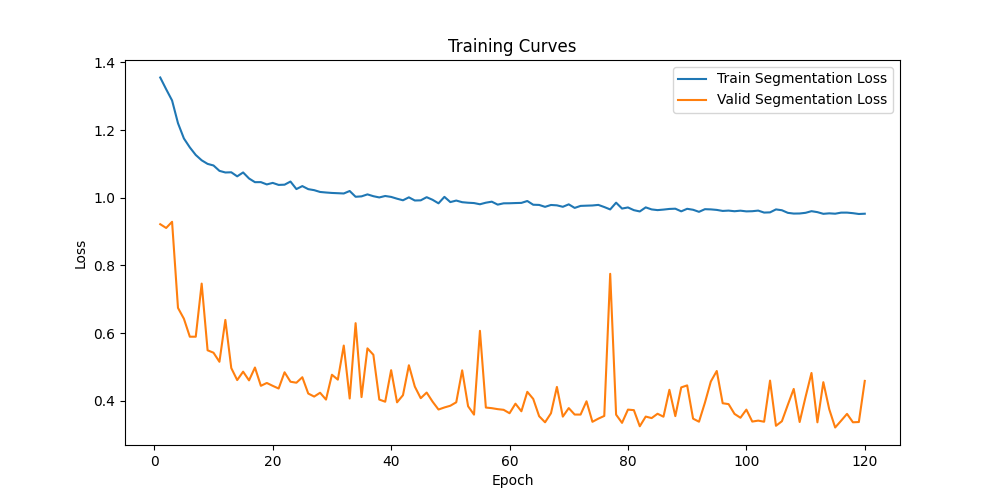}
  \caption*{B$\rightarrow$B + E$\rightarrow$E + PMD}
 \end{minipage}
 
 \begin{minipage}[b]{0.48\textwidth}
  \centering
  \includegraphics[width=\textwidth]{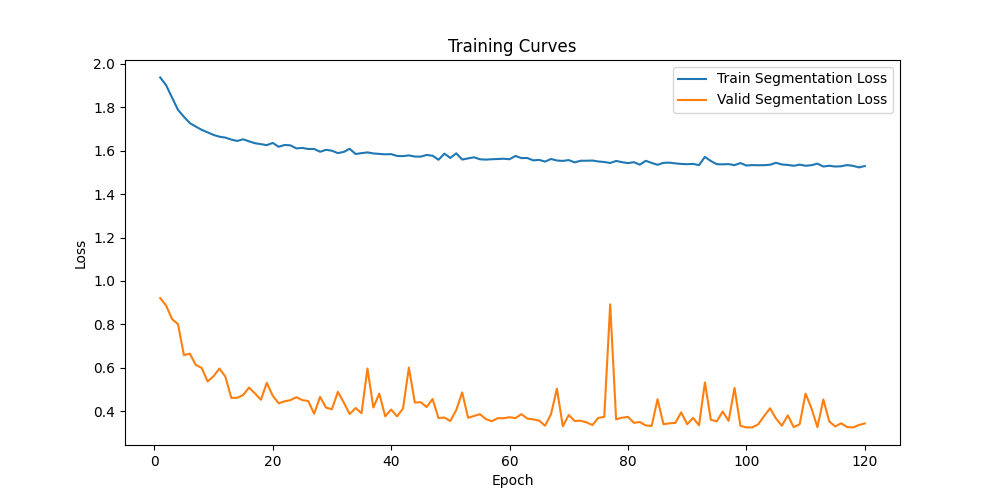}
  \caption*{B$\rightarrow$B + E$\rightarrow$E + D$\rightarrow$D}
 \end{minipage}
 \hfill
 \begin{minipage}[b]{0.48\textwidth}
  \centering
  \includegraphics[width=\textwidth]{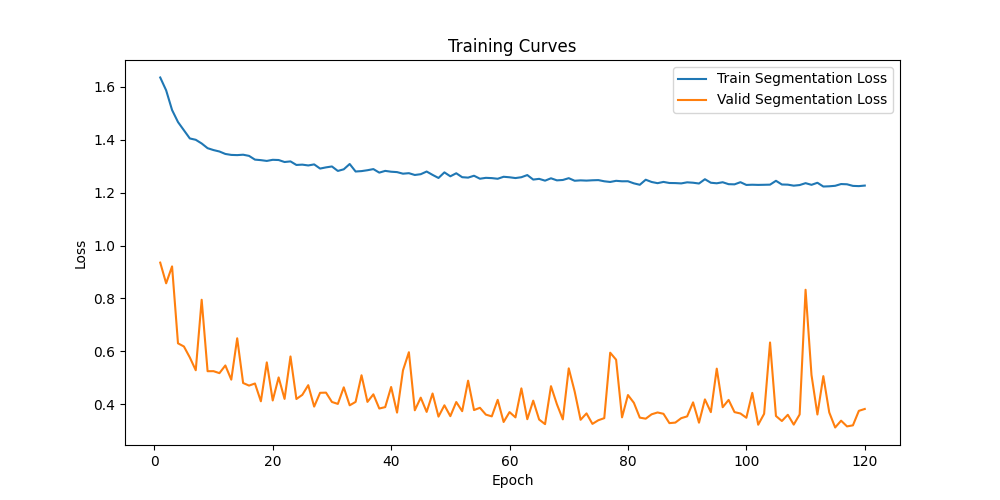}
  \caption*{B$\rightarrow$B + E$\rightarrow$E + D$\rightarrow$D + PMD}
 \end{minipage}
 
 \begin{minipage}[b]{0.48\textwidth}
  \centering
  \includegraphics[width=\textwidth]{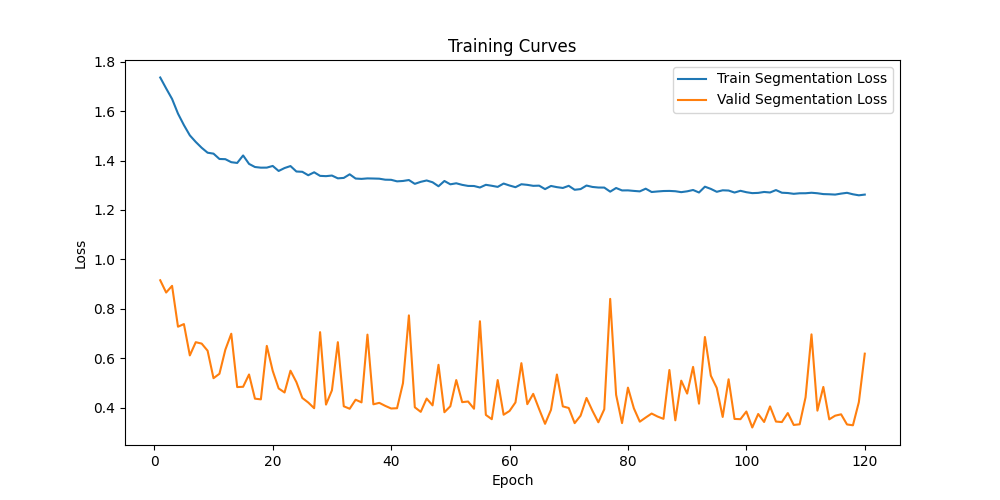}
  \caption*{E$\rightarrow$E}
 \end{minipage}
 \hfill
 \begin{minipage}[b]{0.48\textwidth}
  \centering
  \includegraphics[width=\textwidth]{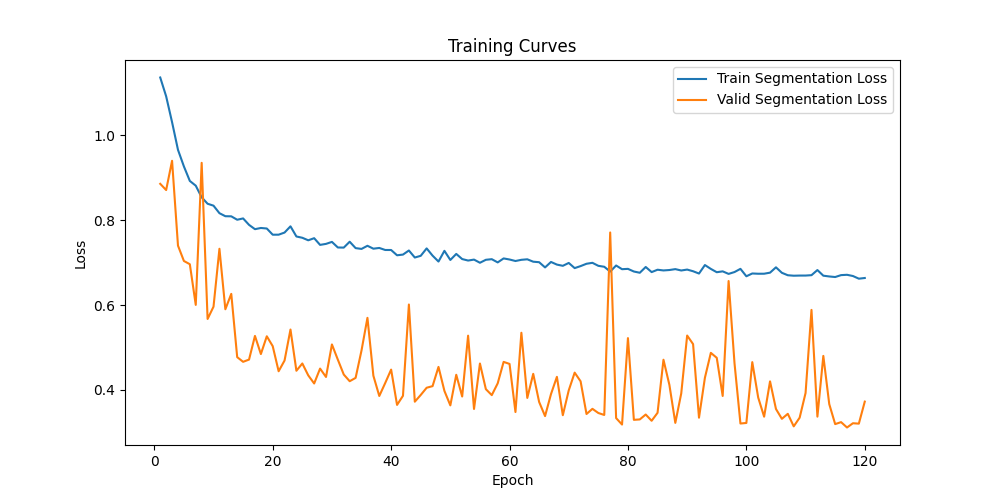}
  \caption*{E$\rightarrow$E + PMD}
 \end{minipage}
 \captionsetup{justification=centering}
 \caption{Training Curves at Various Scales of Knowledge Distillation Between T2 and S1 Using Contrastive Learning.}
 \label{fig:T2-S1_KD_ConLoss_training_curves_2}
\end{figure}
\begin{figure}[htbp]
 \centering
 \begin{minipage}[b]{0.48\textwidth}
  \centering
  \includegraphics[width=\textwidth]{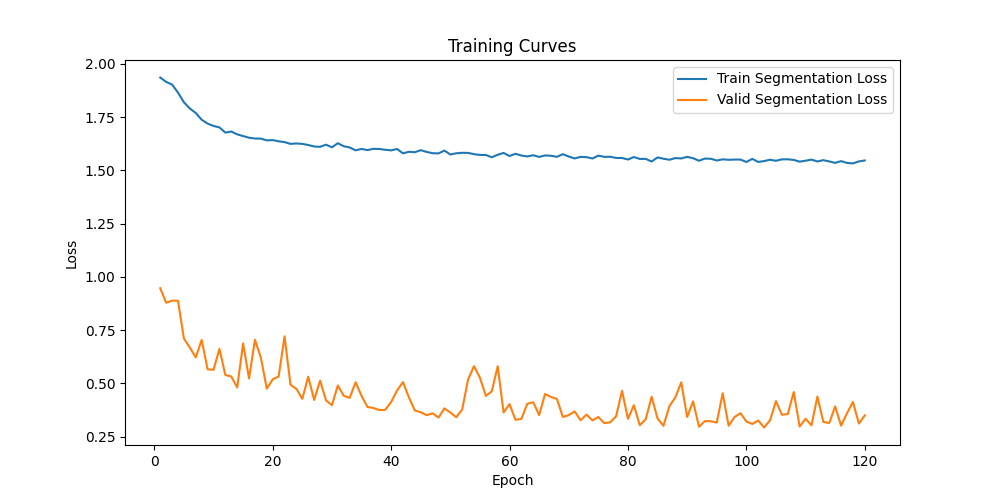}
  \caption*{B$\rightarrow$B}
 \end{minipage}
 \hfill
 \begin{minipage}[b]{0.48\textwidth}
  \centering
  \includegraphics[width=\textwidth]{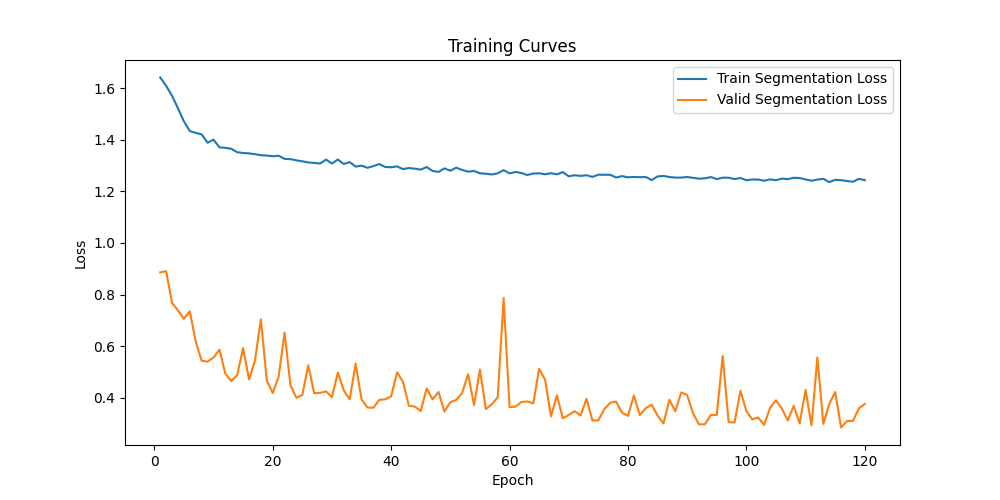}
  \caption*{B$\rightarrow$B + PMD}
 \end{minipage}
 
 \begin{minipage}[b]{0.48\textwidth}
  \centering
  \includegraphics[width=\textwidth]{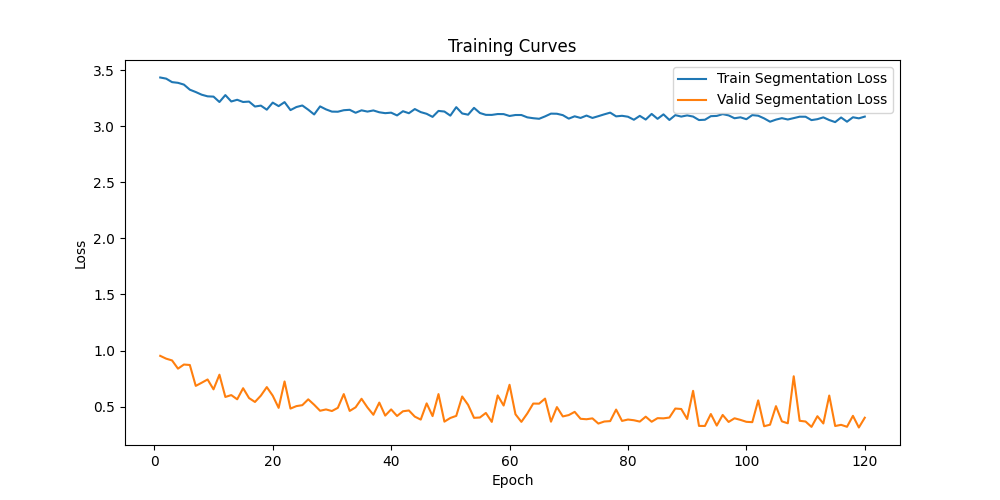}
  \caption*{B$\rightarrow$B + E$\rightarrow$E}
 \end{minipage}
 \hfill
 \begin{minipage}[b]{0.48\textwidth}
  \centering
  \includegraphics[width=\textwidth]{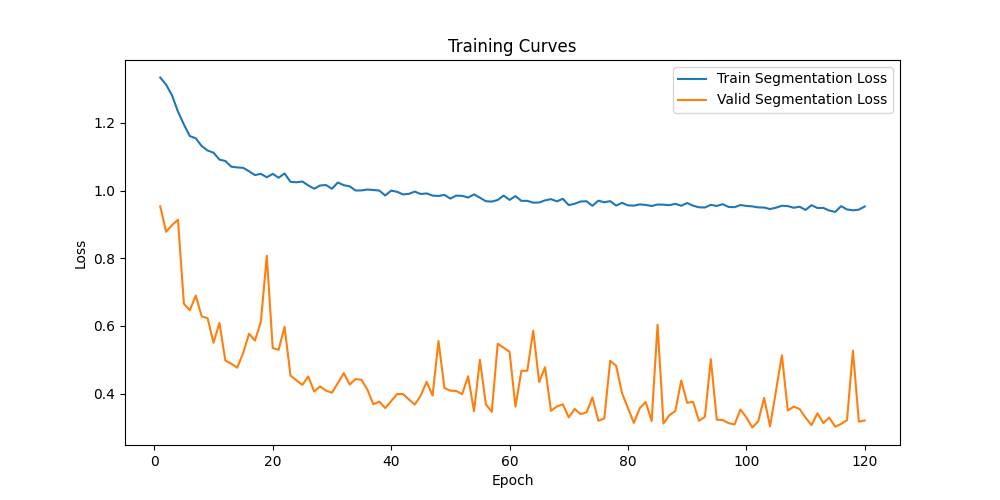}
  \caption*{B$\rightarrow$B + E$\rightarrow$E + PMD}
 \end{minipage}
 
 \begin{minipage}[b]{0.48\textwidth}
  \centering
  \includegraphics[width=\textwidth]{Images/Training_Curves/KD_T1_S2/Training_Curve_KD_B-B+E-E+PMD.png}
  \caption*{B$\rightarrow$B + E$\rightarrow$E + D$\rightarrow$D}
 \end{minipage}
 \hfill
 \begin{minipage}[b]{0.48\textwidth}
  \centering
  \includegraphics[width=\textwidth]{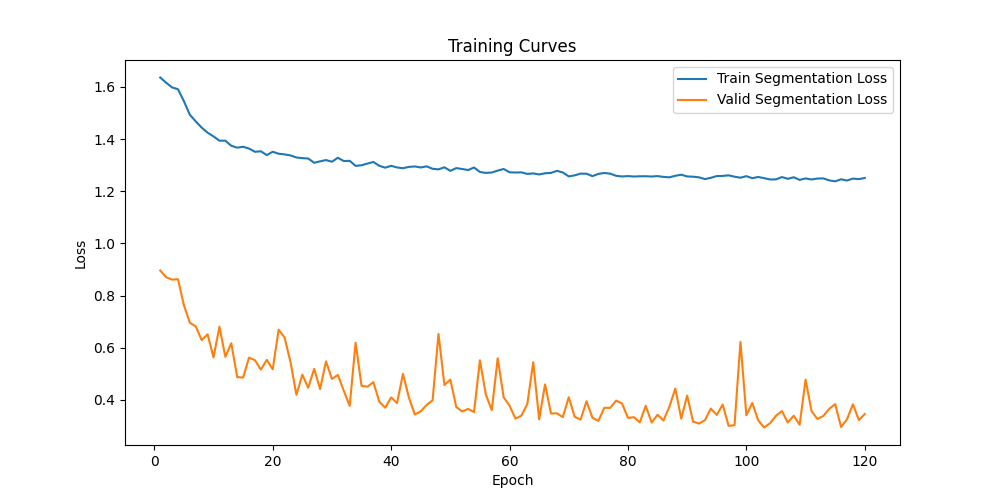}
  \caption*{B$\rightarrow$B + E$\rightarrow$E + D$\rightarrow$D + PMD}
 \end{minipage}
 
 \begin{minipage}[b]{0.48\textwidth}
  \centering
  \includegraphics[width=\textwidth]{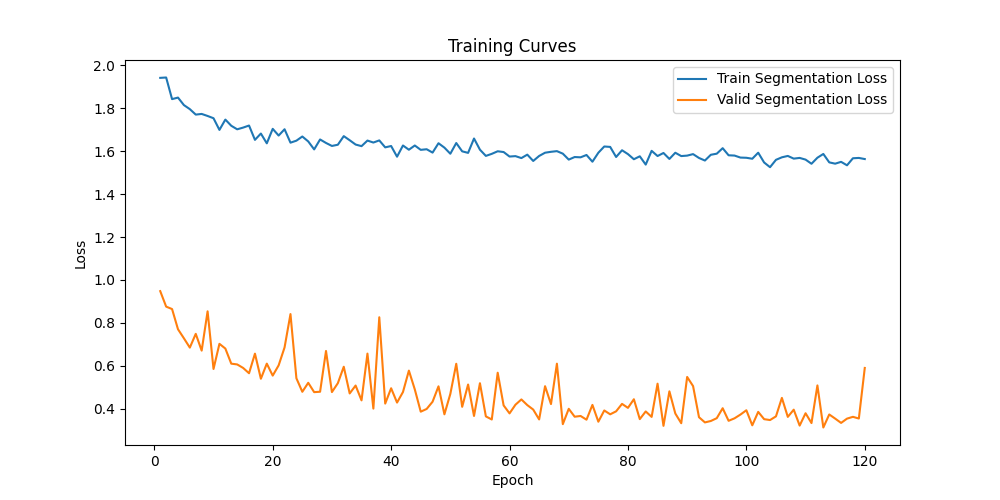}
  \caption*{E$\rightarrow$E}
 \end{minipage}
 \hfill
 \begin{minipage}[b]{0.48\textwidth}
  \centering
  \includegraphics[width=\textwidth]{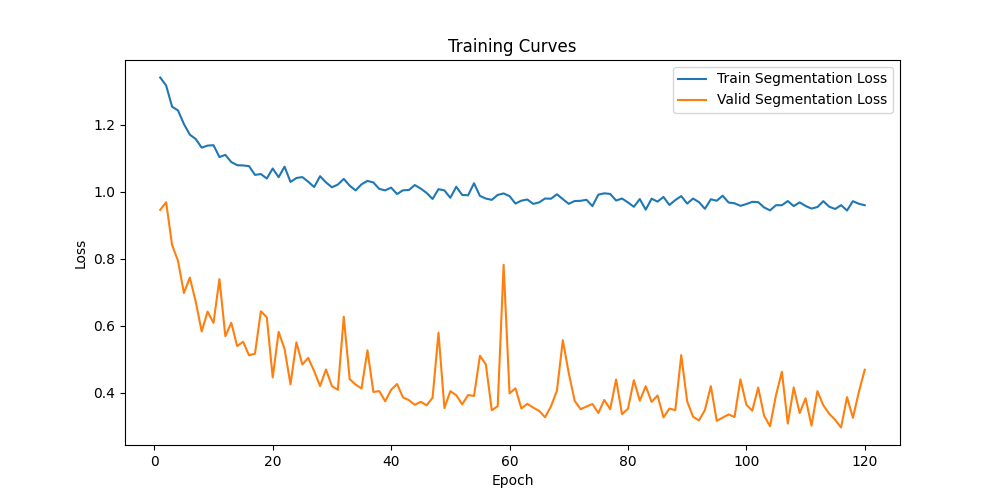}
  \caption*{E$\rightarrow$E + PMD}
 \end{minipage}
 \captionsetup{justification=centering}
 \caption{Training Curves at Various Scales of Knowledge Distillation Between T1 and S2 Using Contrastive Learning.}
 \label{fig:T1-S2_KD_ConLoss_training_curves}
\end{figure}

In Figures \ref{fig:T2-S1_KD_ConLoss_training_curves_1} and \ref{fig:T2-S1_KD_ConLoss_training_curves_2}, we examine the training dynamics of S1 when incorporating knowledge from T2. Additionally, Figure~\ref{fig:T1-S2_KD_ConLoss_training_curves} illustrates the training process of S2 using distilled knowledge from T1.

\begin{figure}[htbp]
 \centering
 \begin{minipage}[b]{0.44\textwidth}
  \centering
  \includegraphics[width=\textwidth]{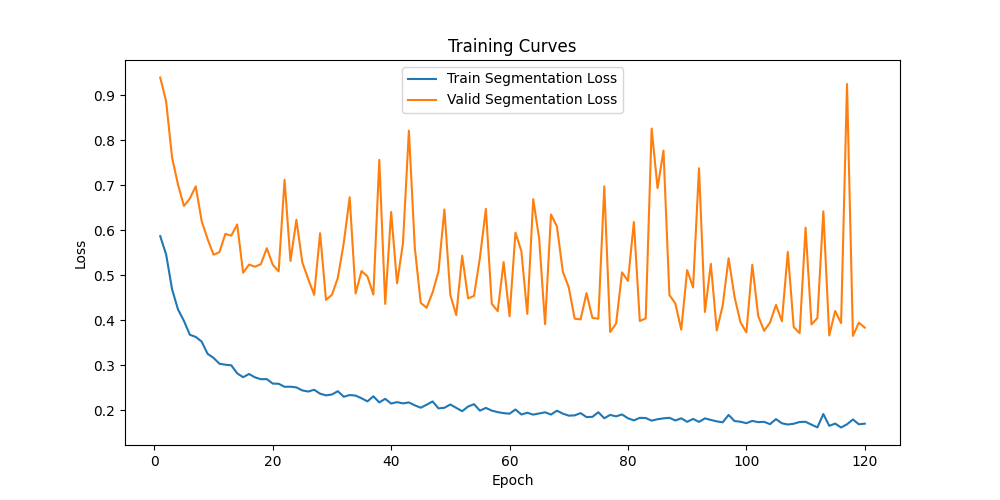}
  \caption*{B$\rightarrow$B}
 \end{minipage}
 \begin{minipage}[b]{0.44\textwidth}
  \centering
  \includegraphics[width=\textwidth]{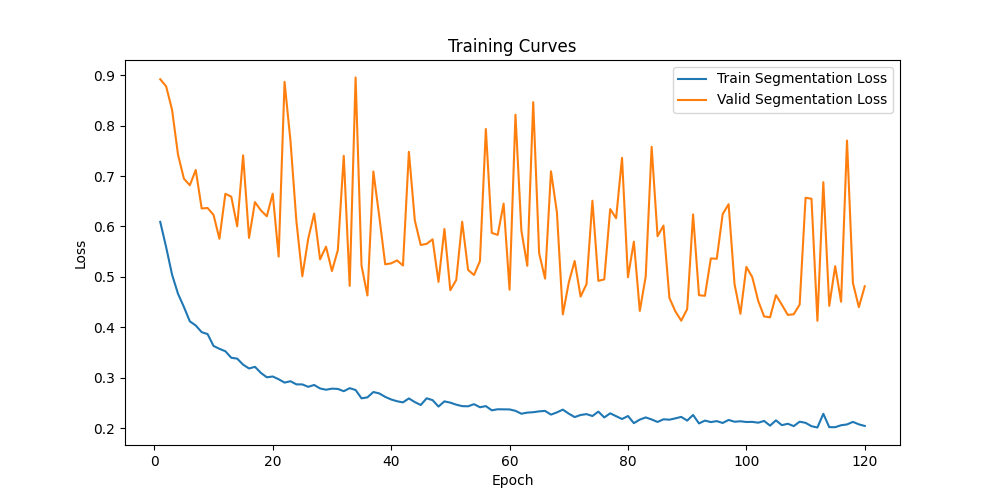}
  \caption*{B$\rightarrow$B + E$\rightarrow$E}
 \end{minipage}
 \vfill
 \begin{minipage}[b]{0.44\textwidth}
  \centering
  \includegraphics[width=\textwidth]{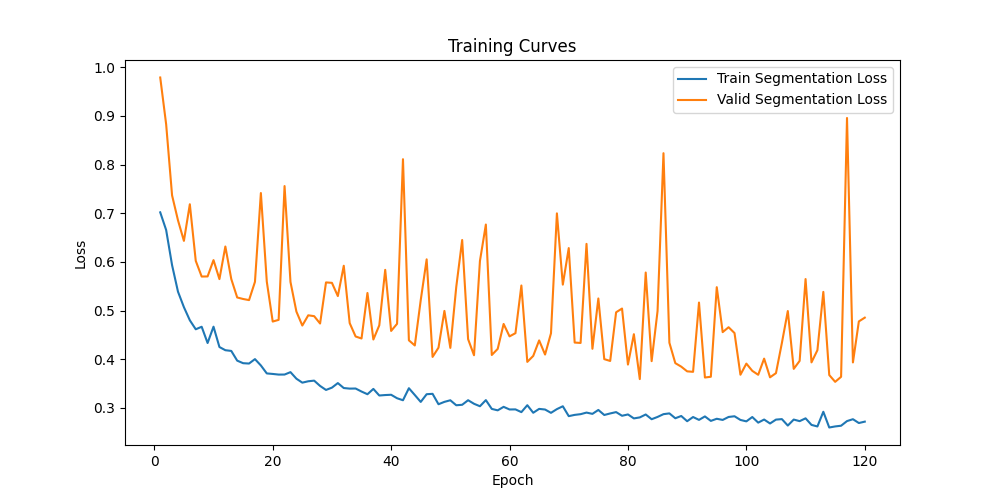}
  \caption*{B$\rightarrow$B + E$\rightarrow$E + D$\rightarrow$D}
 \end{minipage}
 \begin{minipage}[b]{0.44\textwidth}
  \centering
  \includegraphics[width=\textwidth]{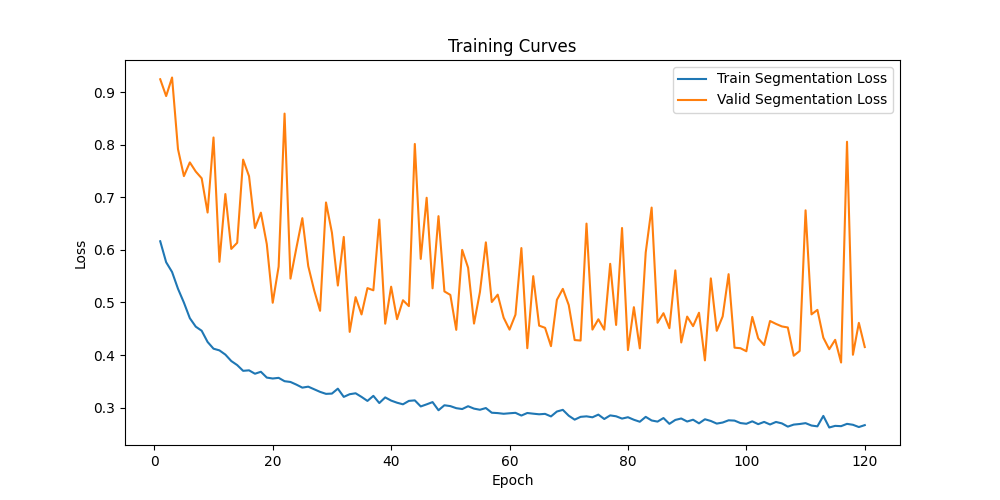}
  \caption*{B$\rightarrow$B+E$\rightarrow$E+D$\rightarrow$D+PMD}
 \end{minipage}
 \vfill
 \begin{minipage}[b]{0.44\textwidth}
  \centering
  \includegraphics[width=\textwidth]{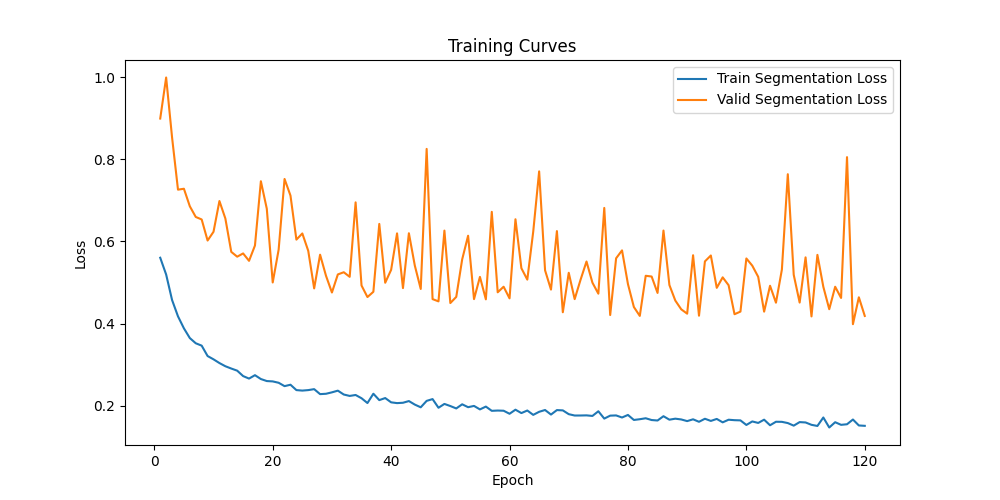}
  \caption*{B$\rightarrow$B + PMD}
 \end{minipage}
 \begin{minipage}[b]{0.44\textwidth}
  \centering
  \includegraphics[width=\textwidth]{Images/Training_Curves/Spleen_KD_Student_SmallerUNet/Training_Curve_KD_D-D.png}
  \caption*{D$\rightarrow$D}
 \end{minipage}
 \captionsetup{justification=centering}
 \caption{Training Curves Illustrating Multi-Scale Knowledge Distillation Between T1 and S1 Using MSE Loss.}
 \label{fig:T1-S1_KD_MSE_training_curves}
\end{figure}

Furthermore, we plotted training curves during knowledge distillation, but instead of contrastive loss, we used MSE loss for these variants of the training. It is evident from the plots in Figure~\ref{fig:T1-S1_KD_MSE_training_curves} that despite knowledge distillation, the learning process has not improved, when compared to and the curves presented in Figure~\ref{fig:T1-S1_KD_ConLoss_training_curves_2}. This provides a strong piece of evidence towards our hypothesis and the aim of this thesis i.e. utilizing contrastive learning for effective knowledge distillation.

\section{Quantitative Results}
\label{section:Quantitative Results}
In this section, we present the quantitative results of our experiments conducted during the master's thesis, providing a detailed assessment of our architecture's performance across multiple metrics for the medical image segmentation task. We also provide insights into the efficacy of knowledge distillation through our quantitative results. These measures provide a solid testimony for supporting our hypothesis help us evaluate the efficacy of our methodologies and allow for direct comparisons with other approaches. We hope to use quantitative metrics to not only assess the performance of our models but also to better understand the impact of knowledge distillation on their capabilities. 
\begin{table}[!ht]
    \centering
    \captionsetup{justification=centering}
    \caption{Performance Comparison of the Teacher Network on Spleen Segmentation Task}
    \begin{adjustbox}{width=\textwidth} 
    \begin{tabular}{lcccccc}
        \toprule
        \textbf{Method} & \textbf{IoU} & \textbf{Dice} & \textbf{Recall} & \textbf{Precision} & \textbf{\#Params (M)} \\
        \midrule
        SpleenSeg-U-Net & 0.943 & 0.971 & 0.976 & 0.965 & 31.04  \\
        T1: Multi-Task U-Net & 0.950 (+0.74\%) & 0.974 (+0.30\%) & 0.981 (+0.51\%) & 0.968 (+0.31\%) & 43.23 \\
        T2: Multi-Task TransUNet & 0.953 (+1.06\%) & 0.976 (+0.51\%) & 0.980 (+0.41\%) & 0.971 (+0.62\%) & 172.62 \\
        \bottomrule
    \end{tabular}
    \end{adjustbox}
    \label{tab:Teacher_segmentation_comparison}
\end{table}

We first examine the performance of the teacher models T1 and T2 on four standard metrics for medical image segmentation. The comparison is presented in Table \ref{tab:Teacher_segmentation_comparison}. Table~\ref{tab:Teacher_segmentation_comparison} compares the performance of two teacher models, T1 (a Multi-Task U-Net) and T2 (a Multi-Task TransUNet~\cite{chen2021transunet}), and another model, SpleenSeg-U-Net (vanilla U-Net~\cite{ronneberger2015u}) on the task of spleen segmentation. The \% change in T1 and T2 from SpleenSeg-U-Net is also illustrated in brackets. 

These two teacher models are well-established architectures for medical image segmentation, having previously achieved state-of-the-art results. Notably, T1 and T2 both include a multi-task framework with a reconstruction head. T2 is also an encoder-decoder architecture nut takes a different approach to feature augmentation using a transformer-based model. The results show that T1 outperform the SpleenSeg-U-Net model across all metrics. This improvement is most noticeable in terms of IoU, Dice coefficient, recall, and precision, highlighting the importance of multi-task learning in improving segmentation results. Specifically, the inclusion of a reconstruction head in T1 vs not including in SpleenSeg-U-Net aids in generalizing the overall architecture, resulting in improved segmentation performance. These findings support the efficacy of multi-task learning frameworks in medical image segmentation tasks and provide useful insights. This also adds support to our previous hypothesis regarding multi-task learning. Interestingly, T2, despite having a larger parameter count than T1, performs slightly better across all metrics. This suggests that transformer-based architectures can effectively capture complex spatial dependencies in medical images. We now shift our focus to the performance comparison of segmentation methods employing knowledge distillation utilizing contrastive learning, as presented in Tab~\ref{tab:Teacher_Student_KD_segmentation_comparison}.
\begin{table}[htbp]
\centering
\captionsetup{justification=centering}
\caption{Performance Comparison of Segmentation Methods using Knowledge Distillation utilizing Contrastive Learning}
\begin{adjustbox}{width=\textwidth} 
\begin{tabular}{l*{5}{S[table-format=1.3]}} 
\toprule
\textbf{Method} & {\textbf{IoU}} & {\textbf{Dice}} & {\textbf{Recall}} & {\textbf{Precision}} & {\textbf{\#Params (M)}}\\
\midrule
\textcolor{blue}{S1 : U-Net-S} & \textcolor{blue}{0.557} & \textcolor{blue}{0.715} & \textcolor{blue}{0.685} & \textcolor{blue}{0.749} & \textcolor{blue}{0.057}\\
\textcolor{magenta}{S2 : U-Net} & \textcolor{magenta}{0.627} & \textcolor{magenta}{0.770} & \textcolor{magenta}{0.725} & \textcolor{magenta}{0.822} & \textcolor{magenta}{0.227}\\
KD(T1,S1) & {0.629 (+12.92\%)} &{0.772 (+7.97\%)} & {0.721 (+5.25\%)} & {0.831 (+10.94\%)} & 0.057\\
KD(T1,S2) & {0.659 (+5.10\%)} & {0.794 (+3.11\%)} & {0.748 (+3.17\%)} & {0.846 (+2.91\%)} & 0.227\\
KD(T2,S1) & {0.623 (+11.84\%)} &{0.767 (+7.27\%)} &{0.707 (+3.21\%)} &{0.839 (+12.01\%)} & 0.057\\
KD(T2,S2) & {0.658 (+4.94\%)} & {0.793 (+2.98\%)} & {0.750 (+3.44\%)} & {0.841 (+2.31\%)} & 0.227\\
\bottomrule
\end{tabular}
\end{adjustbox}
\label{tab:Teacher_Student_KD_segmentation_comparison}
\end{table}

Table~\ref{tab:Teacher_Student_KD_segmentation_comparison} compares the performance of different segmentation methods, such as baseline U-Net models (S1 and S2) and our proposed teacher-student models with knowledge distillation. KD(T1, S1) denotes a model in which knowledge is distilled from teacher T1 to student S1, while KD(T1, S2) denotes knowledge being distilled from T1 to student S2, and so on. The term ``S1: U-Net-S" refers to a simplified U-Net model as discussed in Chapter~\ref{chapter:Methodology}, Section~\ref{section:Student Network}, whereas ``S2: U-Net" denotes a slightly larger standard U-Net architecture. The number of parameters of these models is also mentioned in the above table in the last column. Our proposed teacher-student models, KD(T1, S1) and KD(T1, S2), outperform the baseline models S1 \& S2 on all evaluation metrics, respectively. KD(T1, S1) achieves a significant increase of 12.9\% in IoU and 7.97\% in Dice coefficient, whereas  KD(T1, S2) improves by 5.1\% and 3.9\% when compared to the corresponding baseline models S1 and S2, respectively. These enhancements demonstrate the effectiveness of knowledge distillation using contrastive learning in improving segmentation performance for the spleen segmentation task. Furthermore, the parameter efficiency of the models is maintained, with the number of parameters remaining consistent with the student models with knowledge distillation. This highlights the practical applicability of our approach in real-world scenarios, where resource constraints are often a concern. Thus providing strong support to our overall research and thesis objectives. 

It is worth noting that the improvement from the model KD(T1, S1) to KD(T1, S2) is not as significant, implying that using a heavier student model may not always be feasible when using knowledge distillation. This observation highlights the effectiveness of our design approach, which prioritizes the use of learning techniques like contrastive representation learning over increasing the complexity of the student model. We can achieve significant performance improvements by leveraging contrastive learning without requiring overly complex student architectures. This strategic approach not only ensures computational efficiency but also emphasizes the potential for significant performance gains via innovative learning methodologies.

Similar to this, if we performed experiments where knowledge is distilled from a much larger model (T2) to the same student model S1 and S2, denoted as KD(T2, S1) and KD(T2, S2), respectively, where T2 is utilizing attention mechanism to extract useful features from the input image. However, it must be noted in the Table~\ref{tab:Teacher_Student_KD_segmentation_comparison} that in spite of using a more parameter heavy model the performance of the student models has not increased significantly (12.9\% vs 11.8\% in S1) and (5.1\% vs 4.9\%), compared to the performance when knowledge is distilled from a lighter teacher model T1. This could be because of the student models' (S1 and S2) ability to apply the condensed knowledge from the larger teacher model (T2) in a way that makes sense for the overall performance gain. The student models S1 and S2 might not have the capacity to effectively leverage these complex features, even though T2 might have more parameters and be able to capture more intricate features from the input images using attention mechanisms. Moreover, the positive and negative feature pairs may not be fully utilized during contrastive learning. Therefore, it is crucial to keep in mind that having a larger model not always guarantees increase in performance for the student model. Rather, it largely depends on both the models, the choice of knowledge distillation loss, the sclaes in which the knowledge is distilled etc. 

In summary, the results presented in Table~\ref{tab:Teacher_Student_KD_segmentation_comparison} provide compelling evidence of the effectiveness of knowledge distillation utilizing contrastive learning in improving segmentation performance while maintaining parameter efficiency. These findings reinforce the potential of our proposed methodology in advancing medical image segmentation techniques and hold promise for future research in the field. In the following section, we will discuss the qualitative results of our experiments.

\section{Qualitative Results}
To fully comprehend the results of our experiment and their implications for medical image segmentation, we need to examine the qualitative outcomes. Qualitative analysis provides a more intuitive understanding of the model by examining actual segmentation results from CT images taken from the test set. We can learn a great deal about the functionality and behaviour of our models in practical situations by visually examining these segmentation outputs. This qualitative investigation rounds out the quantitative analysis by providing a comprehensive viewpoint on the effectiveness and robustness of our suggested methodologies. 

\label{section:Qualitative Results}
\begin{figure}[htbp]
    \centering
    \begin{subfigure}[b]{0.16\textwidth} 
        \centering
        \includegraphics[width=\textwidth]{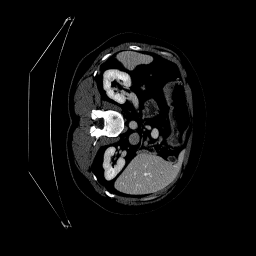}
    \end{subfigure}
    \begin{subfigure}[b]{0.16\textwidth}
        \centering
        \includegraphics[width=\textwidth]{Images/Result_For_Thesis/Reconstruction_Teacher/Recon_Input/image_spleen_22_61.png}
    \end{subfigure}
    \begin{subfigure}[b]{0.16\textwidth}
        \centering
        \includegraphics[width=\textwidth]{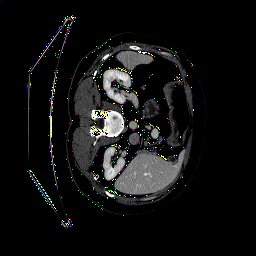}
    \end{subfigure}
    \begin{subfigure}[b]{0.16\textwidth}
        \centering
        \includegraphics[width=\textwidth]{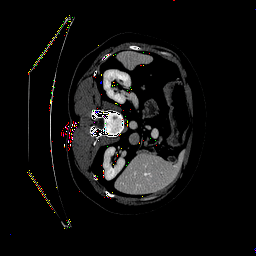}
    \end{subfigure} \\
    \begin{subfigure}[b]{0.16\textwidth} 
        \centering
        \includegraphics[width=\textwidth]{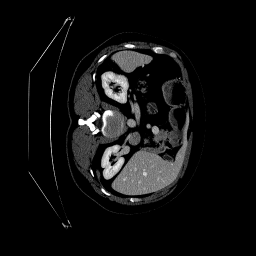}
    \end{subfigure}
    \begin{subfigure}[b]{0.16\textwidth}
        \centering
        \includegraphics[width=\textwidth]{Images/Result_For_Thesis/Reconstruction_Teacher/Recon_Input/image_spleen_22_67.png}
    \end{subfigure}
    \begin{subfigure}[b]{0.16\textwidth}
        \centering
        \includegraphics[width=\textwidth]{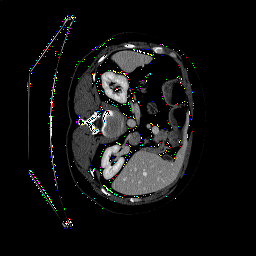}
    \end{subfigure}
    \begin{subfigure}[b]{0.16\textwidth}
        \centering
        \includegraphics[width=\textwidth]{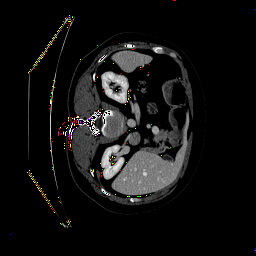}
    \end{subfigure} \\
    \begin{subfigure}[b]{0.16\textwidth} 
        \centering
        \includegraphics[width=\textwidth]{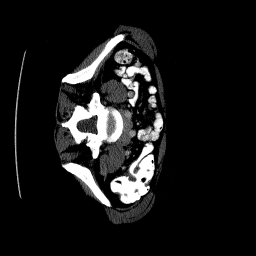}
    \end{subfigure}
    \begin{subfigure}[b]{0.16\textwidth}
        \centering
        \includegraphics[width=\textwidth]{Images/Result_For_Thesis/Reconstruction_Teacher/Recon_Input/image_spleen_62_34.png}
    \end{subfigure}
    \begin{subfigure}[b]{0.16\textwidth}
        \centering
        \includegraphics[width=\textwidth]{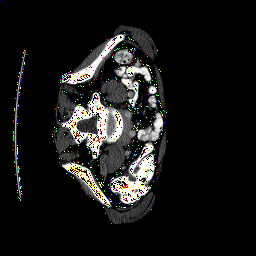}
    \end{subfigure}
    \begin{subfigure}[b]{0.16\textwidth}
        \centering
        \includegraphics[width=\textwidth]{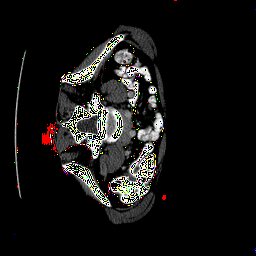}
    \end{subfigure} \\
    \begin{subfigure}[b]{0.16\textwidth} 
    \centering
    \includegraphics[width=\textwidth]{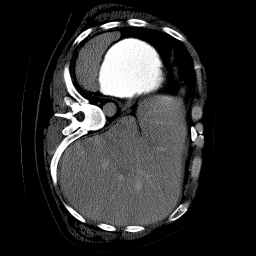}
    \caption*{Input}
    \end{subfigure}
    \begin{subfigure}[b]{0.16\textwidth}
        \centering
        \includegraphics[width=\textwidth]{Images/Result_For_Thesis/Reconstruction_Teacher/Recon_Input/image_spleen_63_41.png}
    \caption*{GT}
    \end{subfigure}
    \begin{subfigure}[b]{0.16\textwidth}
        \centering
        \includegraphics[width=\textwidth]{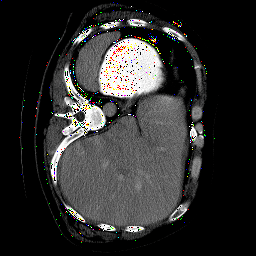}
    \caption*{Teacher (T1)}
    \end{subfigure}
    \begin{subfigure}[b]{0.16\textwidth}
        \centering
        \includegraphics[width=\textwidth]{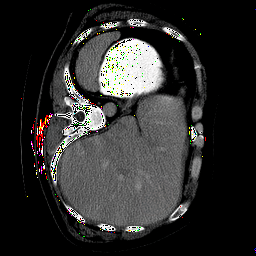}
    \caption*{Teacher (T2)}
    \end{subfigure}
    \captionsetup{justification=centering}
    \caption{Visualising Spleen Image Reconstructions of T1 \& T2}
\label{fig:Teacher_Qual_Perfrmance_Recon}
\end{figure}

\begin{figure}[ht!]
    \centering
    \begin{subfigure}[b]{0.16\textwidth} 
        \centering
        \includegraphics[width=\textwidth]{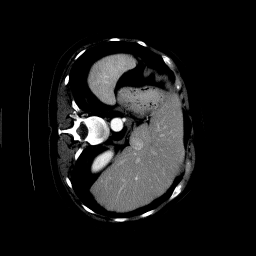}
    \end{subfigure}
    \begin{subfigure}[b]{0.16\textwidth}
        \centering
        \includegraphics[width=\textwidth]{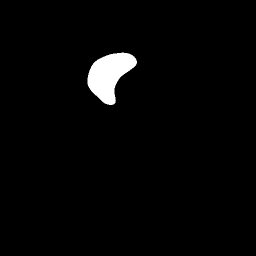}
    \end{subfigure}
    \begin{subfigure}[b]{0.16\textwidth}
        \centering
        \includegraphics[width=\textwidth]{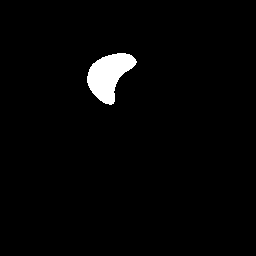}
    \end{subfigure}
    \begin{subfigure}[b]{0.16\textwidth}
        \centering
        \includegraphics[width=\textwidth]{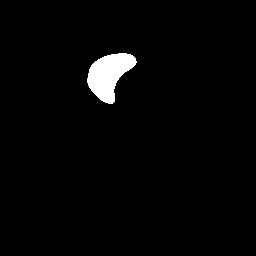}
    \end{subfigure} \\
    \begin{subfigure}[b]{0.16\textwidth} 
        \centering
        \includegraphics[width=\textwidth]{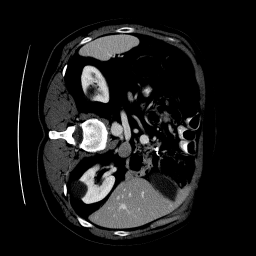}
    \end{subfigure}
    \begin{subfigure}[b]{0.16\textwidth}
        \centering
        \includegraphics[width=\textwidth]{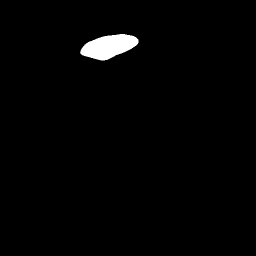}
    \end{subfigure}
    \begin{subfigure}[b]{0.16\textwidth}
        \centering
        \includegraphics[width=\textwidth]{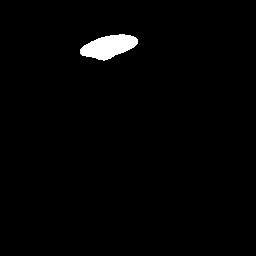}
    \end{subfigure}
    \begin{subfigure}[b]{0.16\textwidth}
        \centering
        \includegraphics[width=\textwidth]{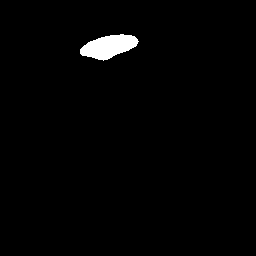}
    \end{subfigure}\\
    \begin{subfigure}[b]{0.16\textwidth} 
        \centering
        \includegraphics[width=\textwidth]{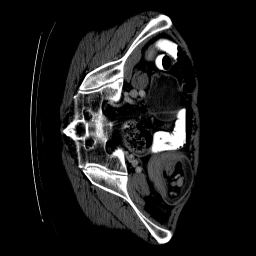}
    \end{subfigure}
    \begin{subfigure}[b]{0.16\textwidth}
        \centering
        \includegraphics[width=\textwidth]{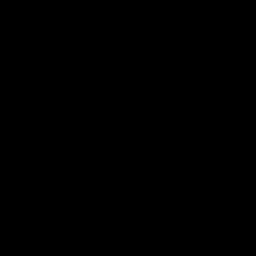}
    \end{subfigure}
    \begin{subfigure}[b]{0.16\textwidth}
        \centering
        \includegraphics[width=\textwidth]{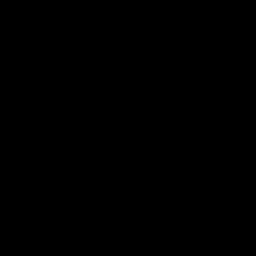}
    \end{subfigure}
    \begin{subfigure}[b]{0.16\textwidth}
        \centering
        \includegraphics[width=\textwidth]{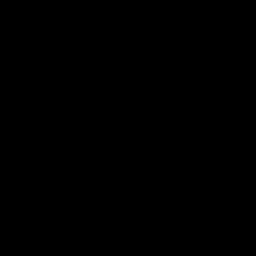}
    \end{subfigure}\\
    \begin{subfigure}[b]{0.16\textwidth} 
    \centering
    \includegraphics[width=\textwidth]{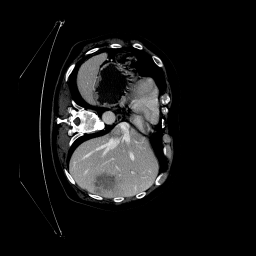}
    \caption*{Input}
    \end{subfigure}
    \begin{subfigure}[b]{0.16\textwidth}
        \centering
        \includegraphics[width=\textwidth]{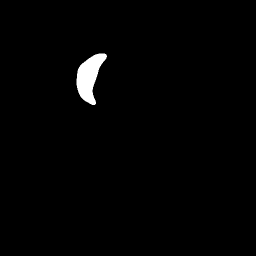}
    \caption*{GT}
    \end{subfigure}
    \begin{subfigure}[b]{0.16\textwidth}
        \centering
        \includegraphics[width=\textwidth]{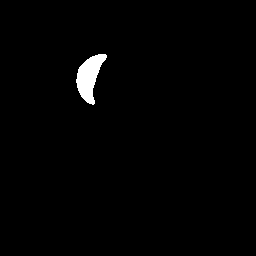}
    \caption*{Teacher (T1)}
    \end{subfigure}
    \begin{subfigure}[b]{0.16\textwidth}
        \centering
        \includegraphics[width=\textwidth]{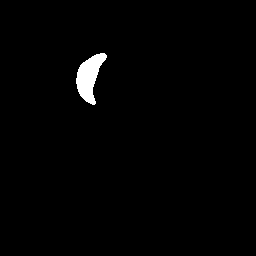}
    \caption*{Teacher (T2)}
    \end{subfigure}
    \captionsetup{justification=centering}
    \caption{Visualising Spleen Segmentation of T1 \& T2}
\label{fig:Teacher_Qual_Perfrmance_Seg}
\end{figure}

We begin by looking at the outcomes of our Teacher models T1 and T2 on both reconstruction and segmentation tasks. The results of which can be found in Figure~\ref{fig:Teacher_Qual_Perfrmance_Recon} and Figure~\ref{fig:Teacher_Qual_Perfrmance_Seg} on reconstruction and segmentation tasks, respectively on examples taken from the test set. The reconstruction results exhibit that both models are susceptible to capturing noise during the reconstruction step. The pixel-level segmentation maps are represented by the black region, which represents the background, and the white region, which represents the region of interest—in this case, the spleen. It is noteworthy that either of the multi-task teacher models was not specially optimized for a particular task, as it would be beyond the scope of our thesis investigation. We aim to evaluate the effectiveness of knowledge distillation via a comparative analysis rather than trying to create a state-of-the-art model on reconstruction or segmentation. 

To demonstrate the effectiveness of knowledge distillation, we then provide a comparative analysis of segmentation outcomes. We compare the outputs of the student model and the KD(T1, S1) model, which incorporates knowledge distillation, with the ground truth segmentation. This comparison sheds light on how the student model's (S1) segmentation performance is impacted by the distilled knowledge from the teacher model (T1). Through a qualitative analysis of these results, we hope to clarify how knowledge distillation using contrastive learning enhances segmentation quality and accuracy. 
\begin{figure}[ht!]
    \centering
    \begin{subfigure}[b]{0.16\textwidth} 
        \centering
        \includegraphics[width=\textwidth]{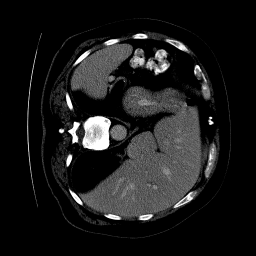}
    \end{subfigure}
    \begin{subfigure}[b]{0.16\textwidth}
        \centering
        \includegraphics[width=\textwidth]{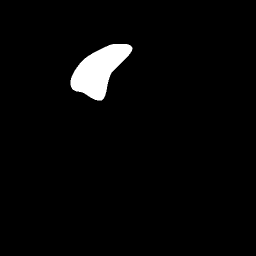}
    \end{subfigure}
    \begin{subfigure}[b]{0.16\textwidth}
        \centering
        \includegraphics[width=\textwidth]{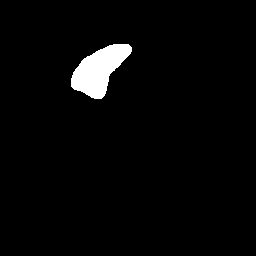}
    \end{subfigure}
    \begin{subfigure}[b]{0.16\textwidth}
        \centering
        \includegraphics[width=\textwidth]{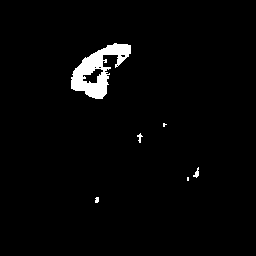}
    \end{subfigure}
    \begin{subfigure}[b]{0.16\textwidth}
        \centering
        \includegraphics[width=\textwidth]{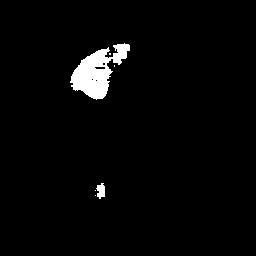}
    \end{subfigure}
    \begin{subfigure}[b]{0.16\textwidth} 
        \centering
        \includegraphics[width=\textwidth]{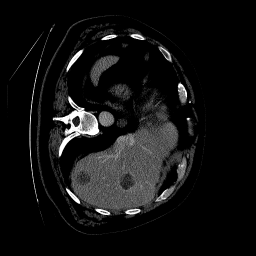}
    \end{subfigure}
    \begin{subfigure}[b]{0.16\textwidth}
        \centering
        \includegraphics[width=\textwidth]{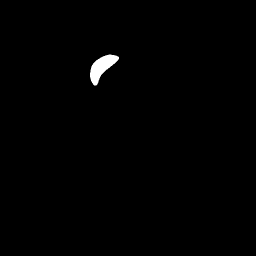}
    \end{subfigure}
    \begin{subfigure}[b]{0.16\textwidth}
        \centering
        \includegraphics[width=\textwidth]{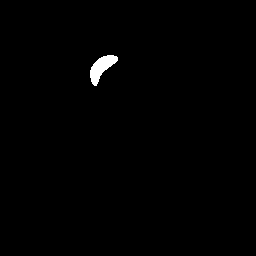}
    \end{subfigure}
    \begin{subfigure}[b]{0.16\textwidth}
        \centering
        \includegraphics[width=\textwidth]{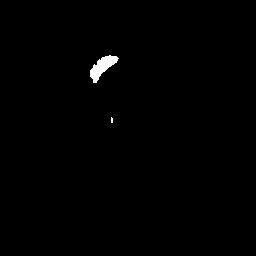}
    \end{subfigure}
    \begin{subfigure}[b]{0.16\textwidth}
        \centering
        \includegraphics[width=\textwidth]{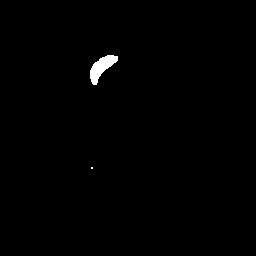}
    \end{subfigure}
    \begin{subfigure}[b]{0.16\textwidth} 
        \centering
        \includegraphics[width=\textwidth]{Images/Result_For_Thesis/Segmentation_Teacher/Seg_Input/image_spleen_9_19.png}
    \end{subfigure}
    \begin{subfigure}[b]{0.16\textwidth}
        \centering
        \includegraphics[width=\textwidth]{Images/Result_For_Thesis/Seg_GT/image_spleen_9_19.png}
    \end{subfigure}
    \begin{subfigure}[b]{0.16\textwidth}
        \centering
        \includegraphics[width=\textwidth]{Images/Result_For_Thesis/Segmentation_Teacher/Seg_Output_MT_UNet/image_spleen_9_19.png}
    \end{subfigure}
    \begin{subfigure}[b]{0.16\textwidth}
        \centering
        \includegraphics[width=\textwidth]{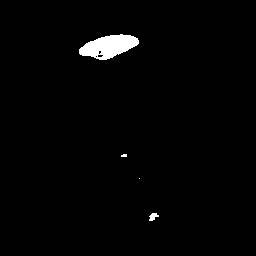}
    \end{subfigure}
    \begin{subfigure}[b]{0.16\textwidth}
        \centering
        \includegraphics[width=\textwidth]{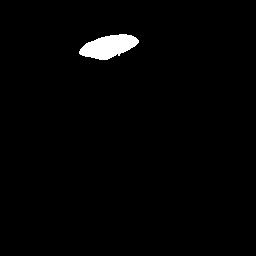}
    \end{subfigure}
    \begin{subfigure}[b]{0.16\textwidth} 
        \centering
        \includegraphics[width=\textwidth]{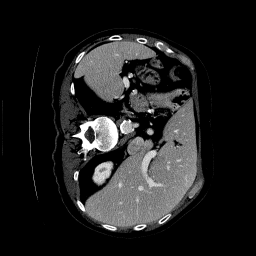}
    \end{subfigure}
    \begin{subfigure}[b]{0.16\textwidth}
        \centering
        \includegraphics[width=\textwidth]{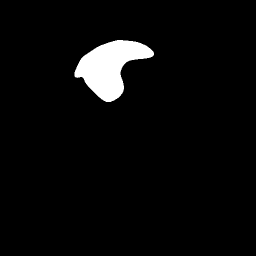}
    \end{subfigure}
    \begin{subfigure}[b]{0.16\textwidth}
        \centering
        \includegraphics[width=\textwidth]{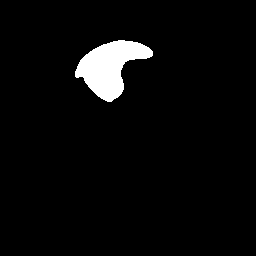}
    \end{subfigure}
    \begin{subfigure}[b]{0.16\textwidth}
        \centering
        \includegraphics[width=\textwidth]{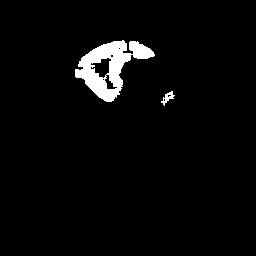}
    \end{subfigure}
    \begin{subfigure}[b]{0.16\textwidth}
        \centering
        \includegraphics[width=\textwidth]{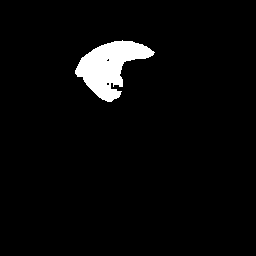}
    \end{subfigure}
        \begin{subfigure}[b]{0.16\textwidth} 
        \centering
        \includegraphics[width=\textwidth]{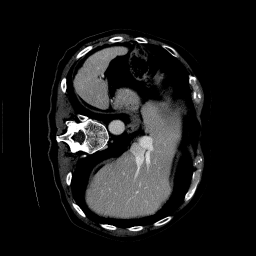}
    \end{subfigure}
    \begin{subfigure}[b]{0.16\textwidth}
        \centering
        \includegraphics[width=\textwidth]{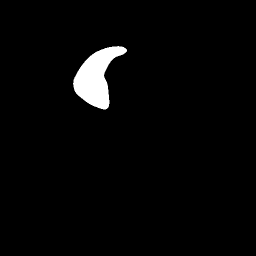}
    \end{subfigure}
    \begin{subfigure}[b]{0.16\textwidth}
        \centering
        \includegraphics[width=\textwidth]{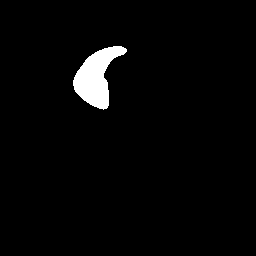}
    \end{subfigure}
    \begin{subfigure}[b]{0.16\textwidth}
        \centering
        \includegraphics[width=\textwidth]{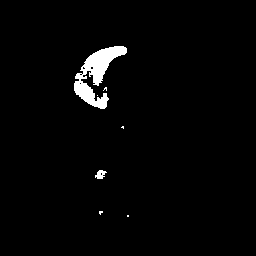}
    \end{subfigure}
    \begin{subfigure}[b]{0.16\textwidth}
        \centering
        \includegraphics[width=\textwidth]{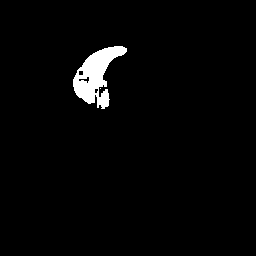}
    \end{subfigure}
        \begin{subfigure}[b]{0.16\textwidth} 
        \centering
        \includegraphics[width=\textwidth]{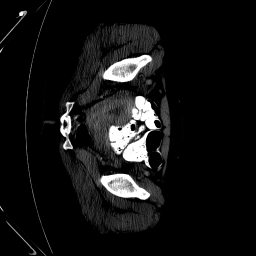}
        \caption*{Input}
    \end{subfigure}
    \begin{subfigure}[b]{0.16\textwidth}
        \centering
        \includegraphics[width=\textwidth]{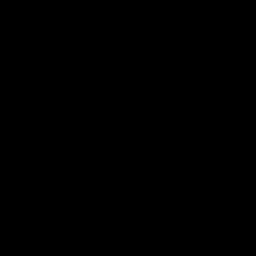}
        \caption*{GT}
    \end{subfigure}
    \begin{subfigure}[b]{0.16\textwidth}
        \centering
        \includegraphics[width=\textwidth]{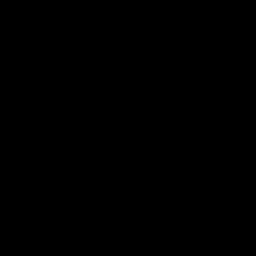}
        \caption*{Teacher (T1)}
    \end{subfigure}
    \begin{subfigure}[b]{0.16\textwidth}
        \centering
        \includegraphics[width=\textwidth]{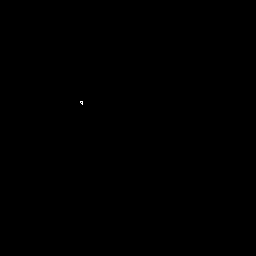}
        \caption*{Student (S1)}
    \end{subfigure}
    \begin{subfigure}[b]{0.16\textwidth}
        \centering
        \includegraphics[width=\textwidth]{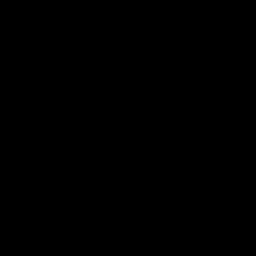}
        \caption*{(KD(T1, S1))}
    \end{subfigure}
    \captionsetup{justification=centering}
    \caption{Comparitive Analysis of Segmentation Results between Ground Truth, Output of Student (S1) and KD(T1, S1) on Spleen CT Images }
\label{fig:Comp_Anal_Seg_v1}
\end{figure}
Figure~\ref{fig:Comp_Anal_Seg_v1} shows a comparison side-by-side for observing the knowledge distillation effects easily. We find that our approach not only fixes the errors in the student model (S1) but also brings its segmentation results closer to ground truth. 
\begin{figure}[ht!]
    \centering
    \begin{subfigure}[b]{0.16\textwidth} 
        \centering
        \includegraphics[width=\textwidth]{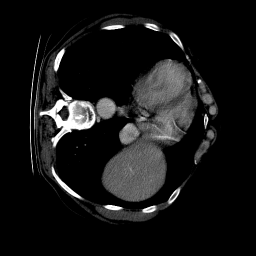}
    \end{subfigure}
    \begin{subfigure}[b]{0.16\textwidth}
        \centering
        \includegraphics[width=\textwidth]{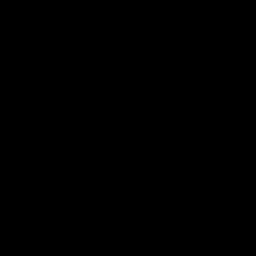}
    \end{subfigure}
    \begin{subfigure}[b]{0.16\textwidth}
        \centering
        \includegraphics[width=\textwidth]{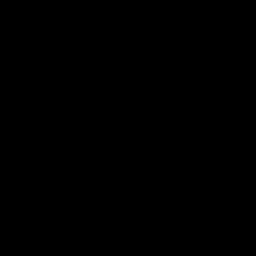}
    \end{subfigure}
    \begin{subfigure}[b]{0.16\textwidth}
        \centering
        \includegraphics[width=\textwidth]{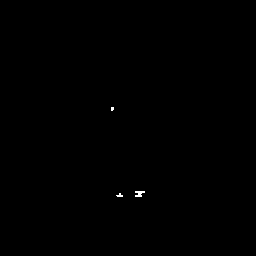}
    \end{subfigure}
    \begin{subfigure}[b]{0.16\textwidth}
        \centering
        \includegraphics[width=\textwidth]{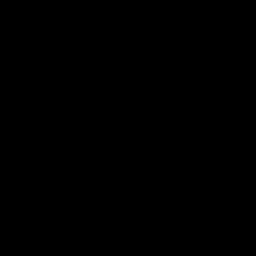}
    \end{subfigure}
    \begin{subfigure}[b]{0.16\textwidth} 
        \centering
        \includegraphics[width=\textwidth]{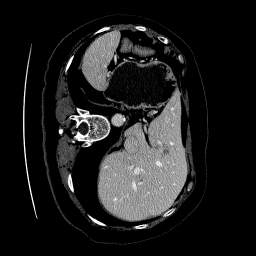}
    \end{subfigure}
    \begin{subfigure}[b]{0.16\textwidth}
        \centering
        \includegraphics[width=\textwidth]{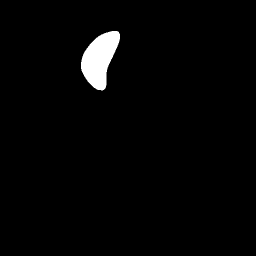}
    \end{subfigure}
    \begin{subfigure}[b]{0.16\textwidth}
        \centering
        \includegraphics[width=\textwidth]{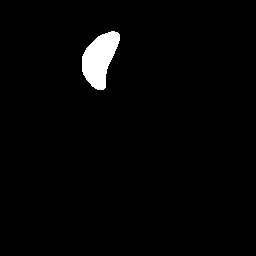}
    \end{subfigure}
    \begin{subfigure}[b]{0.16\textwidth}
        \centering
        \includegraphics[width=\textwidth]{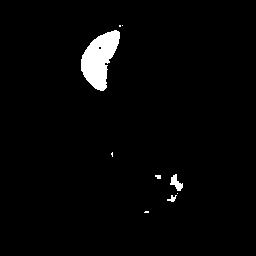}
    \end{subfigure}
    \begin{subfigure}[b]{0.16\textwidth}
        \centering
        \includegraphics[width=\textwidth]{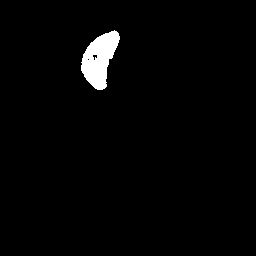}
    \end{subfigure}
    \begin{subfigure}[b]{0.16\textwidth} 
        \centering
        \includegraphics[width=\textwidth]{Images/Result_For_Thesis/Segmentation_Teacher/Seg_Input/image_spleen_12_94.png}
    \end{subfigure}
    \begin{subfigure}[b]{0.16\textwidth}
        \centering
        \includegraphics[width=\textwidth]{Images/Result_For_Thesis/Seg_GT/image_spleen_12_94.png}
    \end{subfigure}
    \begin{subfigure}[b]{0.16\textwidth}
        \centering
        \includegraphics[width=\textwidth]{Images/Result_For_Thesis/Segmentation_Teacher/Seg_Output_MT_UNet/image_spleen_12_94.png}
    \end{subfigure}
    \begin{subfigure}[b]{0.16\textwidth}
        \centering
        \includegraphics[width=\textwidth]{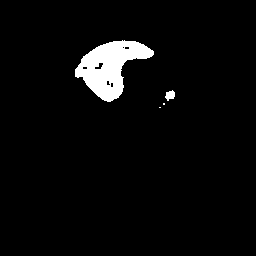}
    \end{subfigure}
    \begin{subfigure}[b]{0.16\textwidth}
        \centering
        \includegraphics[width=\textwidth]{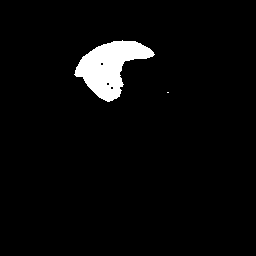}
    \end{subfigure}
        \begin{subfigure}[b]{0.16\textwidth} 
        \centering
        \includegraphics[width=\textwidth]{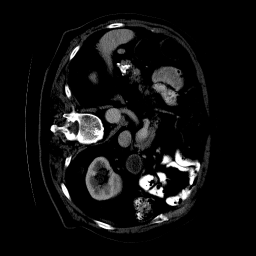}
    \end{subfigure}
    \begin{subfigure}[b]{0.16\textwidth}
        \centering
        \includegraphics[width=\textwidth]{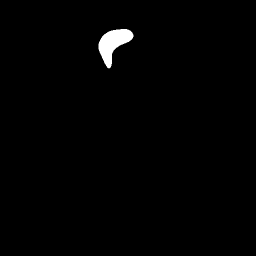}
    \end{subfigure}
    \begin{subfigure}[b]{0.16\textwidth}
        \centering
        \includegraphics[width=\textwidth]{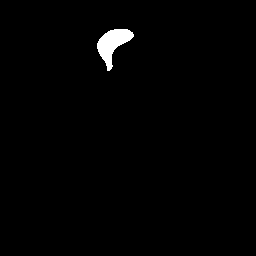}
    \end{subfigure}
    \begin{subfigure}[b]{0.16\textwidth}
        \centering
        \includegraphics[width=\textwidth]{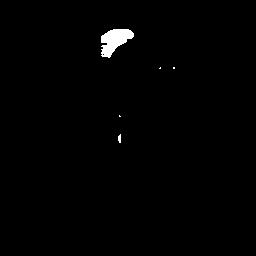}
    \end{subfigure}
    \begin{subfigure}[b]{0.16\textwidth}
        \centering
        \includegraphics[width=\textwidth]{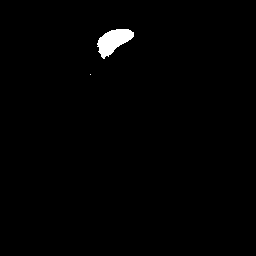}
    \end{subfigure}
        \begin{subfigure}[b]{0.16\textwidth} 
        \centering
        \includegraphics[width=\textwidth]{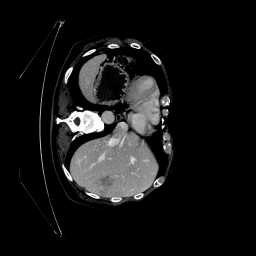}
    \end{subfigure}
    \begin{subfigure}[b]{0.16\textwidth}
        \centering
        \includegraphics[width=\textwidth]{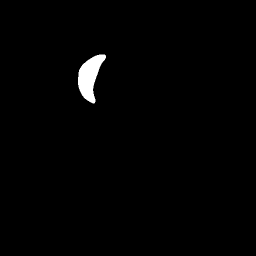}
    \end{subfigure}
    \begin{subfigure}[b]{0.16\textwidth}
        \centering
        \includegraphics[width=\textwidth]{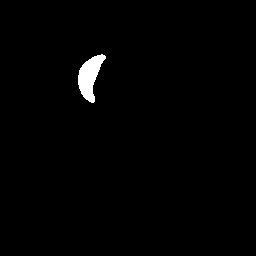}
    \end{subfigure}
    \begin{subfigure}[b]{0.16\textwidth}
        \centering
        \includegraphics[width=\textwidth]{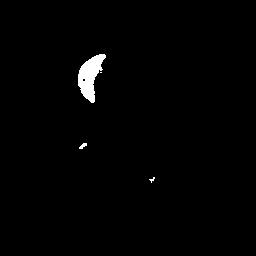}
    \end{subfigure}
    \begin{subfigure}[b]{0.16\textwidth}
        \centering
        \includegraphics[width=\textwidth]{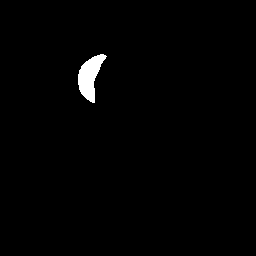}
    \end{subfigure}
        \begin{subfigure}[b]{0.16\textwidth} 
        \centering
        \includegraphics[width=\textwidth]{Images/Result_For_Thesis/Segmentation_Teacher/Seg_Input/image_spleen_46_34.png}
        \caption*{Input}
    \end{subfigure}
    \begin{subfigure}[b]{0.16\textwidth}
        \centering
        \includegraphics[width=\textwidth]{Images/Result_For_Thesis/Seg_GT/image_spleen_46_34.png}
        \caption*{GT}
    \end{subfigure}
    \begin{subfigure}[b]{0.16\textwidth}
        \centering
        \includegraphics[width=\textwidth]{Images/Result_For_Thesis/Segmentation_Teacher/Seg_Output_MT_UNet/image_spleen_46_34.png}
        \caption*{Teacher (T1)}
    \end{subfigure}
    \begin{subfigure}[b]{0.16\textwidth}
        \centering
        \includegraphics[width=\textwidth]{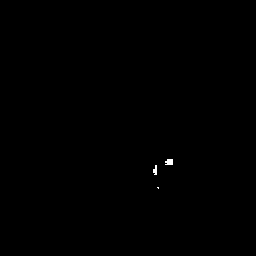}
        \caption*{Student (S2)}
    \end{subfigure}
    \begin{subfigure}[b]{0.16\textwidth}
        \centering
        \includegraphics[width=\textwidth]{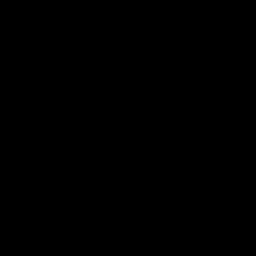}
        \caption*{KD(T1, S2)}
    \end{subfigure}
    \captionsetup{justification=centering}
    \caption{Comparitive Analysis of Segmentation Results between Ground Truth, Output of Student (S2) and KD(T1, S2) on Spleen CT Images }
\label{fig:Comp_Anal_Seg_v2}
\end{figure}

\begin{figure}[ht!]
    \centering
    \begin{subfigure}[b]{0.16\textwidth} 
        \centering
        \includegraphics[width=\textwidth]{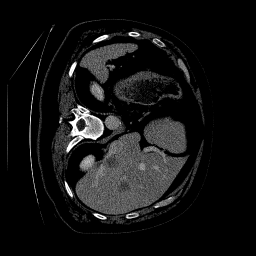}
    \end{subfigure}
    \begin{subfigure}[b]{0.16\textwidth}
        \centering
        \includegraphics[width=\textwidth]{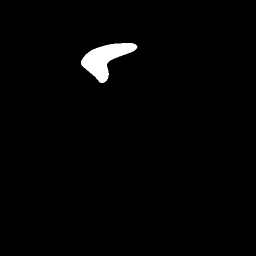}
    \end{subfigure}
    \begin{subfigure}[b]{0.16\textwidth}
        \centering
        \includegraphics[width=\textwidth]{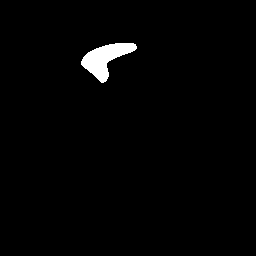}
    \end{subfigure}
    \begin{subfigure}[b]{0.16\textwidth}
        \centering
        \includegraphics[width=\textwidth]{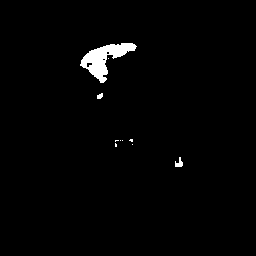}
    \end{subfigure}
    \begin{subfigure}[b]{0.16\textwidth}
        \centering
        \includegraphics[width=\textwidth]{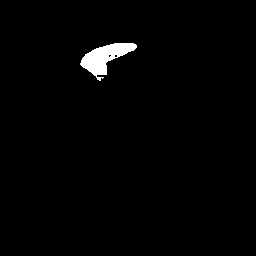}
    \end{subfigure}
    \begin{subfigure}[b]{0.16\textwidth} 
        \centering
        \includegraphics[width=\textwidth]{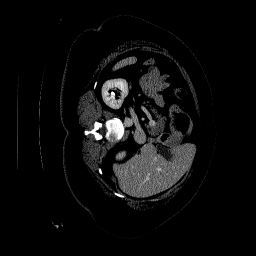}
    \end{subfigure}
    \begin{subfigure}[b]{0.16\textwidth}
        \centering
        \includegraphics[width=\textwidth]{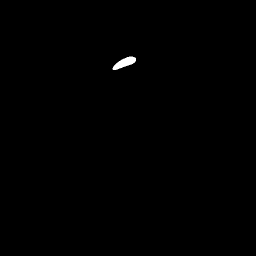}
    \end{subfigure}
    \begin{subfigure}[b]{0.16\textwidth}
        \centering
        \includegraphics[width=\textwidth]{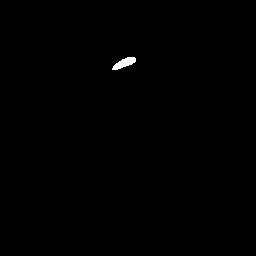}
    \end{subfigure}
    \begin{subfigure}[b]{0.16\textwidth}
        \centering
        \includegraphics[width=\textwidth]{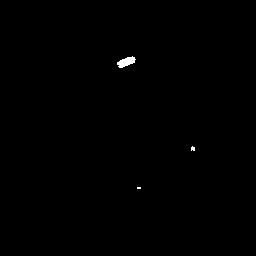}
    \end{subfigure}
    \begin{subfigure}[b]{0.16\textwidth}
        \centering
        \includegraphics[width=\textwidth]{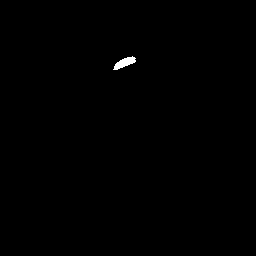}
    \end{subfigure}
    \begin{subfigure}[b]{0.16\textwidth} 
        \centering
        \includegraphics[width=\textwidth]{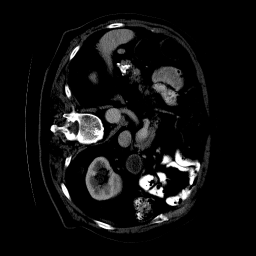}
    \end{subfigure}
    \begin{subfigure}[b]{0.16\textwidth}
        \centering
        \includegraphics[width=\textwidth]{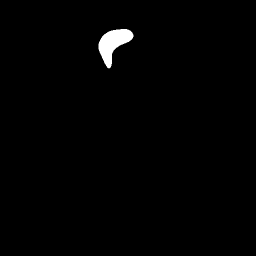}
    \end{subfigure}
    \begin{subfigure}[b]{0.16\textwidth}
        \centering
        \includegraphics[width=\textwidth]{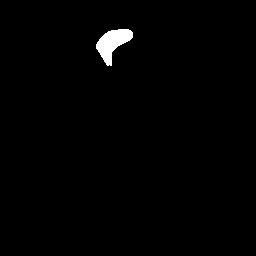}
    \end{subfigure}
    \begin{subfigure}[b]{0.16\textwidth}
        \centering
        \includegraphics[width=\textwidth]{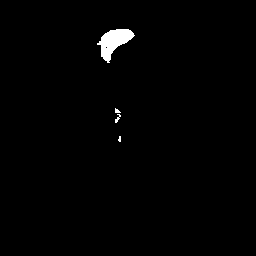}
    \end{subfigure}
    \begin{subfigure}[b]{0.16\textwidth}
        \centering
        \includegraphics[width=\textwidth]{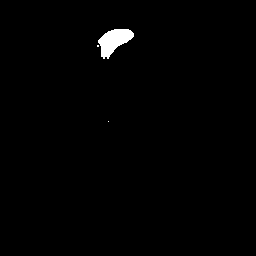}
    \end{subfigure}
    \begin{subfigure}[b]{0.16\textwidth} 
        \centering
        \includegraphics[width=\textwidth]{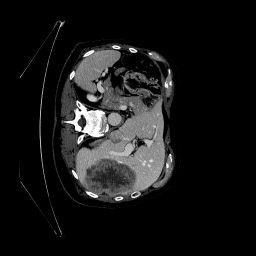}
    \end{subfigure}
    \begin{subfigure}[b]{0.16\textwidth}
        \centering
        \includegraphics[width=\textwidth]{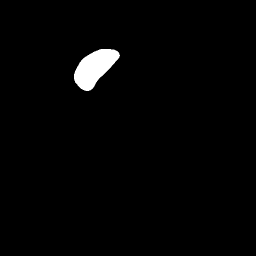}
    \end{subfigure}
    \begin{subfigure}[b]{0.16\textwidth}
        \centering
        \includegraphics[width=\textwidth]{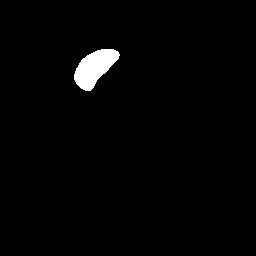}
    \end{subfigure}
    \begin{subfigure}[b]{0.16\textwidth}
        \centering
        \includegraphics[width=\textwidth]{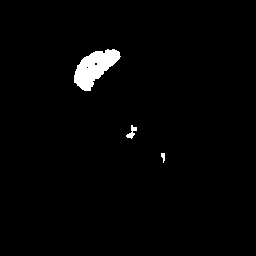}
    \end{subfigure}
    \begin{subfigure}[b]{0.16\textwidth}
        \centering
        \includegraphics[width=\textwidth]{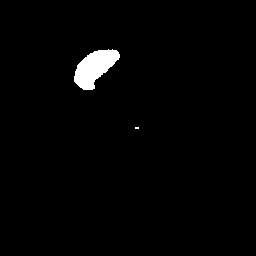}
    \end{subfigure}
    \begin{subfigure}[b]{0.16\textwidth} 
        \centering
        \includegraphics[width=\textwidth]{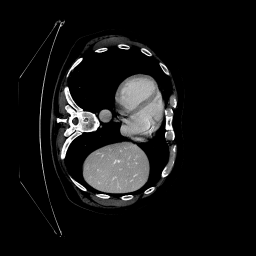}
    \end{subfigure}
    \begin{subfigure}[b]{0.16\textwidth}
        \centering
        \includegraphics[width=\textwidth]{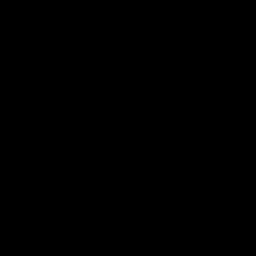}
    \end{subfigure}
    \begin{subfigure}[b]{0.16\textwidth}
        \centering
        \includegraphics[width=\textwidth]{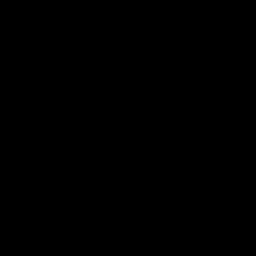}
    \end{subfigure}
    \begin{subfigure}[b]{0.16\textwidth}
        \centering
        \includegraphics[width=\textwidth]{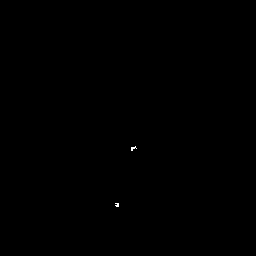}
    \end{subfigure}
    \begin{subfigure}[b]{0.16\textwidth}
        \centering
        \includegraphics[width=\textwidth]{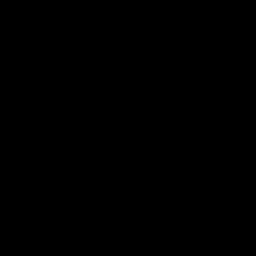}
    \end{subfigure}
        \begin{subfigure}[b]{0.16\textwidth} 
        \centering
        \includegraphics[width=\textwidth]{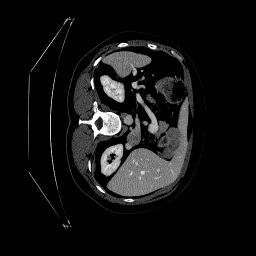}
        \caption*{Input}
    \end{subfigure}
    \begin{subfigure}[b]{0.16\textwidth}
        \centering
        \includegraphics[width=\textwidth]{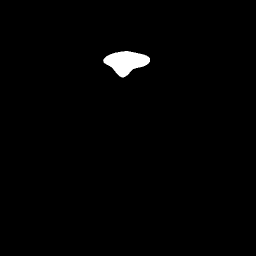}
        \caption*{GT}
    \end{subfigure}
    \begin{subfigure}[b]{0.16\textwidth}
        \centering
        \includegraphics[width=\textwidth]{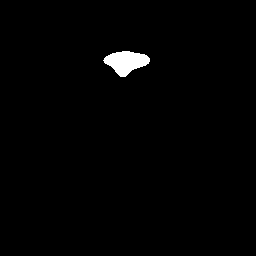}
        \caption*{Teacher (T1)}
    \end{subfigure}
    \begin{subfigure}[b]{0.16\textwidth}
        \centering
        \includegraphics[width=\textwidth]{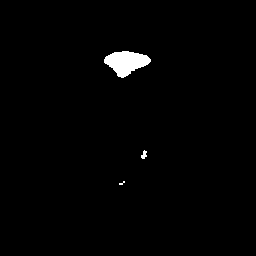}
        \caption*{Student (S1)}
    \end{subfigure}
    \begin{subfigure}[b]{0.16\textwidth}
        \centering
        \includegraphics[width=\textwidth]{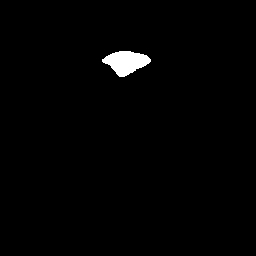}
        \caption*{KD(T2, S1)}
    \end{subfigure}
    \captionsetup{justification=centering}
    \caption{Comparitive Analysis of Segmentation Results between Ground Truth, Output of Student (S1) and KD(T2, S1) on Spleen CT Images }
\label{fig:Comp_Anal_T2_S1}
\end{figure}

\begin{figure}[ht!]
    \centering
    \begin{subfigure}[b]{0.16\textwidth} 
        \centering
        \includegraphics[width=\textwidth]{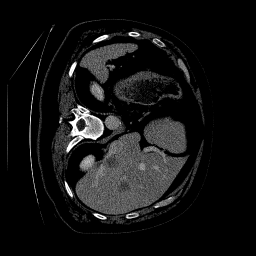}
    \end{subfigure}
    \begin{subfigure}[b]{0.16\textwidth}
        \centering
        \includegraphics[width=\textwidth]{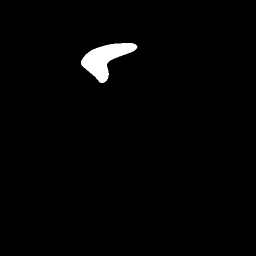}
    \end{subfigure}
    \begin{subfigure}[b]{0.16\textwidth}
        \centering
        \includegraphics[width=\textwidth]{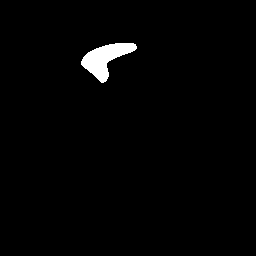}
    \end{subfigure}
    \begin{subfigure}[b]{0.16\textwidth}
        \centering
        \includegraphics[width=\textwidth]{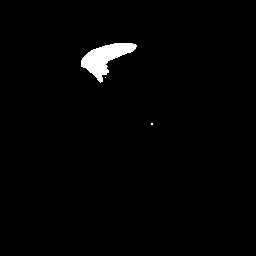}
    \end{subfigure}
    \begin{subfigure}[b]{0.16\textwidth}
        \centering
        \includegraphics[width=\textwidth]{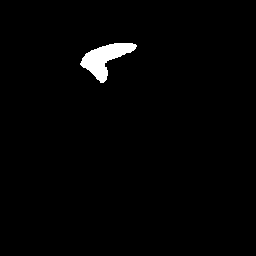}
    \end{subfigure}
    \begin{subfigure}[b]{0.16\textwidth} 
        \centering
        \includegraphics[width=\textwidth]{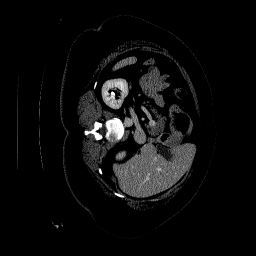}
    \end{subfigure}
    \begin{subfigure}[b]{0.16\textwidth}
        \centering
        \includegraphics[width=\textwidth]{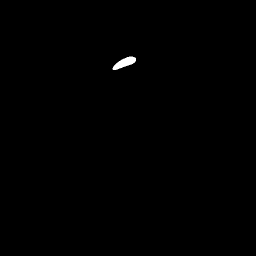}
    \end{subfigure}
    \begin{subfigure}[b]{0.16\textwidth}
        \centering
        \includegraphics[width=\textwidth]{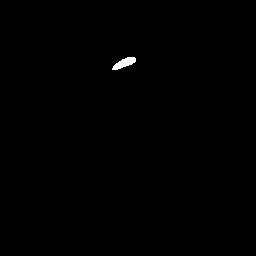}
    \end{subfigure}
    \begin{subfigure}[b]{0.16\textwidth}
        \centering
        \includegraphics[width=\textwidth]{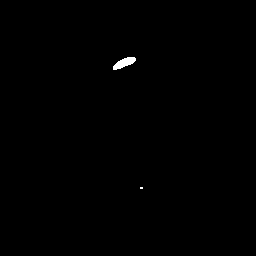}
    \end{subfigure}
    \begin{subfigure}[b]{0.16\textwidth}
        \centering
        \includegraphics[width=\textwidth]{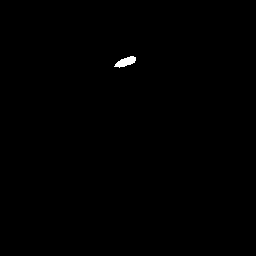}
    \end{subfigure}
    \begin{subfigure}[b]{0.16\textwidth} 
        \centering
        \includegraphics[width=\textwidth]{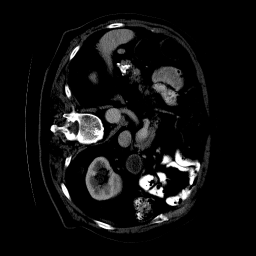}
    \end{subfigure}
    \begin{subfigure}[b]{0.16\textwidth}
        \centering
        \includegraphics[width=\textwidth]{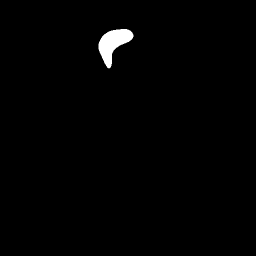}
    \end{subfigure}
    \begin{subfigure}[b]{0.16\textwidth}
        \centering
        \includegraphics[width=\textwidth]{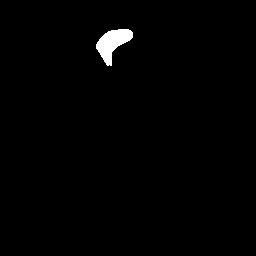}
    \end{subfigure}
    \begin{subfigure}[b]{0.16\textwidth}
        \centering
        \includegraphics[width=\textwidth]{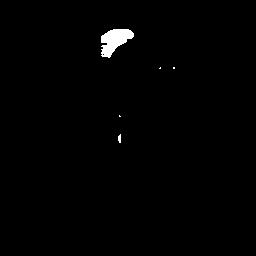}
    \end{subfigure}
    \begin{subfigure}[b]{0.16\textwidth}
        \centering
        \includegraphics[width=\textwidth]{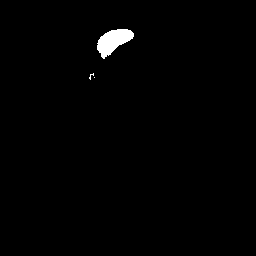}
    \end{subfigure}
    \begin{subfigure}[b]{0.16\textwidth} 
        \centering
        \includegraphics[width=\textwidth]{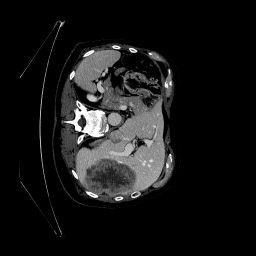}
        \caption*{Input}
    \end{subfigure}
    \begin{subfigure}[b]{0.16\textwidth}
        \centering
        \includegraphics[width=\textwidth]{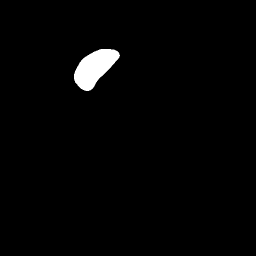}
        \caption*{GT}
    \end{subfigure}
    \begin{subfigure}[b]{0.16\textwidth}
        \centering
        \includegraphics[width=\textwidth]{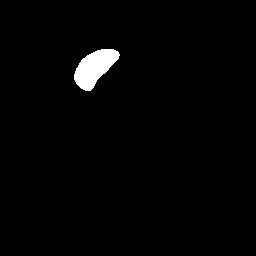}
        \caption*{Teacher (T1)}
    \end{subfigure}
    \begin{subfigure}[b]{0.16\textwidth}
        \centering
        \includegraphics[width=\textwidth]{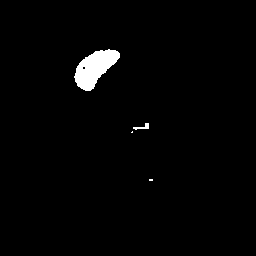}
        \caption*{Student (S2)}
    \end{subfigure}
    \begin{subfigure}[b]{0.16\textwidth}
        \centering
        \includegraphics[width=\textwidth]{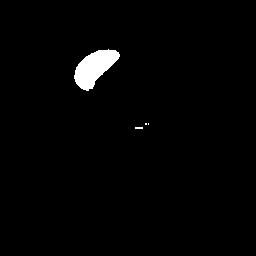}
        \caption*{KD(T2, S2)}
    \end{subfigure}
    \captionsetup{justification=centering}
    \caption{Comparitive Analysis of Segmentation Results between Ground Truth, Output of Student (S2) and KD(T2, S2) on Spleen CT Images }
\label{fig:Comp_Anal_T2_S2}
\end{figure}

An intuitive illustration of how distilled knowledge aids in removing artefacts or false positives from the segmented mask, for example, is given in the first row of the figure. Likewise, the example in the third row shows how knowledge distillation can increase segmentation accuracy. The fourth row should be looked at to further illustrate how the knowledge distillation process improves the segmentation mask. Similarly, we performed a comparison analysis on the student model S2, using teacher T1 as the distillation source, on a few images from the test set. Figure~\ref{fig:Comp_Anal_Seg_v2} depicts a side-by-side comparison analysis. Since S2 is relatively larger than S1, the distillation effect is not as significant when compared to KD(T1, S1). While knowledge distillation is still effective in improving segmentation performance, the larger capacity of S2 means that it already possesses a higher level of expressiveness and capacity to learn from the training data. Consequently, the additional guidance provided by the teacher model (T1) may not result in as drastic improvements as observed with the smaller student model (S1).  Nevertheless, the example shown in the second row of Figure~\ref{fig:Comp_Anal_Seg_v2}, demonstrates how knowledge distillation aided in removing artefacts captured as noise during the segmentation process. Similar examples can be observed in the fourth and fifth rows, respectively. Furthermore, Figure~\ref{fig:Comp_Anal_T2_S1} shows the outcomes generated by S1 when knowledge is being distilled from T2 instead of T1. Though T2, is larger than compared to T1 in terms of parameters and also utilises the self-attention mechanism to extract features from the input image, the performance of S1 is not enhanced significantly. Based on the provided examples, it can be conclusively seen that for most of the cases, the student model has improved in terms of removing noise only, rather than improving the precision of the predicted mask. Nevertheless, S1 does have notable improvements when using T2 as the teacher. Similarly, Figure~\ref{fig:Comp_Anal_T2_S2}, shows the combination KD(T2, S2) on the same examples chosen from the test set, as the KD(T2, S1) model. The figure is illustrative on how a larger and more capable student model's performs better with a larger teacher network. 

Moreover, the third example conclusively shows that distillation not only helps in eliminating false positives, such as artefacts but also contributes to obtaining a more precise segmentation mask. 

It's important to be mindful that, even though knowledge distillation has often led to promising results, there are still circumstances in which the student model's performance does not substantially improve. This can be attributed to the inherent challenges in optimizing the knowledge distillation network, which is itself a field of research that would be too vast to accommodate within the scope of this thesis. To illustrate this point, Figure~\ref{fig:Failed_Cases_KD(T1, S1)} and Figure~\ref{fig:Failed_Cases_KD(T1, S2)} showcase a few examples of both KD(T1, S1) and KD(T1, S2) where the performance of the student model after knowledge distillation is not significant. Several factors, including model complexity, data distribution, and optimization challenges, may prevent the student model from making significant progress even with the guidance given by the teacher model. Moreover, compared to other loss functions like cross-entropy loss, dice loss may not penalize sudden changes in predictions as much. Occasionally, this results in excessively noisy segmentation outcomes.

\begin{figure}[ht!]
    \centering
    \begin{subfigure}[b]{0.16\textwidth} 
        \centering
        \includegraphics[width=\textwidth]{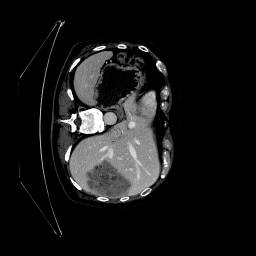}
    \end{subfigure}
    \begin{subfigure}[b]{0.16\textwidth}
        \centering
        \includegraphics[width=\textwidth]{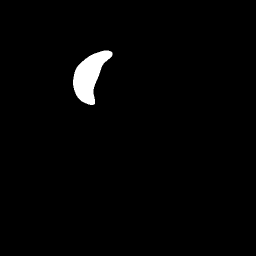}
    \end{subfigure}
    \begin{subfigure}[b]{0.16\textwidth}
        \centering
        \includegraphics[width=\textwidth]{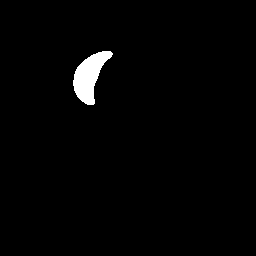}
    \end{subfigure}
    \begin{subfigure}[b]{0.16\textwidth}
        \centering
        \includegraphics[width=\textwidth]{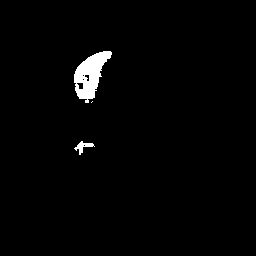}
    \end{subfigure}
    \begin{subfigure}[b]{0.16\textwidth}
        \centering
        \includegraphics[width=\textwidth]{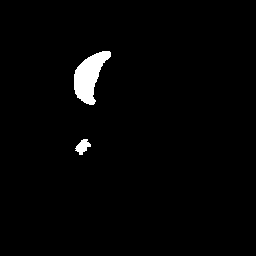}
    \end{subfigure}
        \begin{subfigure}[b]{0.16\textwidth} 
        \centering
        \includegraphics[width=\textwidth]{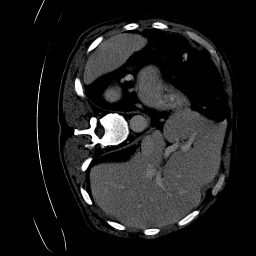}
        \caption*{Input}
    \end{subfigure}
    \begin{subfigure}[b]{0.16\textwidth}
        \centering
        \includegraphics[width=\textwidth]{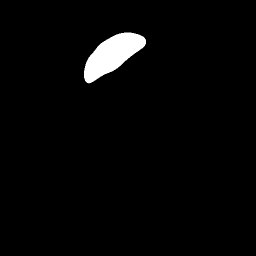}
        \caption*{GT}
    \end{subfigure}
    \begin{subfigure}[b]{0.16\textwidth}
        \centering
        \includegraphics[width=\textwidth]{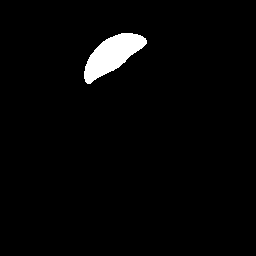}
        \caption*{Teacher (T1)}
    \end{subfigure}
    \begin{subfigure}[b]{0.16\textwidth}
        \centering
        \includegraphics[width=\textwidth]{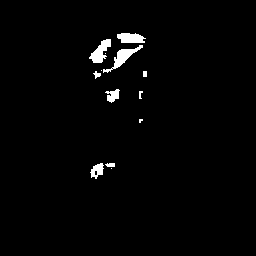}
         \caption*{Student (S1)}
    \end{subfigure}
    \begin{subfigure}[b]{0.16\textwidth}
        \centering
        \includegraphics[width=\textwidth]{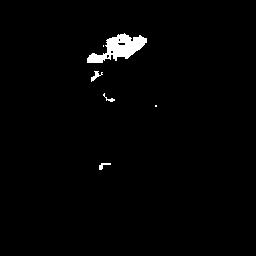}
         \caption*{KD(T1, S1)}
    \end{subfigure}
    \captionsetup{justification=centering}
    \caption{Instances of Limited Improvement in S1 Segmentation via Knowledge Distillation using Contrastive Learning}
\label{fig:Failed_Cases_KD(T1, S1)}
\end{figure}
\begin{figure}[ht!]
    \centering
    \begin{subfigure}[b]{0.16\textwidth} 
        \centering
        \includegraphics[width=\textwidth]{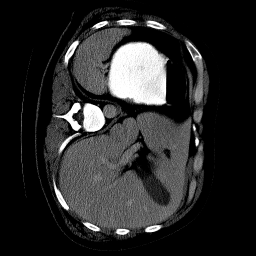}
    \end{subfigure}
    \begin{subfigure}[b]{0.16\textwidth}
        \centering
        \includegraphics[width=\textwidth]{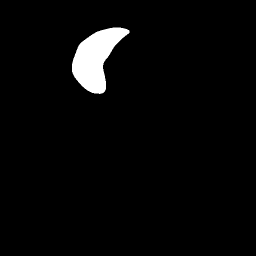}
    \end{subfigure}
    \begin{subfigure}[b]{0.16\textwidth}
        \centering
        \includegraphics[width=\textwidth]{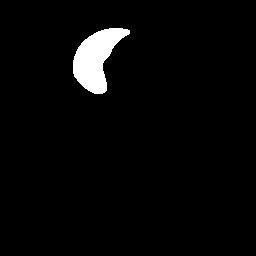}
    \end{subfigure}
    \begin{subfigure}[b]{0.16\textwidth}
        \centering
        \includegraphics[width=\textwidth]{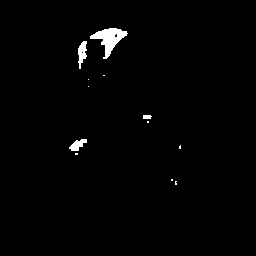}
    \end{subfigure}
    \begin{subfigure}[b]{0.16\textwidth}
        \centering
        \includegraphics[width=\textwidth]{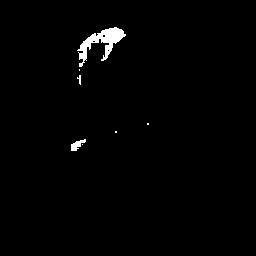}
    \end{subfigure}
        \begin{subfigure}[b]{0.16\textwidth} 
        \centering
        \includegraphics[width=\textwidth]{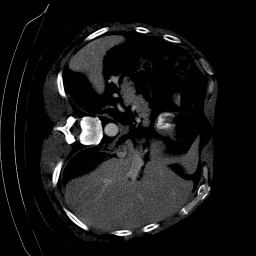}
        \caption*{Input}
    \end{subfigure}
    \begin{subfigure}[b]{0.16\textwidth}
        \centering
        \includegraphics[width=\textwidth]{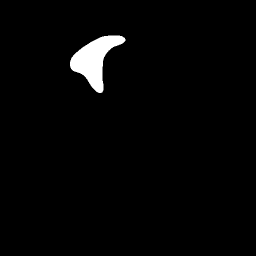}
        \caption*{GT}
    \end{subfigure}
    \begin{subfigure}[b]{0.16\textwidth}
        \centering
        \includegraphics[width=\textwidth]{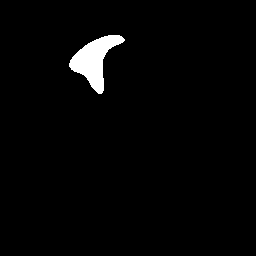}
        \caption*{Teacher (T1)}
    \end{subfigure}
    \begin{subfigure}[b]{0.16\textwidth}
        \centering
        \includegraphics[width=\textwidth]{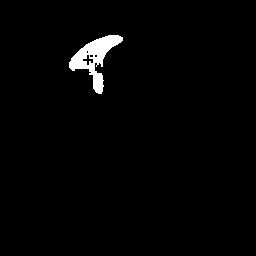}
         \caption*{Student (S2)}
    \end{subfigure}
    \begin{subfigure}[b]{0.16\textwidth}
        \centering
        \includegraphics[width=\textwidth]{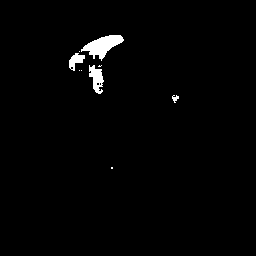}
         \caption*{KD(T1, S2)}
    \end{subfigure}
    \captionsetup{justification=centering}
    \caption{Instances of Limited Improvement in S2 Segmentation via Knowledge Distillation using Contrastive Learning}
\label{fig:Failed_Cases_KD(T1, S2)}
\end{figure}
In summary, the qualitative results provided an in-depth analysis of segmentation outcomes obtained from the knowledge distillation experimentation. The evaluation includes assessments of teacher models T1 and T2, as well as comparisons of student models S1 and S2 under T1's guidance. While significant improvements in segmentation accuracy and artefact reduction are observed, particularly with S1, challenges arise with the larger size of S2 and optimizing the distillation network. These findings demonstrate the effectiveness of knowledge distillation in improving segmentation results, while also emphasizing the need for additional research to address its limitations and optimise its performance.

In the next section, we will go over the ablation studies that were conducted in this thesis in detail. These studies aimed to investigate the impact of a variety of factors, including multi-scale knowledge distillation, the choice between data and knowledge distillation, and the incorporation of contrastive learning.

\section{Ablation Study}
\label{section:Ablation Studies}
The ablation studies were conducted with three primary objectives in mind. Firstly, to evaluate the effects of multi-scale knowledge distillation across various teacher and student network variants. Secondly, to investigate the implications of data versus knowledge distillation on performance. And finally, to observe the impact of contrastive learning in facilitating the overall knowledge transfer process. Through these studies, we aimed to gain insights into the effectiveness and interactions of these different techniques in improving model performance and knowledge transfer in the context of our task. For all the ablation experiments we use the same spleen dataset from \cite{simpson2019large}.

To begin with, an ablation study was conducted to investigate the efficacy of multi-scale knowledge distillation in improving segmentation performance, the results of which can be found in Table~\ref{tab:multi_scale_kd_s1}. The study focused on the teacher-student pair KD(T1, S1), intending to evaluate various configurations of knowledge distillation across different layers of the network architecture. Specifically, distillation was performed from the bottleneck layer (B), encoder (E), and decoder (D) layers individually, as well as in combination (B $\rightarrow$ B + E $\rightarrow$ E + D $\rightarrow$ D). The results indicate that all variants of multi-scale knowledge distillation lead to improvements in segmentation performance compared to the baseline student model (S1). However, the most significant performance improvement in segmentation is seen in encoder-to-encoder knowledge distillation using contrastive learning, which is highlighted in the table as well. This is primarily because, in a UNet-like architecture, during the feature decomposition low-level features are extracted which is useful for the segmentation task. 
\begin{table}[ht!]
    \centering
    \captionsetup{justification=centering}
    \caption{Multi-Scale Knowledge Distillation Results on KD(T1, S1)}
    \begin{adjustbox}{width=\textwidth} 
    \begin{tabular}{lcccc}
        \toprule
        \textbf{Method} & \textbf{IoU} & \textbf{Dice} & \textbf{Recall} & \textbf{Precision} \\
        \midrule
        \textcolor{blue}{Base (S1)} & \textcolor{blue}{0.557} & \textcolor{blue}{0.715} & \textcolor{blue}{0.685} & \textcolor{blue}{0.749} \\
        \( \text{B} \rightarrow \text{B} \) & 0.617 (+10.8\%) & 0.763 (+6.7\%) & 0.704 (+2.7\%) & 0.833 (+11.2\%)\\
        \textbf{\( \text{E} \rightarrow \text{E} \)} & \textbf{0.621 (+11.5\%)} & \textbf{0.766 (+7.1\%)} & \textbf{0.711 (+3.7\%)} & \textbf{0.811 (+8.2\%)} \\
        \( \text{D} \rightarrow \text{D} \) & 0.612 (+9.9\%) & 0.759 (+6.1\%) & 0.695 (+1.4\%) & 0.835 (+11.4\%) \\
        \( \text{B} \rightarrow \text{B} + \text{E} \rightarrow \text{E} + \text{D} \rightarrow \text{D} \) & 0.593 (+6.5\%) & 0.744 (+4.0\%) & 0.684 (-0.1\%) & 0.817 (+9.0\%) \\
        \bottomrule
    \end{tabular}
    \end{adjustbox}
    \label{tab:multi_scale_kd_s1}
\end{table}
\begin{table}[ht!]
    \centering
    \captionsetup{justification=centering}
    \caption{Multi-Scale Knowledge Distillation Results on KD(T2, S1)}
    \begin{adjustbox}{width=\textwidth} 
    \begin{tabular}{lcccc}
        \toprule
        \textbf{Method} & \textbf{IoU} & \textbf{Dice} & \textbf{Recall} & \textbf{Precision} \\
        \midrule
        \textcolor{blue}{Base (S1)} & \textcolor{blue}{0.557} & \textcolor{blue}{0.715} & \textcolor{blue}{0.685} & \textcolor{blue}{0.749} \\
        \( \text{B} \rightarrow \text{B} \) & \( {0.622} \) (+11.7\%) & \( {0.767} \) (+7.2\%) & \( {0.697} \) (+1.7\%) & \( {0.853} \) (+13.8\%) \\
        \( \text{E} \rightarrow \text{E} \) & \(0.605\) (+8.6\%) & \(0.754 \) (+5.4\%) & \(0.725\) (+5.8\%) & \(0.785\) (+4.8\%) \\
        \( \text{D} \rightarrow \text{D} \) & \(0.614\) (+10.2\%) & \(0.761 \) (+6.4\%) & \(0.703\) (+2.6\%) & \(0.828\) (+10.5\%) \\
       \textbf{\( \text{B} \rightarrow \text{B} \) + \( \text{E} \rightarrow \text{E} \)} & \(\textbf{0.622}\) \textbf{(+11.6\%)} & \(\textbf{0.767} \) \textbf{(+7.2\%)} & \(\textbf{0.724}\) \textbf{(+5.6\%)} & \(\textbf{0.815}\) \textbf{(+8.8\%)} \\
        \( \text{B} \rightarrow \text{B} + \text{E} \rightarrow \text{E} + \text{D} \rightarrow \text{D} \) & \(0.615\) (+10.4\%) & \(0.762\) (+6.5\%) & \(0.683\) (-0.1\%) & \(0.861\) (+14.9\%) \\
        \bottomrule
    \end{tabular}
    \end{adjustbox}
    \label{tab:multi_scale_kd_T2-S1}
\end{table}

\begin{table}[ht!]
    \centering
    \captionsetup{justification=centering}
    \caption{Multi-Scale Knowledge Distillation Results with Progressive Maps Distillation on KD(T2, S1)}
    \begin{adjustbox}{width=\textwidth} 
    \begin{tabular}{lcccc}
        \toprule
        \textbf{Method} & \textbf{IoU} & \textbf{Dice} & \textbf{Recall} & \textbf{Precision} \\
        \midrule
        \textcolor{blue}{Base (S1)} & \textcolor{blue}{0.557} & \textcolor{blue}{0.715} & \textcolor{blue}{0.685} & \textcolor{blue}{0.749} \\
        \( \text{B} \rightarrow \text{B} \) + PMD & \( 0.613 \) (+10.0\%) & \( 0.760 \) (+6.2\%) & \( 0.707 \) (+3.2\%) & \( 0.822 \) (+9.7\%) \\
        \textbf{\( \text{E} \rightarrow \text{E} \) + PMD} & \(\textbf{0.623}\) \textbf{(+11.8\%)} & \(\textbf{0.767} \) \textbf{(+7.2\%)} & \(\textbf{0.707}\) \textbf{(+3.2\%)} & \(\textbf{0.839}\) \textbf{(+12.0\%)} \\
        \( \text{B} \rightarrow \text{B} + \text{D} \rightarrow \text{D} \) + PMD & \( 0.620 \) (+11.3\%) & \( 0.766 \) (+7.1\%) & \( 0.705 \) (+2.9\%) & \( 0.837 \) (+11.7\%) \\
        \( \text{B} \rightarrow \text{B} + \text{E} \rightarrow \text{E}\) + PMD &0.614 (+10.2\%) & 0.761 (+6.4\%) & 0.690 (+0.7\%) & 0.849 (+13.3\%) \\
        \( \text{B} \rightarrow \text{B} + \text{E} \rightarrow \text{E} + \text{D} \rightarrow \text{D} \) + PMD &0.622  (+11.6\%) &0.766  (+7.1\%) &0.706 (+3.0\%) &0.838 (+11.8\%) \\
        \midrule
    \end{tabular}
    \end{adjustbox}
    \label{tab:multi_scale_kd_pmd_T2-S1}
\end{table}

The improvements we have observed suggest that the overall architecture that leverages information from different layers of the network is effective in improving feature representation and segmentation accuracy. When distilled from individual layers, such as bottleneck (B $\rightarrow$ B), encoder (E $\rightarrow$ E), and decoder (D $\rightarrow$ D) layers, significant gains in performance are achieved across all metrics. This supports our initial hypothesis regarding multi-scale information transfer. However, it is worth noting that distilling information from all layers simultaneously results in slightly lower performance. Distilling knowledge from all layers simultaneously presents a significant challenge for contrastive learning to optimize effectively. Each layer contains distinct and diverse information that should be treated separately. Attempting to distil knowledge from all layers concurrently could lead to information overload for the student model, which can impact its ability to learn effectively. Contrastive learning relies on distinguishing between positive and negative examples to learn representations effectively. However, when the student model is confronted with a plethora of diverse information from all layers simultaneously, it becomes challenging to discern which information is most relevant and informative. 
\begin{figure}[htbp]
    \centering
    \begin{subfigure}[b]{0.16\textwidth} 
        \centering
        \includegraphics[width=\textwidth]{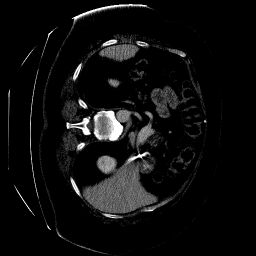}
    \end{subfigure}
    \begin{subfigure}[b]{0.16\textwidth}
        \centering
        \includegraphics[width=\textwidth]{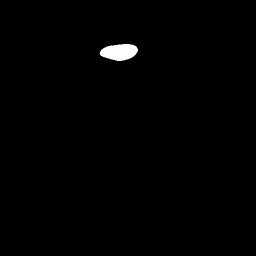}
    \end{subfigure}
    \begin{subfigure}[b]{0.16\textwidth}
        \centering
        \includegraphics[width=\textwidth]{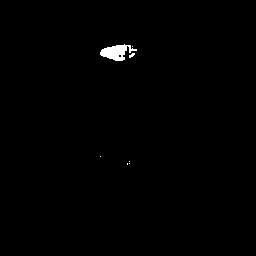}
    \end{subfigure} 
    \begin{subfigure}[b]{0.16\textwidth}
        \centering
        \includegraphics[width=\textwidth]{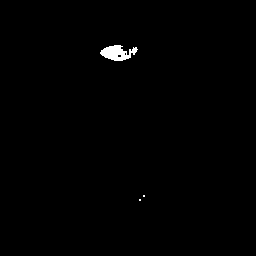}
    \end{subfigure} \\
    \begin{subfigure}[b]{0.16\textwidth} 
        \centering
        \includegraphics[width=\textwidth]{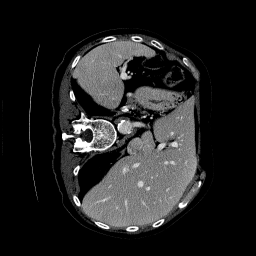}
    \end{subfigure}
    \begin{subfigure}[b]{0.16\textwidth}
        \centering
        \includegraphics[width=\textwidth]{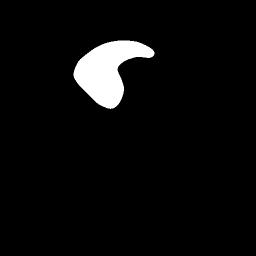}
    \end{subfigure}
    \begin{subfigure}[b]{0.16\textwidth}
        \centering
        \includegraphics[width=\textwidth]{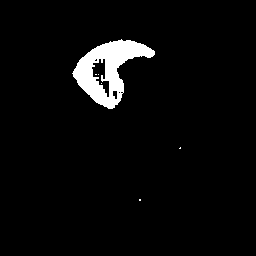}
    \end{subfigure}     
    \begin{subfigure}[b]{0.16\textwidth}
        \centering
        \includegraphics[width=\textwidth]{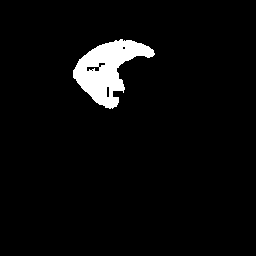}
    \end{subfigure} \\
    \begin{subfigure}[b]{0.16\textwidth} 
    \centering
    \includegraphics[width=\textwidth]{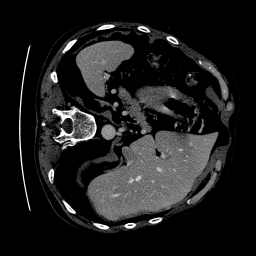}
    \caption*{Input}
    \end{subfigure}
    \begin{subfigure}[b]{0.16\textwidth}
        \centering
        \includegraphics[width=\textwidth]{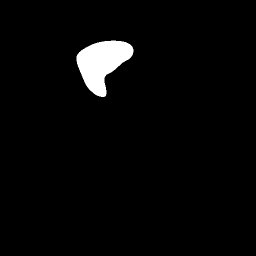}
        \caption*{GT}
    \end{subfigure}
    \begin{subfigure}[b]{0.16\textwidth}
        \centering
        \includegraphics[width=\textwidth]{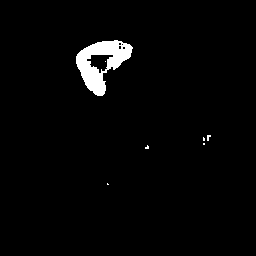}
        \caption*{(B$\rightarrow$B)}
    \end{subfigure}
    \begin{subfigure}[b]{0.16\textwidth}
    \centering
    \includegraphics[width=\textwidth]{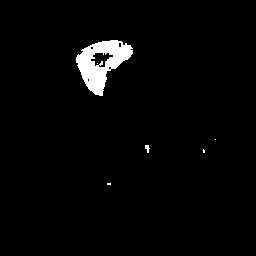}
    \caption*{(E$\rightarrow$E)}
\end{subfigure}
    \captionsetup{justification=centering}
    \caption{Results on Different Scale of Knowledge Distillation in (T1, S1)}
\label{fig:E-EvsB-B_MultiScale_KD}
\end{figure}
Therefore, it is crucial to consider each layer's unique information separately to avoid information overload and improve the student model's accuracy. We took a few examples to provide the same observation qualitatively. The results shown in Figure~\ref{fig:E-EvsB-B_MultiScale_KD} provide support to our argument on this ablation study. 

\begin{table}[ht!]
    \centering
    \captionsetup{justification=centering}
    \caption{Selected Scales of Multi-Scale Knowledge Distillation Results on KD(T1, S1)}
    \begin{adjustbox}{width=\textwidth} 
    \begin{tabular}{lcccc}
        \toprule
        \textbf{Method} & \textbf{IoU} & \textbf{F1 Score} & \textbf{Recall} & \textbf{Precision} \\
        \midrule
        \textcolor{blue}{Base (S1)} & \textcolor{blue}{0.557} & \textcolor{blue}{0.715} & \textcolor{blue}{0.685} & \textcolor{blue}{0.749} \\
        \( \text{B} \rightarrow \text{B} \)+\( \text{E} \rightarrow \text{E} \) & 0.610 (+9.5\%) & 0.757 (+5.8\%) & 0.718 (+4.8\%) & 0.801 (+6.9\%) \\
        \( \text{E} \rightarrow \text{E} \)+\( \text{D} \rightarrow \text{D} \) & 0.616 (+10.5\%) & 0.762 (+6.5\%) & 0.722 (+5.4\%) & 0.808 (+7.8\%) \\
        \( \text{B} \rightarrow \text{B} \)+\( \text{D} \rightarrow \text{D} \) & 0.616 (+10.5\%) & 0.762 (+6.5\%) & 0.703 (+2.6\%) & 0.833 (+11.2\%) \\
        \bottomrule
    \end{tabular}
    \end{adjustbox}
    \label{tab:selected_multi_scale_kd_pmd}
\end{table}

Table~\ref{tab:selected_multi_scale_kd_pmd} shows the results of selected scales of multi-scale knowledge distillation experiments performed on the teacher-student pair KD(T1, S1). These findings highlight the importance of multi-scale knowledge distillation in improving the segmentation performance of a student model. Distillation from specific scales, such as bottleneck to bottleneck combined with encoder to encoder, encoder to encoder combined with a decoder to decoder, and bottleneck to bottleneck combined with a decoder to decoder, consistently results in improvements across all metrics when compared to the baseline student model (S1), however, based on what we have investigated in this section, the most significant increase remains the encoder-to-encoder variant.

The next ablation was done to investigate the efficacy of PMD in the knowledge distillation process. Table~\ref{tab:multi_scale_kd_pmd} summarises the results of multi-scale knowledge distillation experiments with PMD on the teacher-student pair KD(T1, S1). The results showcase various configurations of knowledge distillation, incorporating PMD along with distillation from different layers of the network architecture. Overall, the introduction of PMD in the knowledge distillation process yields promising improvements with consistent enhancements in segmentation performance across various configurations compared to the baseline student model (S1), however, not all scales see a performance improvement. This is mostly because during contrastive learning not all combinations of features complement each other, we can see a decline in the performance in selected cases as well. Nevertheless, the variant involving a bottleneck-to-bottleneck knowledge distillation with PMD produced the best results in the (T1, S1) pair. This combination leverages the rich semantic information captured by the bottleneck layer, which represents high-level abstract features essential for accurate segmentation. This result is also visualized in Figure~\ref{fig:Res_MultiScale_KD}. 
\begin{table}[ht!]
    \centering
    \captionsetup{justification=centering}
    \caption{Multi-Scale Knowledge Distillation Results with Progressive Maps Distillation on KD(T1, S1)}
    \begin{adjustbox}{width=\textwidth} 
    \begin{tabular}{lcccc}
        \toprule
        \textbf{Method}                & \textbf{IoU}       & \textbf{Dice}  & \textbf{Recall}    & \textbf{Precision} \\
        \midrule
        \textcolor{blue}{Base (S1)} & \textcolor{blue}{0.557} & \textcolor{blue}{0.715} & \textcolor{blue}{0.685} & \textcolor{blue}{0.749} \\
        \textbf{\( \text{B} \rightarrow \text{B} \) + PMD }          & \textbf{0.629 (+12.9\%) }   & \textbf{0.772 (+7.9\%)}    & \textbf{0.721 (+5.2\%) }   & \textbf{0.831 (+10.9\%) }   \\
        \( \text{B} \rightarrow \text{B} \)+\( \text{E} \rightarrow \text{E} \) + PMD       & 0.607 (+8.9\%)    & 0.756 (+5.7\%)    & 0.694 (+1.3\%)    & 0.829 (+10.6\%)    \\
        \( \text{E} \rightarrow \text{E} \) + PMD           & 0.617 (+10.7\%)   & 0.763 (+6.57\%)    & 0.733 (+7.0\%)    & 0.796 (+6.2\%)    \\
        \( \text{B} \rightarrow \text{B} \)+\( \text{E} \rightarrow \text{E} \)+\( \text{D} \rightarrow \text{D} \) + PMD   & 0.604 (+8.4\%)    & 0.753 (+5.3\%)    & 0.702 (+2.4\%)    & 0.812 (+8.4\%)    \\
        \( \text{D} \rightarrow \text{D} \) + PMD           & 0.541 (-2.8\%)    & 0.702 (-1.8\%)    & 0.653 (-4.6\%)    & 0.759 (+1.3\%)    \\
        \( \text{E} \rightarrow \text{E} \)+\( \text{D} \rightarrow \text{D} \) + PMD       & 0.628 (+12.7\%)    & 0.771 (+7.8\%)    & 0.722 (+5.4\%)    & 0.828 (+10.5\%)    \\
        \textbf{\( \text{B} \rightarrow \text{B} \)+\( \text{D} \rightarrow \text{D} \) + PMD} & \textbf{0.628 (+12.7\%)} & \textbf{0.771 (+7.8\%)} & \textbf{0.722 (+5.4\%)} & \textbf{0.828 (+10.5\%)} \\
        \bottomrule 
    \end{tabular} 
    \end{adjustbox}
    \label{tab:multi_scale_kd_pmd}
\end{table}

\begin{figure}[ht!]
    \centering
    \begin{subfigure}[b]{0.16\textwidth} 
        \centering
        \includegraphics[width=\textwidth]{Images/Ablation_Images/Table_1/image_spleen_10_25.png}
    \end{subfigure}
    \begin{subfigure}[b]{0.16\textwidth}
        \centering
        \includegraphics[width=\textwidth]{Images/Ablation_Images/Table_1/image_spleen_10_25_GT.png}
    \end{subfigure}
    \begin{subfigure}[b]{0.16\textwidth}
        \centering
        \includegraphics[width=\textwidth]{Images/Ablation_Images/Table_1/mask_spleen_10_25_predicted_B-B.png}
    \end{subfigure}
    \begin{subfigure}[b]{0.16\textwidth}
        \centering
        \includegraphics[width=\textwidth]{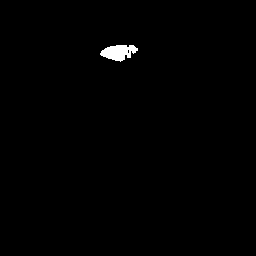}
    \end{subfigure} \\
    \begin{subfigure}[b]{0.16\textwidth} 
        \centering
        \includegraphics[width=\textwidth]{Images/Ablation_Images/Table_1/image_spleen_12_101.png}
    \end{subfigure}
    \begin{subfigure}[b]{0.16\textwidth}
        \centering
        \includegraphics[width=\textwidth]{Images/Ablation_Images/Table_1/image_spleen_12_101_GT.png}
    \end{subfigure}
    \begin{subfigure}[b]{0.16\textwidth}
        \centering
        \includegraphics[width=\textwidth]{Images/Ablation_Images/Table_1/mask_spleen_12_101_predicted_B-B.png}
    \end{subfigure}
    \begin{subfigure}[b]{0.16\textwidth}
        \centering
        \includegraphics[width=\textwidth]{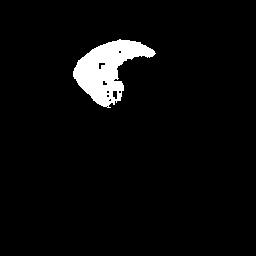}
    \end{subfigure} \\
        \begin{subfigure}[b]{0.16\textwidth} 
        \centering
        \includegraphics[width=\textwidth]{Images/Ablation_Images/Table_1/image_spleen_13_51.png}
        \caption*{Input}
    \end{subfigure}
    \begin{subfigure}[b]{0.16\textwidth}
        \centering
        \includegraphics[width=\textwidth]{Images/Ablation_Images/Table_1/image_spleen_13_51_GT.png}
        \caption*{GT}
    \end{subfigure}
    \begin{subfigure}[b]{0.16\textwidth}
        \centering
        \includegraphics[width=\textwidth]{Images/Ablation_Images/Table_1/mask_spleen_13_51_predicted_B-B.png}
        \caption*{(B$\rightarrow$B)}
    \end{subfigure}
    \begin{subfigure}[b]{0.16\textwidth}
        \centering
        \includegraphics[width=\textwidth]{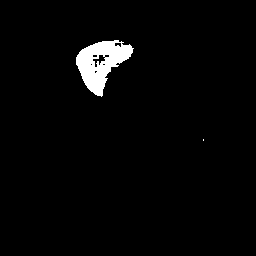}
        \caption*{(B$\rightarrow$B)+PMD}
    \end{subfigure}
    \captionsetup{justification=centering}
    \caption{Experimental Result Showcasing the Significance of PMD with Multi-Scale Knowledge Distillation in (T1, S1)}
\label{fig:Res_MultiScale_KD}
\end{figure}
Similar results were obtained when experiments were performed on the (S2) student model using T1 as the distillation source. The results of these experiments can be found in Table~\ref{tab:multi_scale_kd_s2} and \ref{tab:multi_scale_kd_pmd_s2}, respectively. 
\begin{table}[ht!]
    \centering
    \captionsetup{justification=centering}
    \caption{Multi-Scale Knowledge Distillation Results on KD(T1, S2)}
    \begin{adjustbox}{width=\textwidth} 
    \begin{tabular}{lcccc}
        \toprule
        \textbf{Method} & \textbf{IoU} & \textbf{Dice} & \textbf{Recall} & \textbf{Precision} \\
        \midrule
        \textcolor{magenta}{Base (S2)} & \textcolor{magenta}{0.627} & \textcolor{magenta}{0.770} & \textcolor{magenta}{0.725} & \textcolor{magenta}{0.822} \\
        \textbf{\( \text{B} \rightarrow \text{B} \)} &\textbf{0.649 (+3.5\%) } &\textbf{0.787 (+2.2\%)} &\textbf{0.724 (-0.1\%)} &\textbf{0.862 (+4.8\%)} \\
        \( \text{E} \rightarrow \text{E} \) &0.621 (-0.9\%) &0.766 (-0.5\%) &0.681 (-6.0\%) &0.875 (+6.4\%) \\
        \( \text{D} \rightarrow \text{D} \) &0.647 (+3.1\%) &0.786 (+2.0\%) &0.733 (+1.1\%) &0.846 (+2.9\%) \\
        \( \text{B} \rightarrow \text{B} + \text{E} \rightarrow \text{E} \) &0.620 (-1.1\%) &0.765 (-0.6\%) &0.716 (-1.2\%) &0.821 (-0.1\%) \\
        \( \text{B} \rightarrow \text{B} + \text{E} \rightarrow \text{E} + \text{D} \rightarrow \text{D} \) &0.643 (+2.5\%) &0.783 (+1.6\%) &0.760 (+4.8\%) &0.807 (-1.8\%) \\
        \bottomrule
    \end{tabular}
    \end{adjustbox}
    \label{tab:multi_scale_kd_s2}
\end{table}
\begin{table}[ht!]
    \centering
    \captionsetup{justification=centering}
    \caption{Multi-Scale Knowledge Distillation Results with Progressive Maps Distillation on KD(T1, S2)}
    \begin{adjustbox}{width=\textwidth} 
    \begin{tabular}{lcccc}
        \toprule
        \textbf{Method} & \textbf{IoU} & \textbf{Dice} & \textbf{Recall} & \textbf{Precision} \\
        \midrule
        \textcolor{magenta}{Base (S2)} & \textcolor{magenta}{0.627} & \textcolor{magenta}{0.770} & \textcolor{magenta}{0.725} & \textcolor{magenta}{0.822} \\
        \textbf{\( \text{B} \rightarrow \text{B} \) + PMD} & \textbf{0.659 (+5.1\%)} & \textbf{0.794 (+3.1\%)} & \textbf{0.748 (+3.1\%)} & \textbf{0.846 (+2.9\%)} \\
        \( \text{E} \rightarrow \text{E} \) + PMD & 0.638 (+1.7\%) & 0.779 (+1.1\%) & 0.756 (+4.2\%) & 0.803 (-2.3\%) \\
        \( \text{D} \rightarrow \text{D} \) + PMD & 0.656 (+4.6\%) & 0.792 (+2.8\%) & 0.759 (+4.6\%) & 0.828 (+0.7\%) \\
        \( \text{B} \rightarrow \text{B} + \text{E} \rightarrow \text{E} \) + PMD & 0.643 (+2.5\%) & 0.783 (+1.6\%) & 0.753 (+3.8\%) & 0.815 (-0.8\%) \\
        \( \text{B} \rightarrow \text{B} + \text{E} \rightarrow \text{E} + \text{D} \rightarrow \text{D} \) + PMD & 0.642 (+2.3\%) & 0.782 (+1.5\%) & 0.723 (+0.2\%) & 0.850 (+3.4\%) \\
        \bottomrule
    \end{tabular}
    \end{adjustbox}
    \label{tab:multi_scale_kd_pmd_s2}
\end{table}

We carried out an ablation study to demonstrate the superiority of contrastive learning over knowledge distillation using mean squared error (MSE) loss. Our study compared the performance of these two techniques on the pair (T1, S1) at each scale. As evident from the results presented in Table~\ref{tab:kd_results_MSE}, the performance increase achieved through simple knowledge distillation using MSE loss was not as significant as the results obtained with contrastive loss. Thus, our study highlights the effectiveness of contrastive learning in achieving better performance outcomes. Figure~\ref{fig:Res_MultiScale_KD_MSE} shows a comparison of the segmentations obtained using the MSE and contrastive losses on some CT images from the test set.

\begin{table}[ht!]
    \centering
    \caption{Results of Knowledge Distillation using MSE Loss on KD(T1, S1)}
    \begin{adjustbox}{width=\textwidth} 
    \begin{tabular}{lccccc}
        \toprule
        \textbf{Method} & \textbf{IoU} & \textbf{Dice} & \textbf{Recall} & \textbf{Precision} \\
        \midrule
        \textcolor{blue}{Base (S1)} & \textcolor{blue}{0.557} & \textcolor{blue}{0.715} & \textcolor{blue}{0.685} & \textcolor{blue}{0.749} \\
        \( \text{B} \rightarrow \text{B} \) & 0.566 (+1.6\%) & 0.723 (+1.1\%) & 0.678 (-1.0\%) & 0.774 (+3.3\%) \\
        \( \text{B} \rightarrow \text{B} \)+PMD & 0.531 (-4.6\%) & 0.693 (-3.0\%) & 0.654 (-4.5\%) & 0.738 (-1.4\%) \\
        \( \text{B} \rightarrow \text{B} \)+\( \text{E} \rightarrow \text{E} \) & 0.519 (-6.8\%) & 0.683 (-4.4\%) & 0.613 (-10.5\%) & 0.772 (+3.0\%) \\
        \( \text{B} \rightarrow \text{B} \)+\( \text{E} \rightarrow \text{E} \) + PMD & 0.545 (+2.1\%) & 0.705 (-1.3\%) & 0.668 (-2.4\%) & 0.747 (-0.2\%) \\
        \textbf{\( \text{B} \rightarrow \text{B} + \text{E} \rightarrow \text{E} + \text{D} \rightarrow \text{D} \)}  &\textbf{0.587 (+5.3\%)} &\textbf{0.740 (+3.4\%)} &\textbf{0.714 (+4.2\%)} &\textbf{0.786 (+4.9\%)} \\
        \( \text{B} \rightarrow \text{B} + \text{E} \rightarrow \text{E} + \text{D} \rightarrow \text{D} \) + PMD  & 0.548 (-1.6\%) &0.708 (-0.9\%)  &0.626 (-8.6\%) &0.816 (+8.9\%) \\
        \bottomrule
    \end{tabular}
    \end{adjustbox}
    \label{tab:kd_results_MSE}
\end{table}

\begin{figure}[ht!]
    \centering
    \begin{subfigure}[b]{0.16\textwidth} 
        \centering
        \includegraphics[width=\textwidth]{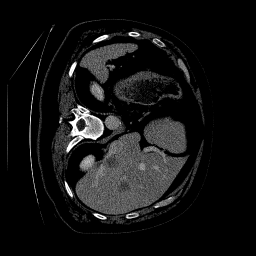}
    \end{subfigure}
    \begin{subfigure}[b]{0.16\textwidth}
        \centering
        \includegraphics[width=\textwidth]{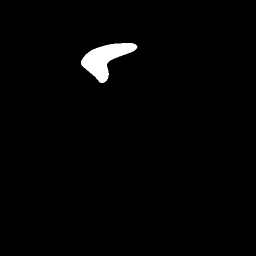}
    \end{subfigure}
    \begin{subfigure}[b]{0.16\textwidth}
        \centering
        \includegraphics[width=\textwidth]{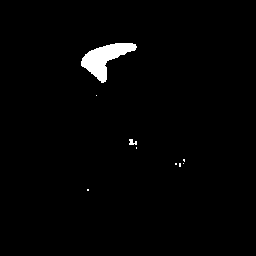}
    \end{subfigure}
    \begin{subfigure}[b]{0.16\textwidth}
        \centering
        \includegraphics[width=\textwidth]{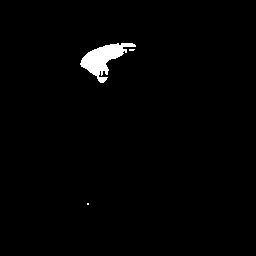}
    \end{subfigure} \\
    \begin{subfigure}[b]{0.16\textwidth} 
        \centering
        \includegraphics[width=\textwidth]{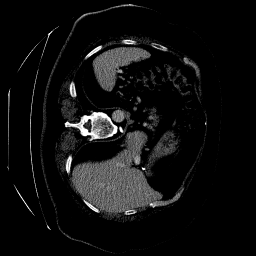}
    \end{subfigure}
    \begin{subfigure}[b]{0.16\textwidth}
        \centering
        \includegraphics[width=\textwidth]{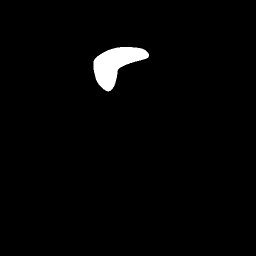}
    \end{subfigure}
    \begin{subfigure}[b]{0.16\textwidth}
        \centering
        \includegraphics[width=\textwidth]{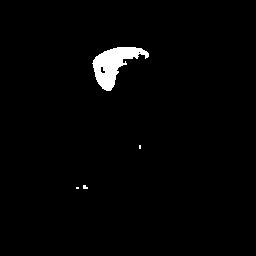}
    \end{subfigure}
    \begin{subfigure}[b]{0.16\textwidth}
        \centering
        \includegraphics[width=\textwidth]{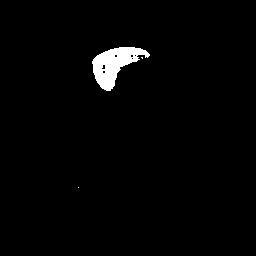}
    \end{subfigure} \\
    \begin{subfigure}[b]{0.16\textwidth} 
        \centering
        \includegraphics[width=\textwidth]{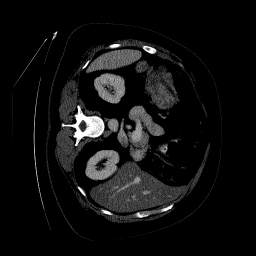}
    \end{subfigure}
    \begin{subfigure}[b]{0.16\textwidth}
        \centering
        \includegraphics[width=\textwidth]{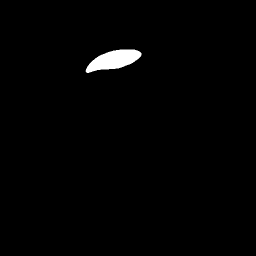}
    \end{subfigure}
    \begin{subfigure}[b]{0.16\textwidth}
        \centering
        \includegraphics[width=\textwidth]{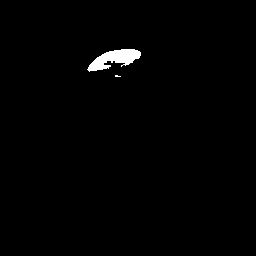}
    \end{subfigure}
    \begin{subfigure}[b]{0.16\textwidth}
        \centering
        \includegraphics[width=\textwidth]{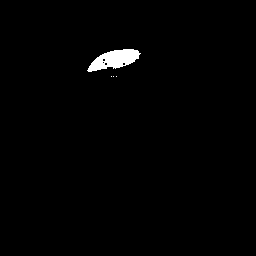}
    \end{subfigure} \\
        \begin{subfigure}[b]{0.16\textwidth} 
        \centering
        \includegraphics[width=\textwidth]{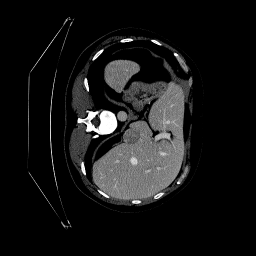}
        \caption*{Input}
    \end{subfigure}
    \begin{subfigure}[b]{0.16\textwidth}
        \centering
        \includegraphics[width=\textwidth]{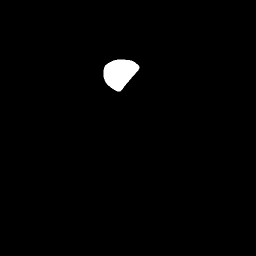}
        \caption*{GT}
    \end{subfigure}
    \begin{subfigure}[b]{0.16\textwidth}
        \centering
        \includegraphics[width=\textwidth]{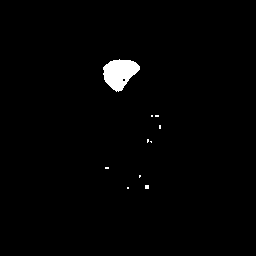}
        \caption*{MSE Loss}
    \end{subfigure}
    \begin{subfigure}[b]{0.16\textwidth}
        \centering
        \includegraphics[width=\textwidth]{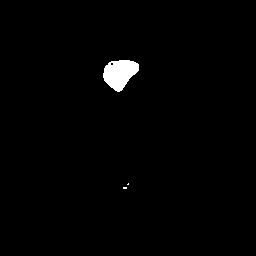}
        \caption*{Cont. Loss}
    \end{subfigure}
    \captionsetup{justification=centering}
    \caption{Experimental Result on Multi-Scale Knowledge Distillation in (T1, S1) using MSE vs Contrastive Loss}
\label{fig:Res_MultiScale_KD_MSE}
\end{figure} 
In this ablation study, we aim to explore the impact of data volume on a model's performance on a segmentation task. As mentioned earlier in Chapter~\ref{chapter:introduction}, the student model is trained with only half the data of the teacher model. This is done intentionally to ensure that the student's performance after knowledge distillation is not solely reliant on the volume of data it has been trained on. We will investigate what happens when a student model undergoes knowledge distillation with the same volume of data as the teacher ($S1_{\text{teacher}}$) and with half the volume of the same (S1). If the latter scenario yields better results, it indicates that performance improvement is not solely due to the quantity of data. Thus, it is advantageous if a smaller model can perform comparably or even better with less data, thanks to effective knowledge distillation.
\begin{table}[ht!]
    \centering
    \caption{Comparitive Study of Knowledge Distillation on Data Volume}
    \begin{adjustbox}{width=\textwidth} 
    \begin{tabular}{lcccc}
        \toprule
        \textbf{Method} & \textbf{IoU} & \textbf{Dice} & \textbf{Recall} & \textbf{Precision} \\
        \midrule
        \multicolumn{5}{c}{\textbf{With Same Training Volume as Teacher Network KD(T1, S1)}} \\
        \midrule
       \textcolor{blue}{\ensuremath{S1_{\text{teacher}}}} & \textcolor{blue}{0.617} & \textcolor{blue}{0.763} & \textcolor{blue}{0.727} & \textcolor{blue}{0.803} \\
        \( \text{B} \rightarrow \text{B} \) & 0.615 (-0.3\%) & 0.761 (-0.2\%) & 0.707 (-2.7\%) & 0.824 (+2.6\%) \\
        \( \text{B} \rightarrow \text{B} \) + PMD & 0.622 (+0.8\%) & 0.767 (+0.5\%) & 0.714 (-1.7\%) & 0.829 (+3.2\%) \\
        \( \text{E} \rightarrow \text{E} \) & 0.607 (-1.6\%) & 0.756 (-0.9\%) & 0.718 (-1.2\%) & 0.797 (-0.7\%) \\
        \( \text{E} \rightarrow \text{E} \) + PMD & 0.607 (-1.6\%) & 0.756 (-0.9\%) & 0.718 (-1.2\%) & 0.797 (-0.7\%) \\
        \( \text{B} \rightarrow \text{B} \)+\( \text{E} \rightarrow \text{E} \) & 0.609 (-1.2\%) & 0.757 (-0.7\%) & 0.697 (-4.1\%) & 0.828 (+3.1\%) \\
        \( \text{B} \rightarrow \text{B} \)+\( \text{E} \rightarrow \text{E} \) + PMD & 0.616 (-0.1\%) & 0.762 (-0.1\%) & 0.710 (-2.3\%) & 0.822 (+2.3\%) \\
        \textbf{\( \text{B} \rightarrow \text{B} \)+\( \text{E} \rightarrow \text{E} \)+\( \text{D} \rightarrow \text{D} \)} & \textbf{0.623 (+0.9\%)} & \textbf{0.767 (+0.5\%)} & \textbf{0.718 (-1.2\%)} & \textbf{0.825 (+2.7\%)} \\
        \( \text{B} \rightarrow \text{B} \)+\( \text{E} \rightarrow \text{E} \)+\( \text{D} \rightarrow \text{D} \) + PMD & 0.618 (+0.1\%) & 0.764 (+0.1\%) & 0.711 (-2.2\%) & 0.826 (+2.8\%) \\
        \midrule
        \multicolumn{5}{c}{\textbf{With 50\% Training Volume and Knowledge Distillation KD(T1, S1)}} \\
        \midrule
        \textcolor{blue}{Base (S1)} & \textcolor{blue}{0.557} & \textcolor{blue}{0.715} & \textcolor{blue}{0.685} & \textcolor{blue}{0.749} \\
        \( \text{B} \rightarrow \text{B} \) & 0.617 (+10.7\%) & 0.763 (+6.7\%) & 0.704 (+2.7\%) & 0.833 (+11.2\%) \\
        \( \text{E} \rightarrow \text{E} \) & 0.621 (+11.4\%) & 0.766 (+7.1\%) & 0.711 (+3.7\%) & 0.811 (+8.2\%) \\
        \( \text{D} \rightarrow \text{D} \) & 0.612 (+9.8\%) & 0.759 (+6.1\%) & 0.695 (+1.4\%) & 0.835 (+11.4\%) \\
        \( \text{B} \rightarrow \text{B} + \text{E} \rightarrow \text{E} + \text{D} \rightarrow \text{D} \)  & 0.593 (+6.4\%) & 0.744 (+4.0\%) & 0.684 (-0.1\%) & 0.817 (+9.0\%) \\
        \( \text{B} \rightarrow \text{B} \)+\( \text{E} \rightarrow \text{E} \)           & 0.610 (+9.5\%) & 0.757 (+5.8\%) & 0.718 (+4.1\%) & 0.801 (+6.9\%) \\
        \( \text{E} \rightarrow \text{E} \)  +  PMD           & 0.616 (+10.5\%) & 0.762 (+6.5\%) & 0.722 (+5.4\%) & 0.808 (+7.8\%) \\
        \textbf{\( \text{B} \rightarrow \text{B} \) + PMD} & \textbf{0.629 (+12.9\%)} & \textbf{0.772 (+7.9\%)} & \textbf{0.721 (+5.2\%)} & \textbf{0.831 (+10.9\%)} \\
        \( \text{B} \rightarrow \text{B} \)+\( \text{E} \rightarrow \text{E} \) + PMD       & 0.607 (+8.9\%) & 0.756 (+5.7\%) & 0.694 (+1.3\%) & 0.829 (+10.6\%) \\
        \( \text{E} \rightarrow \text{E} \) + PMD           & 0.617 (+10.7\%) & 0.763 (+6.7\%) & 0.733 (+7.0\%) & 0.796 (+6.2\%) \\
        \( \text{B} \rightarrow \text{B} \)+\( \text{E} \rightarrow \text{E} \)+\( \text{D} \rightarrow \text{D} \)+PMD   & 0.604 (+8.4\%) & 0.753 (+5.3\%) & 0.702 (+2.4\%) & 0.812 (+8.4\%) \\
        \( \text{D} \rightarrow \text{D} \) + PMD           & 0.541 (-2.8\%) & 0.702 (-1.8\%) & 0.653 (-4.6\%) & 0.759 (+1.3\%) \\
        \(\textbf{E} \rightarrow \textbf{E}\)+\(\textbf{D} \rightarrow \textbf{D}\) + \textbf{PMD} & \textbf{0.628 (+12.7\%)} & \textbf{0.771 (+7.8\%)} & \textbf{0.722 (+5.4\%)} & \textbf{0.828 (+10.5\%)}\\
        \bottomrule
    \end{tabular}
    \end{adjustbox}
    \label{tab:kd_comparison_data_volume}
\end{table}
The results of the study are presented in Table~\ref{tab:kd_comparison_data_volume}. The best performance within each experiment category is highlighted in bold. Interestingly, a simple bottleneck-to-bottleneck contrastive knowledge distillation method produced better results (0.617 vs. 0.615 IoU) with a student model that was trained on half the data compared to its counterpart. This suggests that using knowledge distillation with a smaller model can help to overcome data volume challenges, especially in medical imaging tasks.

\section{Statistical Significance of the Model}
\label{section:Statistical Significance of the Model} 
In analyzing the findings of our investigation, we conducted a thorough analysis to determine the impact of knowledge distillation on improving the performance of the Student model (S1). The student model (S1) baseline performance (IoU) was determined to be 0.55 in terms of the mean IoU score. While the mean IoU score of all the scales tested achieved through knowledge distillation KD(T1, S1)was notably higher at 0.603.  Variance analysis revealed that the results obtained through knowledge distillation had a significantly lower variance than those obtained through the student model, highlighting the consistency and stability achieved by this approach. The ANOVA test used to compare the two sets of results resulted in an F-statistic of 5.48 and a p-value of 0.039. With the p-value falling below the conventional significance level of 0.05, we can confidently conclude that there is a statistically significant difference in the performance of the student model and the knowledge distillation technique and is highly unlikely to be due to random chance. This finding emphasizes the significance and efficacy of knowledge distillation as a method for improving model performance in my thesis research.

\section{Summary}
In this chapter,  we delved into a comprehensive analysis of experiments conducted to support our hypothesis made at the beginning of the thesis, focusing on the efficacy of knowledge distillation using contrastive learning. The chapter provided details on the experimental setup, including the dataset split, hardware configurations, and the neural network optimization strategies used. The choice of evaluation metrics used to assess the performance of segmentation models is also discussed in detail. The chapter also illustrates the progress of the model during training and validation through visualization of training and validation curves. These curves provide insights into the model’s training progress and ability to generalize. This is important for understanding the training dynamics and identifying potential issues such as overfitting or underfitting.

The main focus of this chapter is on the quantitative results obtained from different experiments, which involve comparing various model architectures and configurations of knowledge distillation between the teacher and the student. One important aspect that is discussed in detail is the use of teacher-student pairs with knowledge distillation, which has been shown to significantly improve segmentation performance. The results of these experiments are presented in a clear and organized manner, with a comprehensive understanding of the performance gains achieved through knowledge distillation across different evaluation metrics. The qualitative results provide additional information to supplement the quantitative analysis. They offer a visual perspective on the effectiveness of knowledge distillation in improving segmentation accuracy and reducing artefacts in segmented images. The chapter discusses the implications of these qualitative findings in detail, providing valuable insights into the practical implications of the experimental results.

Ablation studies are conducted to further investigate the impact of various factors such as multi-scale knowledge distillation, contrastive learning, and data volume on model performance. These studies shed light on how different techniques can improve segmentation performance, while also revealing insight into the underlying mechanisms driving model enhancements. Finally, the chapter concludes with a thorough analysis of the statistical significance of the model improvements achieved through knowledge distillation. This analysis provides robust evidence supporting the efficacy of knowledge distillation in enhancing segmentation results.

The experimental results strongly support our hypothesis, demonstrating that transferring knowledge from the teacher to the student network improves the student model's performance in medical image segmentation tasks. These findings support our initial assumptions about the efficacy of knowledge distillation for improving segmentation accuracy, especially in scenarios with limited data and compute availability.

\chapter{Discussion}
\label{chapter:discussion}
In this chapter, we will discuss the findings, implications, and future directions that emerged from our investigation into knowledge distillation techniques for medical image segmentation. We will analyze the key insights we gained from our experiments, shedding light on the effectiveness of multi-scale knowledge distillation, the impact of contrastive learning, and the role of data volume in shaping segmentation performance. Additionally, we will critically examine the limitations of our model and identify potential avenues for future research aimed at addressing these challenges and advancing the field of medical image analysis. We aim to provide a comprehensive understanding of the nuances, opportunities, and complexities inherent in leveraging knowledge distillation to enhance medical image segmentation accuracy.

\section{Key Findings}
Our experiments have revealed several key findings that contribute to a better understanding of knowledge distillation techniques in the context of medical image segmentation. Firstly, our ablation studies have demonstrated the effectiveness of multi-scale knowledge distillation in improving segmentation performance. By distilling knowledge from different layers of the network architecture, we observed significant gains in segmentation accuracy across various metrics. Notably, we found that encoder-to-encoder knowledge distillation emerged as the most impactful variant, highlighting the importance of leveraging high-level features for accurate segmentation. Additionally, contrastive learning played a crucial role in facilitating the knowledge transfer process. Our results have shown that leveraging contrastive learning in combination with knowledge distillation led to substantial improvements in segmentation accuracy. However, it was also evident that simultaneously distilling knowledge from all layers presented challenges, emphasizing the need for a nuanced approach to information transfer. Furthermore, our experiments revealed that effective knowledge distillation can mitigate the impact of data volume on segmentation performance. By training the student model with only half the data of the teacher model, we demonstrated that knowledge distillation can help overcome data scarcity challenges, enabling smaller models to achieve comparable or even superior performance. Finally, statistical analysis has underscored the significance of the improvements achieved through knowledge distillation. By comparing the variance and performance of the student model with and without knowledge distillation, we have established the statistical significance of our findings, providing robust evidence of the efficacy of knowledge distillation in enhancing segmentation results.

\section{Limitations}
Although our experiments have shown promising results, several limitations need to be considered. Firstly, our findings are primarily based on experiments conducted on a specific dataset and segmentation task, so generalizing these results to other domains or datasets may require additional validation. Secondly, the dataset used in our experiments may exhibit inherent biases or limitations, which could impact the generalizability of our findings to real-world scenarios with more diverse data distributions. Although there have been notable enhancements in the accuracy of segmentation and reduction of artefacts, especially with S1, some difficulties are still associated with the larger size of S2 and the optimization of the distillation network. These results illustrate the effectiveness of knowledge distillation in enhancing segmentation outcomes, highlighting the necessity for further research to tackle its limitations and optimize its performance.

\section{Future Prospects}
Looking ahead, our work opens up several avenues for future research and development. As we look forward, various avenues hold great promise for further research and development in the field of knowledge distillation for medical image segmentation. Firstly, we could explore the effectiveness of more lightweight models and conduct experiments to compare their performance against parameter-heavy architectures. This could offer valuable insights into the trade-offs between model complexity and segmentation accuracy. Secondly, we could extend the application of our proposed architecture to multi-class segmentation datasets like ACDC or Synapse. This could provide a more comprehensive understanding of its effectiveness across diverse medical imaging tasks. Additionally, we could investigate alternative types of contrastive loss functions, such as triplet loss, which could offer new ways to improve the knowledge transfer process and enhance segmentation results. Moreover, exploring novel approaches inspired by recent advancements in multimodal learning, like CLIP-like architectures or vision-language models, could yield innovative strategies for distilling knowledge from large-scale pre-trained models into segmentation networks. Furthermore, we could experiment with techniques like SAM (Segment Anything Model) to facilitate knowledge distillation to architectures like UNet. This could lead to further improvements in segmentation performance. By embracing these diverse research directions, we can continue to push the boundaries of knowledge distillation in medical image segmentation and pave the way for more accurate, efficient, and clinically relevant segmentation models.
\chapter{Conclusion}
\label{chapter:conclusion}
In this thesis, we aimed to explore the effectiveness of knowledge distillation techniques, with a particular focus on leveraging contrastive learning, to enhance segmentation performance in the field of medical image analysis. After conducting a thorough literature review, we gained a better understanding of the previous studies done on the topic, as well as their associated challenges and limitations. This comprehensive analysis allowed us to recognize the importance of knowledge distillation in both model compression and transfer learning and as a result, we formulated our hypothesis. Our main goal was to tackle the challenge of improving segmentation accuracy while reducing the computational and resource requirements of deep learning models used in this domain. We proposed the idea that combining knowledge distillation and contrastive learning techniques can be an effective way to transfer knowledge from a larger teacher model to a smaller student model, thus leading to better segmentation accuracy. 

We proposed an architecture that combined multi-scale knowledge distillation, progressive maps distillation (PMD), and supervised contrastive learning techniques. Our experiments showed some important results. Firstly, multi-scale knowledge distillation improved segmentation accuracy in different configurations, especially when distilling information from individual layers like bottleneck, encoder, and decoder layers. Encoder-to-encoder distillation was the most effective variant, which emphasized the significance of utilizing feature representation at different network depths.

We conducted a study to assess the effectiveness of using knowledge distillation techniques to improve segmentation performance. We explored different combinations of teacher and student networks through an extensive literature review. We found that the common belief that larger teacher networks always lead to better performance is not necessarily true. Our experiments showed that sometimes less is more when it comes to knowledge distillation. We conducted a comprehensive assessment of different teacher and student network architectures, including fully convolutional (UNet) and transformer-based (TransUNet) networks. The results of our evaluation were unexpected as we found that simply having a larger teacher network did not necessarily result in an improved performance of the student model. Interestingly, in some cases, the performance of the student model even declined or remained the same when trained using knowledge distilled from a larger teacher network. The experiments we conducted have shown that simply choosing larger teacher networks may not always lead to the desired improvements. It is important to find a balance between model complexity and computational efficiency and to use knowledge distillation techniques carefully to achieve the best segmentation performance. These findings emphasize the need for a thoughtful approach to network design and training.

We tested the effectiveness of progressive map distillation (PMD) in improving segmentation performance. Different configurations were tried, and while some were more successful than others, a bottleneck-to-bottleneck knowledge distillation using PMD between two UNet models yielded the best results. This is because the bottleneck layer captures rich semantic information. Interestingly, we found that a smaller model trained with knowledge distillation can outperform a larger one trained on a larger dataset. This indicates that knowledge distillation has the potential to mitigate challenges posed by limited data in medical imaging tasks. We conducted several experiments and ablation studies to verify this hypothesis and gain a deeper understanding of the factors contributing to model improvements.

Statistical analysis confirmed the effectiveness of knowledge distillation. It demonstrated a consistent and stable approach that outperformed the baseline student model. Our comprehensive quantitative and qualitative studies provide compelling evidence to support our hypothesis. The detailed analysis of experimental results illuminates the efficacy of knowledge distillation techniques in improving segmentation accuracy and reducing artefacts in segmented images.

In summary, the success of our thesis is attributed to the validation of our hypothesis and the creation of a new efficient method to improve the segmentation performance in limited data and compute scenarios. We have addressed the challenges associated with computational resources and data volume by using knowledge distillation techniques, specifically contrastive learning. Our research has contributed to a better understanding of the mechanisms that drive model enhancements and has practical implications for the field of medical imaging. Overall, this thesis represents a significant advancement in segmentation techniques and holds great promise for future research in this area.
\nocite{*} 
\printbibliography[heading=bibintoc] 

\end{document}